\documentclass[%
reprint,
superscriptaddress,
nofootinbib,
 amsmath,amssymb,
 prx,
]{revtex4-2}
\usepackage[utf8]{inputenc} 
\usepackage{graphicx} 
\usepackage{float}
\usepackage{braket}
\usepackage[T1]{fontenc}
\usepackage{xcolor}
\usepackage{hyperref}
\usepackage{bbold}
\usepackage{url}
\usepackage{siunitx}
\usepackage{makecell}
\newcommand{\calciumforty}{$^{40}\mathrm{Ca}^+$}

\usepackage{tikz}
\usetikzlibrary{positioning}
\usetikzlibrary{shadings}

\usepackage[normalem]{ulem}

\let\soutold\sout
\renewcommand{\sout}[1]{\textcolor{red}{\soutold{#1}}}

\begin{document}

\title{Ion trap on borosilicate substrate with integrated femtosecond-laser-written waveguide}

\author{Jakob Wahl}
\altaffiliation[]{These authors contributed equally to this work.}
\affiliation{Institut f\"{u}r Experimentalphysik, Universit\"{a}t Innsbruck, Technikerstraße 25/4, 6020 Innsbruck, Austria}
\affiliation{Infineon Technologies Austria AG, Siemensstraße 2, 9500 Villach, Austria}

\author{Alexander Zesar}
\altaffiliation[]{These authors contributed equally to this work.}
\affiliation{Infineon Technologies Austria AG, Siemensstraße 2, 9500 Villach, Austria}
\affiliation{Institut f\"{u}r Physik, Universität Graz, Universit\"{a}tsplatz 5, 8010 Graz, Austria}

\author{Philipp Hurdax}
\affiliation{MATERIALS - Institute for Sensors, Photonics and Manufacturing Technologies, JOANNEUM RESEARCH, Franz-Pichler-Straße 30, 8160 Weiz, Austria}

\author{Marco Schmauser}
\affiliation{Institut f\"{u}r Experimentalphysik, Universit\"{a}t Innsbruck, Technikerstraße 25/4, 6020 Innsbruck, Austria}

\author{Victoria Schwab}
\affiliation{Institut f\"{u}r Experimentalphysik, Universit\"{a}t Innsbruck, Technikerstraße 25/4, 6020 Innsbruck, Austria}
\affiliation{Infineon Technologies Austria AG, Siemensstraße 2, 9500 Villach, Austria}

\author{Michael Pasquini}
\affiliation{Institut f\"{u}r Experimentalphysik, Universit\"{a}t Innsbruck, Technikerstraße 25/4, 6020 Innsbruck, Austria}

\author{Marco Valentini}
\affiliation{Institut f\"{u}r Experimentalphysik, Universit\"{a}t Innsbruck, Technikerstraße 25/4, 6020 Innsbruck, Austria}

\author{Clemens Rössler}
\affiliation{Infineon Technologies Austria AG, Siemensstraße 2, 9500 Villach, Austria}

\author{Thomas Monz}
\affiliation{Institut f\"{u}r Experimentalphysik, Universit\"{a}t Innsbruck, Technikerstraße 25/4, 6020 Innsbruck, Austria}
\affiliation{Alpine Quantum Technologies GmbH, Technikerstraße 17/1, 6020 Innsbruck, Austria}

\author{Bernhard Lamprecht}
\affiliation{MATERIALS - Institute for Sensors, Photonics and Manufacturing Technologies, JOANNEUM RESEARCH, Franz-Pichler-Straße 30, 8160 Weiz, Austria}

\author{Klemens Schüppert}
\affiliation{Infineon Technologies Austria AG, Siemensstraße 2, 9500 Villach, Austria}

\author{Philipp Schindler}
\affiliation{Institut f\"{u}r Experimentalphysik, Universit\"{a}t Innsbruck, Technikerstraße 25/4, 6020 Innsbruck, Austria}

\begin{abstract}
We present an ion-trap platform on borosilicate glass with an integrated femtosecond-laser-written waveguide 
for on-chip light delivery. 
The optical layer is physically separated from the electrode substrate and bonded atop the trap, 
remaining compatible with silicon-based integration. 
We engineer single-mode low-loss guidance at 729 nm 
with tunable
 mode-field diameter  and achieve low-loss curved waveguides down to
a  radius of curvature of $6~\si{\milli\meter}$.
We also extend single-mode operation to a wavelength  of 405 nm. 
The fabrication process is
compatible with the industrial fabrication of
a single-metal-layer surface-electrode trap, including active fiber alignment and bonding.
We validate the platform in a cryogenic trapped-ion system with $^{40}$Ca$^+$, demonstrating trapping, 
shuttling the ion 
to a zone in front of the waveguide,
and coherent operations driven by 729 nm light delivered through the integrated waveguide. 
We characterize the effect of the exposed dielectric on the ion and measure stray electric fields that show slow drift at a timescale of hours.
The architecture is compatible with hybrid micro-optics (e.g. pick-and-place lenses) to realize single ion addressing and provides a robust, scalable route to integrated light delivery for trapped-ion devices.

\end{abstract}
\maketitle
\tableofcontents

\section{Introduction}
\label{sec: introduction}

\begin{figure*}[ht!]
  \centering
    \begin{tikzpicture}
        \coordinate (xyz) at (3.6,-3);
        \node[inner sep=0pt] (image) at (0,0)
            {
    	\includegraphics[width=0.55\textwidth]{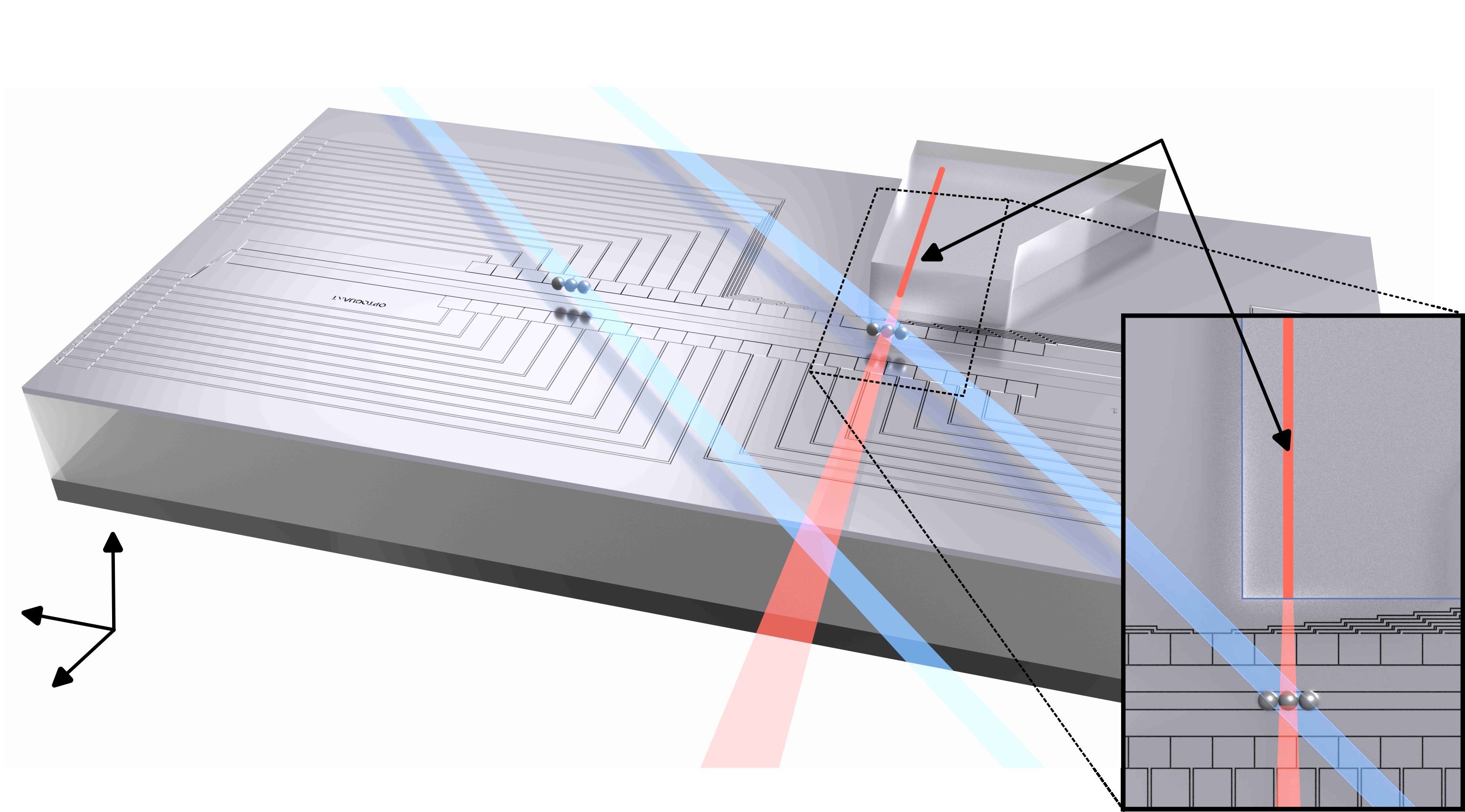}
            };
        \node[align=center] at (2.9,2) {waveguide};
        \node[align=center] at (3.75,-3) {addressing zone};
        \node[align=center] at (-1,1.2) {trapping zone};
        \node[align=center] at (-4.75,-1.59) {x};
        \node[align=center] at (-4.35,-2.) {y};
        \node[align=center] at (-3.95,-1.0) {z};
\end{tikzpicture}
	\caption{
    \label{fig:introduction: trap render}
    Concept of the ion trap. The aluminum trap electrodes are patterned on a borosilicate substrate. 
    Patterned segmented DC electrodes allow for shuttling of the ion between the 
    trapping and the addressing zone. 
    On top of the traps' surface,  another borosilicate glass block is bonded 
    that contains a femtosecond laser-written waveguide for light with $729\si{\nano \meter}$ wavelength (red beam),
    inscribed at a distance to the trap surface that is in-plane with the trapping position of the ion.
    The geometry and size of the optics block is chosen such that ions can be trapped in the \textit{addressing zone} in front of the waveguide, 
    but also at a position millimeter distances away from the glass, 
    allowing for trapping without the influence of the dielectric optics block. 
    The shape of the optics block also allows for free-space laser access at 45° (blue beam), which we use 
    for all wavelengths except the integrated qubit laser, which uses the integrated light path.
	}
\end{figure*}
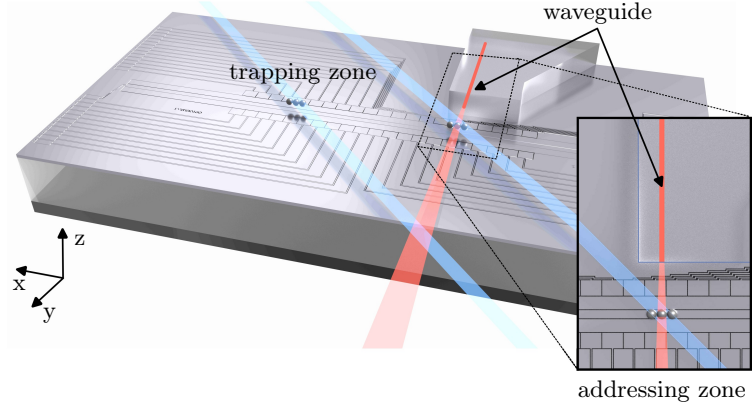

Trapped-ion systems are a proven platform for the realization of universal quantum computers~\cite{Bruzewicz2019}. 
Recent work has demonstrated long coherence times~\cite{Wang2021} and high two-qubit gate fidelities~\cite{ionq2025, smith2025}. 
Currently, trapped-ion systems routinely work with tens  to nearly one hundred of ion qubits, 
and gate operations are predominantly performed using free-space laser light~\cite{quantinuum2025, Wang2021, Pogorelov2021}.
However, useful quantum computation requires scaling to much larger qubit numbers, 
on the order of thousands~\cite{Proctor2025}.

Microfabricated surface ion traps enable to scale these systems to accommodate larger qubit arrays~\cite{kielpinski2002, pino2021}. 
By leveraging mature microfabrication processes, it is possible to achieve the high electrode 
densities and control structures required to host thousands of ions. 
Nevertheless, scaling these chip architectures to large numbers of ions introduces critical engineering challenges for electrical signal routing and optical light delivery.
The challenge for electrical signal routing lies in providing the increasing number of signals required for ion shuttling and micromotion compensation. 
On-chip routing can be relatively straightforwardly implemented with multiple metal layers on the trap chip, but the connection to the signal sources will require the integration of application specific integrated circuits (ASICs)~\cite{malinowski2023} into the ion trap package.

Here, we will focus on integrated optical signal routing,
that ideally allows one to address individual ions in a crystal. This requires
spot sizes on the order of one micrometer, which have been realized with 
focusing optics with a numerical aperture (NA) of approximately 0.2 to 0.3~\cite{Pogorelov2021}. 
Using free-space optics, this necessitates a dedicated objective for each gate zone, 
and the required optical access cannot be provided as qubit numbers increase. 
Integrating light delivery directly into the ion trap structure via waveguides provides means to circumvent these geometrical constraints. 
Furthermore, integrating optics and electrodes on a single substrate ensures that mechanical vibrations act as a common-mode displacement, intrinsically stabilizing laser-ion alignment~\cite{Niffenegger2020}.

State-of-the-art integrated light delivery predominantly relies on planar photonic circuits fabricated beneath the trap electrodes~\cite{Mehta2016, Niffenegger2020}. 
Addressing ions trapped above 2D lithographic waveguides requires out-of-plane focusing elements, 
such as diffractive grating couplers~\cite{Mehta2016}
or microfabricated mirrors with integrated lenses~\cite{badawi2025}.
These out-of-plane coupling elements introduce substantial fabrication challenges. 
Grating couplers rely on periodic structures requiring lithographic resolutions and etching tolerances 
on the order of tens of nanometers to achieve high out-coupling efficiency~\cite{Momenzadeh2026}. 
Furthermore, grating couplers have limited bandwidth which requires multiple grating couplers to deliver all necessary wavelengths. Thus, multi-color optical addressing with gratings requires a large surface area on the trap chip~\cite{Mehta2020}. 

In this work, we propose a scalable platform based on a material stack of multiple borosilicate glasses 
with a silicon wafer on the backside.
The trap, illustrated in Fig.~\ref{fig:introduction: trap render}, 
features a lower substrate consisting of a silicon wafer with a borosilicate glass layer on top,
upon which the surface trap electrodes are patterned. 
Bonded directly onto the trap's surface is a second, 
structured block of borosilicate glass dedicated to optical delivery.
The optics block features integrated 
femtosecond-laser-written (FLW) waveguides~\cite{Eaton2011}. 
These waveguides are formed by tightly focusing a femtosecond laser pulse into the substrate, 
which induces a  local, permanent increase in the refractive index. 
These structures are not restricted to a planar 2D layer but can be 
routed through the bulk material in all three dimensions~\cite{Guan2014}.
In our architecture, the FLW waveguides are inscribed such that they run parallel to the trap surface and emit light at the height of the trapped ion, avoiding the need for out-of-plane couplers. 
While these waveguides can be combined with micro-assembled, 
pick-and-place lenses~\cite{Gruenberg2023} to form a tightly focused addressing spot, 
this work focuses on demonstrating and characterizing the integrated waveguides themselves.

This substrate stack addresses the previously discussed scaling bottlenecks. 
The electrical routing challenge can be addressed with the aid of integrated
ASICs, that
reside on a separate substrate and are connected to the trap 
via through substrate vias~\cite{badawi2025}. 
Our choice of glass substrate material, 
Borosilicate glass,  features a coefficient of thermal expansion (CTE) 
that is closely matched to silicon, allowing for  wafer-level anodic bonding 
and paving the way for direct integration with underlying silicon ASICs. 
Furthermore, ion traps fabricated on a silicon-borosilicate 
stack have proven mechanical stability during cryogenic cooldown~\cite{Auchter2022}.
Second, to address the challenge of optical integration of ion traps, we integrate the light delivery 
by writing FLW waveguides in the borosilicate glass.
We physically separate the optical delivery layer from the trap substrate, 
and by inscribing the waveguides at the height of the ion, 
we eliminate the need for out-of-plane couplers. 

In this paper, we establish the end-to-end functionality of this borosilicate 
platform within a cryogenic trapped-ion system:
In Sec.~\ref{sec: waveguides}, 
we discuss and characterize the engineering of FLW optical waveguides, 
and show that the waveguides can be designed to carry relevant wavelengths for trapping and control of 
\calciumforty ions by tuning the writing parameters. 
In Sec.~\ref{sec: fabrication}, 
we detail the ion trap design process and investigate the impact of the integrated dielectric block on the trapped ion
using finite-element method (FEM) simulations. 
Following this theoretical modeling, we outline a waveguide and trap fabrication and integration process flow which is compatible with ASIC integration. 
We present measured data characterizing the trap's stray electric fields and heating rates to validate its performance. 
In Sec.~\ref{sec: measurements}, 
utilizing an actively aligned optical fiber bonded to the trap chip, 
we route qubit light with a wavelength of $729~\si{\nano\meter}$ from the room-temperature environment directly to the evacuated cryogenic trap. 
We demonstrate the trapping of a \calciumforty ion, 
shuttling into the dedicated optical addressing zone, 
and the successful driving of coherent qubit dynamics via carrier Rabi oscillations using laser light delivered directly through the on-chip integrated optics.

\section{Femtosecond laser written waveguides}
\label{sec: waveguides}

FLW waveguides are optical structures fabricated by tightly focusing ultrashort laser pulses into transparent dielectric materials such as glasses or crystals. In these materials, linear absorption at the writing wavelength is negligible; instead, energy deposition occurs via nonlinear absorption processes confined to the focal region. This leads to highly localized structural modification of the material, typically involving densification, rarefaction, or microstructural damage, which can translate into a permanent change of the local refractive index.

Depending on the material response, femtosecond-laser exposure in glasses can induce both positive and negative refractive index change. While a negative index change requires the construction of waveguides using cladding-like structures, a positive index change enables direct inscription of waveguide cores in a single writing operation. Such positive refractive index changes are frequently achieved in glasses, with reported values in the order of $10^{-3}$--$10^{-2}$~\cite{Ari2025}. These values are sufficient to support weakly guided modes, but are generally associated with significant radiation losses in curved waveguides~\cite{SnyderLove1983}.

A key advantage of this fabrication approach is its inherent three-dimensionality. Waveguides can be embedded at arbitrary depths within the bulk material without the need for lithographic masks or cleanroom processing. This maskless direct-writing capability enables flexible and rapid prototyping of integrated photonic structures in a wide range of dielectric substrates, while maintaining compatibility with standard optical glasses.

In this work, FLW waveguides are used as integrated optical interfaces for trapped-ion quantum systems. The objective is to realize low-loss single-mode waveguides at the relevant wavelengths of 405~nm and 729~nm at a well-defined trapping depth, while enabling efficient in- and out-coupling to trapped ions.

\subsection{Experimental setup and waveguide fabrication}
\label{subsec: waveguides experimental}

Waveguides were fabricated in borosilicate \SI{400}{\micro\meter}- thick glass substrates (SCHOTT BF33) using a femtosecond laser direct-writing system (Spectra-Physics Spirit). The laser provides radiation at 1040 nm and its second harmonic at 520 nm. All results presented in this work were obtained using the 520 nm wavelength. The system allows tuning of the repetition rate from the single-shot regime up to 1 MHz and pulse durations between 209 fs and several picoseconds. All waveguides presented in this work were fabricated using circular polarization of the writing beam, a pulse duration of 209 fs, and a repetition rate of 1 MHz.

\begin{figure}[ht!]
  \centering
  \includegraphics[width=0.5\textwidth]{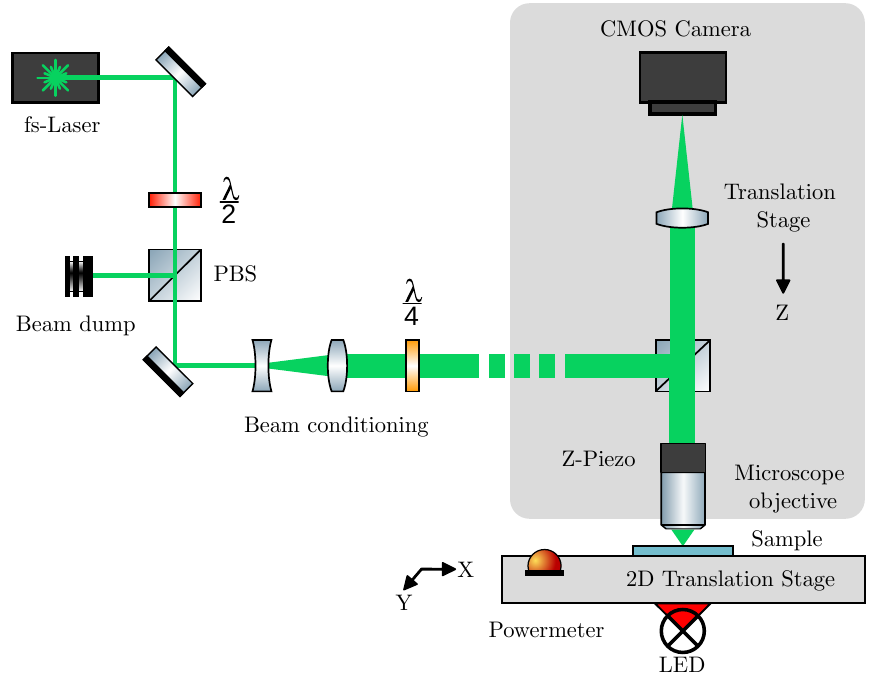}
	\caption{Schematic of the femtosecond laser setup used for waveguide writing and laser ablation.
	}
  \label{fig:waveguides:setup}
\end{figure}

As shown schematically in Fig.~\ref{fig:waveguides:setup}, the writing beam was routed through a modular free-space beam delivery system. An electronically controlled attenuation of the pulse energy at the laser source was combined with a polarization-based variable attenuator (half-wave plate and polarizing beam splitter (PBS) with beam dump) in the beam path. The latter was used to coarsely reduce the pulse energy, enabling finer and more stable control via electronic adjustment. The beam diameter was expanded prior to focusing to approximately match the entrance pupil of the objective. 
A quarter-wave plate was used to generate circular polarization at the sample.

The beam  is focussed into the glass using a microscope objectives with numerical apertures of 0.5 and 0.6 (Zeiss Epiplan Neofluar and Zeiss LD Plan-Neofluar), the latter equipped with an adjustable correction collar for spherical aberration compensation. Waveguides were written by translating the sample along predefined trajectories using a high-precision x–y translation stage (Newport XML210) with an uni-directional repeatability of \SI{0.05}{\micro\meter}  at a constant depth of \SI{170}{\micro\meter} below the glass surface, which corresponds to the ion height in our trap design (see section~\ref{subsec: design}). Writing paths were generated and executed via a LabVIEW-based control interface with PVT (position–velocity–time) motion profiles enabling continuous trajectory execution. 
For sample positioning and alignment, a small fraction of light reflected from the sample surface was imaged onto a CMOS camera, providing real-time visual feedback for locating the surface and aligning the writing position.

We apply femtosecond-laser-assisted cleaving~\cite{Li2019}
to provide well defined surfaces for light in- and out-coupling: 
two parallel tracks were written across the substrate surface by laser ablation, removing material and forming narrow grooves with a typical depth of approximately \SI{80}{\micro\meter}. The surrounding laser-modified region acts as predefined fracture guide, enabling controlled cleaving along the written line. Manual cleaving was then performed along the trenches to obtain clean fracture surfaces and well-defined waveguide facets, as demonstrated in section~\ref{subsec: nosetip characterization}. For optical characterization, samples were cleaved to produce waveguides with a length of 10 mm. Ablation was performed using a repetition rate of approximately 125 kHz, pulse energies approximately \SI{5}{\micro\joule}, a pulse duration of 209 fs, and 4 overscans (i.e., passes along the same trajectory).

The guided mode profiles were characterized by imaging the output facet of the waveguide onto a CMOS camera using a microscope objective with an NA of 0.5. From the recorded near-field intensity distributions, the mode-field diameter (MFD) was determined via $D4\sigma$~\cite{Siegman98}. 
The far-field divergence of the emitted beam was determined from beam profile measurements recorded at different distances from the waveguide facet using a beam profiler. From the measured beam diameters at two propagation distances, the divergence angle was extracted.
If not stated otherwise, all measurements of the transmitted mode of the waveguides in this chapter were done using light with a wavelength of 729~nm.

\subsection{Waveguide formation and guiding properties}
\label{subsec: waveguide formation}

Under a broad range of writing conditions, femtosecond-laser exposure of borosilicate glass in this work leads to light guiding in regions that are spatially displaced from the center of the laser-modified volume. Specifically, we observe that guiding can occur above and/or below the elongated modification track, giving rise to one or two distinct guiding regions separated by a distance on the order of \SI{10}{\micro\meter}. In the following, these are referred to as the \emph{upper} and \emph{lower} waveguide, respectively.

\begin{figure}[ht!]
  \centering
  \includegraphics[width=0.5\textwidth]{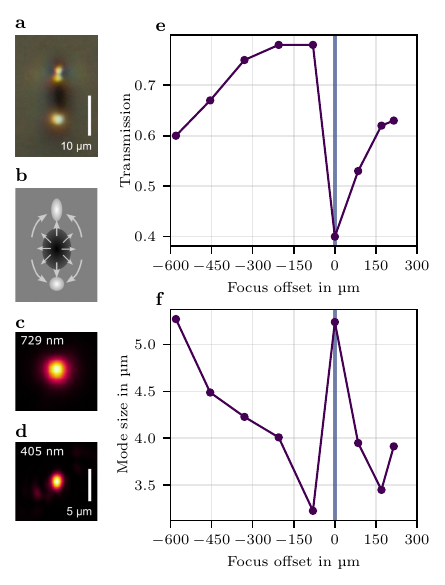}
	\caption{
    Overview of waveguide morphology, mode profiles, and writing-depth optimization in borosilicate glass.
    (a) Transmission microscope image showing two spatially separated guiding regions (“upper” and “lower”) formed within a single written track.
    (b) Schematic illustration of the inferred refractive-index distribution and dominant material redistribution during waveguide formation.
    (c,d) Near-field intensity profiles of a lower waveguide optimized and characterized at (c) 729~nm and (d) 405~nm.
    (e) Measured transmission at a wavelength of $729~\si{\nano \meter}$ as a function of the focus offset, defined as the difference between writing depth and objective correction depth. 
    (f) Effective mode diameter, defined as the diameter of a circular mode with equal effective area, plotted versus the focus offset. For each focus offset in (e) and (f), the plotted values correspond to the best guiding performance obtained by varying the pulse energy.
	}
  \label{fig:waveguides:overview}
\end{figure}

Fig.~\ref{fig:waveguides:overview}a shows a transmission microscope image of the resulting morphology, obtained under writing conditions optimized for single-mode waveguiding and high transmission. The image reveals two localized regions of enhanced light transmission separated by a darker central zone corresponding to the laser-modified volume. Depending on the chosen writing parameters, single-mode waveguiding can be realized in either one of these regions or simultaneously in both, as confirmed by the optical characterization discussed below.

Spatially displaced guiding cores have been discussed in FLW waveguides as a consequence of non-local material response to energy deposition, including temperature and stress gradients and laser-induced mass transport. In particular, Tan et al.~\cite{Tan2020} proposed a temperature-gradient-assisted writing mechanism in which a densified guiding core forms outside the region of maximum energy deposition due to thermally driven matter flow. A schematic illustration of such an off-focus core formation mechanism, adapted from Ref.~\citenum{Tan2020}, is shown in Fig.~\ref{fig:waveguides:overview}b.

While waveguiding in FLW glass structures is more commonly observed within or close to the laser-modified region itself, Ref.~\citenum{Tan2020} reports the formation of a single displaced guiding core in Eagle XG glass. The borosilicate glass structures investigated here can exhibit two vertically separated guiding regions within a single written track. 
This suggests that 
thermally driven material redistribution transfers material away from the region of strongest energy deposition. The elongated focal geometry introduces a preferred direction for this redistribution along the beam axis, resulting in net material transport toward regions above and below the center of the modification track. This leads to localized densification outside the central modified region and thereby enables the formation of spatially displaced guiding cores. The guiding regions are therefore associated with locally densified material, whereas the central track corresponds to a comparatively depleted region and is thus expected to exhibit a reduced refractive index.

Experimentally, the guiding regions support stable single-mode propagation, as verified by the invariance of the output mode profile under lateral displacement of the coupling fiber. Figs.~\ref{fig:waveguides:overview}(c)\&(d) show representative near-field intensity profiles recorded at the output facet of a lower waveguide with light with wavelengths of 729~nm and 405~nm.

The observed spatial separation between the laser-modified region and the actual guiding regions suggests that the relationship between the writing geometry and the resulting mode-field diameter is not straightforward. This raises the question to what extent the mode size can be deliberately influenced in this waveguide formation regime. In the following, we show that the mode-field diameter can nevertheless be systematically tuned by appropriate adjustment of the writing conditions.

\subsection{Mode size engineering and performance implications}

\subsubsection{Spherical aberration as a process parameter}

Spherical aberration arising from refractive index mismatch during focusing into transparent substrates depends on the writing depth and leads to a distortion of the focal volume~\cite{Hnatovsky2005}, which can significantly influence the resulting waveguide properties~\cite{Huang2016, Bisch2019}. 
We employ 
a microscope objective with adjustable aberration correction,  allowing the aberration-free focal depth to be tuned independently of the actual writing position. 
In the following, we investigate the impact of  tuning the free parameters of the aberration correction on the  waveguide formation.

We define the focus offset as the difference between the actual waveguide writing depth and the aberration-free correction depth. A non-zero focus offset leads to an increased focal volume.
Fig.~\ref{fig:waveguides:overview}(e) and (f) summarize the influence of the focus offset on the guiding characteristics of the lower waveguide, as the upper waveguide was only observed within a narrow parameter range and is therefore not considered in this analysis. The analysis was carried out using light with a wavelength of 729~nm. For each focus offset, the pulse energy was varied and the plotted values correspond to the highest observed optical transmission and the smallest achieved effective mode-field diameter at that offset. 

At nominally compensated aberration (zero focus offset), both transmission and mode confinement show a marked deterioration. This behavior is qualitatively consistent with the waveguide formation model proposed in section~\ref{subsec: waveguide formation}. As the focus offset approaches zero, aberration compensation reduces the elongation of the focal volume and thereby decreases the directional asymmetry driving material redistribution. This is expected to result in broader and less localized guiding cores, leading to lower transmission and larger MFD. However, this qualitative picture alone does not explain why the degradation remains strongly localized around zero focus offset rather than changing more gradually with focus offset.
In contrast, optimal guiding performance is generally obtained for moderate negative focus offsets, typically within a range of \SIrange{170}{240}{\micro\meter} across different writing conditions.
This behavior indicates that the aberration-induced deformation of the focal volume plays an essential role in establishing favorable conditions for waveguide formation in borosilicate glass. Deliberately introduced aberration has previously been reported to improve waveguide performance in FLW waveguides in Eagle XG glass.~\cite{Ferreira2021}. Based on the experimentally observed optimum in both transmission and mode confinement, a focus offset of \SI{80}{\micro\meter} is therefore used in the following experiments unless stated otherwise.

\subsubsection{Writing strategies for mode size tuning}

\begin{figure}[ht!]
  \centering
  \includegraphics[width=0.5\textwidth]{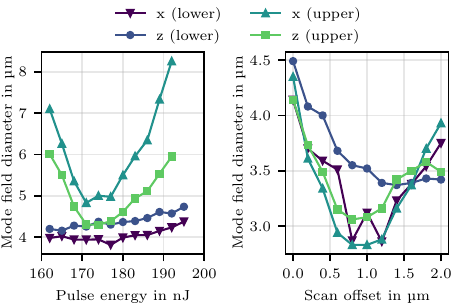}
	\caption{
    Mode field diameter of waveguides located above and below the focal plane, measured along the x- and z-directions.
    The measurement are indicated with symbols, and are connected with lines as a guide for the eye.
    (a) Mode field diameter (MFD) as a function of the pulse energy at a scan offset $\Delta x = 1.5~\si{\micro\meter}$ and 3 overscans.
    (b) MFD as a function of $\Delta x$ at a fixed pulse energy of 177 nJ and N = 8 overscans.
    All structures were written at a scan speed of 50 mm/s. 
	}
  \label{fig:waveguides:mfd-tuning}
\end{figure}

During process development, an extensive parameter space was explored, 
including variations in pulse duration, repetition rate, polarization, numerical aperture, beam overfilling, writing speed, number of overscans, and correction collar position. 
An optimized parameter set was identified to achieve robust,
low-loss single-mode guidance for efficient coupling into the ion trap chips.
This configuration yields stable single-mode propagation with high relative transmission, typically on the order of 80\%.
In the following, we focus on those parameters that were found to exert the most direct influence on the mode-field diameter.
Fig.~\ref{fig:waveguides:mfd-tuning}a shows the measured MFD as a function of pulse energy for both upper and lower waveguides. The MFD was extracted independently along the two transverse axes, corresponding to the directions perpendicular to the writing direction. All data points shown correspond to waveguides with consistently high transmission (on the order of $70\,\%$ or higher when accounting for mode-matching losses).

For both waveguide types, a clear dependence of the MFD on pulse energy is observed. In particular, the upper waveguide exhibits a pronounced dependence of the mode field diameter on the pulse energy, with the smallest mode-field diameters occurring at 171-177~nJ, while larger values are found toward both lower and higher energies. The lower waveguide exhibits a slightly increased mode field diameter towards increasing pulse energy.

The lower waveguide shows a similar qualitative trend, albeit with a reduced tuning range. Here, the mode-field diameter increases slightly with pulse energy and remains systematically larger than that of the upper waveguide over the investigated parameter range.
This behavior indicates that the degree of mode confinement is sensitive to the strength of the laser-induced modification, even though the guiding region itself is spatially separated from the focal volume. Importantly, the observed MFD values fall in a range that is smaller than those typically reported for FLW waveguides in borosilicate glasses~\cite{Eaton2008a, Eaton2008b, Chen2008}, underscoring the strong localization achievable with the present writing approach.

Analogous to pulse energy, the relative positioning of successive overscans can control the deposited energy per volume, providing an additional degree of control over the guided mode. 
We investigate this effect, by introducing a lateral offset $\Delta x$ between individual overscans, such that the modification tracks were distributed over a defined transverse extent. Here, $\Delta x$ denotes the total lateral span covered by the overscans rather than the incremental shift between successive scans. Intuitively, increasing the lateral offset might be expected to broaden the effective modification region and thus lead to a larger mode-field diameter. However, the experimental results reveal a more complex behavior. As shown in Fig.~\ref{fig:waveguides:mfd-tuning}b, the MFD initially decreases with increasing $\Delta x$, reaching values as small as approximately \SI{3}{\micro\meter} for offsets in the range of $1$–\SI{1.5}{\micro\meter}. Only for larger offsets does the mode-field diameter increase again.

This non-monotonic dependence is observed consistently for the upper waveguide and, to a lesser extent, for the lower waveguide. The pronounced reduction of the MFD at intermediate offsets highlights that lateral offset writing can enhance mode confinement beyond what is achieved by pulse energy tuning alone.
A possible interpretation of this behavior is that small lateral offsets during the writing process lead to a more extended positive refractive-index region, effectively increasing the spatial extent of the guiding region. For very small guiding regions, the optical mode is expected to extend substantially beyond the modified volume, resulting in comparatively large mode-field diameters. Increasing the effective size of the guiding region therefore initially improves the confinement of the mode and reduces the mode-field diameter.

For larger offsets, however, the spatial extent of the modified region becomes comparable to or exceeds the intrinsic mode size supported by the refractive-index contrast, such that further increases in the guiding-region size tend to lead to an increase of the mode-field diameter itself. While other effects, such as changes in refractive-index contrast caused by material transport or stress redistribution, may also contribute, the observed behavior is consistent with standard weak-guidance waveguide theory~\cite{Marcuse1978, SnyderLove1983}. In this picture, the smallest mode-field diameter is obtained when the effective guiding region is comparable to the intrinsic modal extent supported by the refractive-index contrast. This provides a simple qualitative picture for mode-field engineering in laser-written waveguides.

Overall,
the combination of pulse energy and lateral-offset writing provides controlled
tuning of the mode-field diameter over a wide range,
from approximately \SIrange{3}{8}{\micro\meter} under the present conditions.
This tunability 
provides the basis for systematically investigating how waveguide properties
depend on the mode-field diameter.
In the following,
we use this capability to study 
far-field divergence in Sec.~\ref{subsubsec: farfield}
and 
bending loss in Sec.~\ref{subsubsec: bendingloss}.

\subsubsection{Far-field divergence versus mode size}
\label{subsubsec: farfield}

\begin{figure}[ht!]
  \centering
  \includegraphics[]{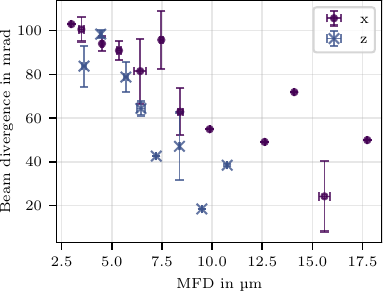}
	\caption{
    Beam divergence as a function of the mode field diameter (MFD) for the two transverse directions.
    The data represents multiple measurements  within  
    an MFD of $1~\si{\micro\meter}$ and indicates the mean value and standard error.
	}
  \label{fig:waveguides:divergence}
\end{figure}

In the presented experiments, the trapped ions are addressed via the waveguide's far-field emission, we investigated how the mode-field diameter affects the far-field characteristics of the emitted beam 
by measuring the divergence of waveguides with different mode sizes.
Smaller guided modes are expected to exhibit larger far-field divergence due to the inverse relationship between spatial confinement and angular spread. Fig.~\ref{fig:waveguides:divergence}(a) summarizes the resulting relationship between the mode-field diameter and the beam divergence angle. 
The measured divergence ranges from 84-103~mrad for the smallest investigated modes 
between 
$2.5~\si{\micro\meter}$
and
$5~\si{\micro\meter}$
to about 40-60~mrad for mode-field diameters of about \SI{10}{\micro\meter}. 
Beyond this point, a further increase in the mode-field diameter does not lead to a further notable reduction in the beam divergence. This saturation of the beam divergence at larger mode-field diameters coincides with a progressive deviation of the near-field mode profile from an ideal Gaussian mode structure
and is attributed to the mode expanding into the surrounding laser-modified refractive index landscape, which distorts the field profile.

The far-field intensity distributions of the waveguide modes exhibit a consistent qualitative behavior: 
less divergent modes show a compact central intensity peak with weak, diffuse surrounding features, whereas more strongly divergent modes show increasingly extended low-intensity structures around the main peak, whereas with increasing divergence this separation becomes less distinct and the intensity distribution evolves into a single broader peak with extended tails. 
These results demonstrate that the far-field divergence can be tuned over a substantial range through controlled adjustment of the MFD. 
This behavior is relevant for ion addressing, since it determines the effective spatial mode profile at the ion position.
The measurement indicates that,
for an ion trap without focusing optics,
where smaller divergence is preferred,
the divergence can be tuned through the MFD up to approximately
$10~\si{\micro\meter}$,
above which larger MFDs do not further reduce the divergence.

\subsubsection{Bending loss versus mode size}
\label{subsubsec: bendingloss}

The mode-field diameter not only determines the divergence of the emitted beam but also strongly influences the sensitivity of a waveguide to bending-induced loss, i.e., radiation losses arising from curvature of the propagation path. Larger optical modes extend further into the surrounding lower-index material and are therefore more susceptible to radiation losses in curved waveguide sections, whereas stronger mode confinement improves bend tolerance~\cite{Marcuse1976}. 
Small bending radii are particularly relevant for integrated photonic layouts, where waveguides must be routed within a limited footprint.

\begin{figure}[ht!]
  \centering
  \includegraphics[width=0.5\textwidth]{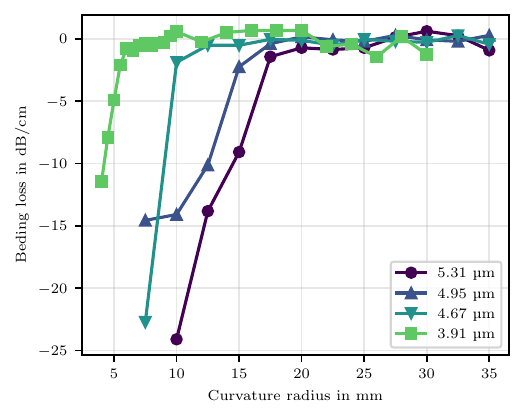}
	\caption{Bending-induced loss of the lower laser-written waveguide as a function of the bending radius for different mode field diameters (MFDs).
	}
  \label{fig:waveguides:bending-loss}
\end{figure}

Waveguides were written along curved trajectories with constant radius of curvature
to investigate the influence of mode size on bending-induced loss. 
The bending radius was varied for several sets of waveguides with different mode-field diameters obtained by adjusting the writing parameters. For each set, the bending loss was determined by comparing the transmission of curved waveguides to that of straight reference waveguides fabricated under otherwise identical conditions. Fig.~\ref{fig:waveguides:bending-loss} summarizes the resulting losses as a function of the bending radius. 

As expected, for large bending radii, the measured losses remain 
below $1~\si{dB\per\centi\meter}$
for all investigated mode-field diameters. 
However, transmission drops rapidly due to bending-induced radiation losses below a critical radius that depends strongly on the mode-field diameter.
For waveguides with a mode-field diameter of  
$5.3~\si{\micro\meter}$ pronounced bending losses appear at radii around $17~\si{\milli\meter}$, 
whereas structures with smaller mode-field diameters of $3.9~\si{\micro\meter}$ exhibit 
bending losses 
below $1~\si{dB\per\centi\meter}$
down to radii of $\approx6~\si{\milli\meter}$.

The achievable bending radii observed here are 
smaller than those commonly reported for FLW single-mode waveguides in glass, 
where characteristic bending radii on the order of several tens of millimeters are reported~\cite{Arriola2013, Dyakonov2016, Wang2024}. Even smaller bend radii have recently been demonstrated using specially engineered multi-pass waveguide morphologies designed for enhanced confinement~\cite{RossAdams2024}. The present results highlight the benefit of strong mode confinement for realizing more compact waveguide geometries in integrated photonic devices.

\subsubsection{Single-mode guiding at 405 nm}

Single-mode waveguiding over a wide range of the spectrum is of particular relevance for integrated ion trap quantum processors, where laser light in the blue and violet spectral range is commonly used for Doppler cooling, state preparation, and fluorescence readout of trapped ions. Integrating such wavelengths in on-chip photonic routing structures is therefore an important requirement for scalable ion trap platforms.

However, achieving single-mode guidance at short wavelengths is challenging for FLW waveguides in glass. For a given refractive-index contrast, decreasing the wavelength increases the number of supported spatial modes, which leads to multimode propagation unless the waveguide dimensions are sufficiently small. The comparatively small mode-field diameters obtained with the present writing strategy therefore suggest that single-mode guiding may remain possible even in the blue spectral range. 
We examine waveguides at a wavelength of $405~\si{\nano\meter}$. 
to investigate this possibility.

Parameter sets yielding small mode-field diameters at 
$729~\si{\nano\meter}$.
were used as a starting point and subsequently refined to identify conditions supporting stable single-mode propagation at 
$405~\si{\nano\meter}$.
Single-mode guiding at $405~\si{\nano\meter}$ was observed for several parameter sets, as illustrated by the near-field intensity profile shown in Fig.~\ref{fig:waveguides:overview}(d). The measured mode profile remains stable under lateral displacement of the coupling fiber, indicating that the observed output corresponds to a guided single mode. The corresponding fabrication parameters 
that differ from the ones used in Sec.~\ref{subsec: waveguides experimental} are
a 
pulse energy of  
$180~\si{\nano J}$, 
a
writing speed of 
$35~\si{\milli\meter\per\second}$.
We do one overscan and use the 
objective Zeiss Epiplan-Neofluar 20x 0.5.
These results show that the strong mode confinement achievable with the present writing strategy allows single-mode guiding to be extended into the blue spectral range.

\section{Surface ion trap}
\label{sec: fabrication}

After describing the waveguides used to address the ion in the previous section, 
we introduce the ion trap that was designed to host the waveguide.
First, we describe how we designed the structure of the electrodes
that are responsible for generating the trapping potential. 
After that, we describe how the ion trap was fabricated at the semiconductor facilities 
at Infineon Technologies Austria AG.

\subsection{Design and layout}
\label{subsec: design}

The  purpose of the presented ion trap is to  demonstrate interaction of the integrated femtosecond laser written waveguide with the ion. 
We are thus using a simple stack where a optics block with a single waveguide is bonded on top of the surface ion trap.
To mitigate risks arising from prior unknown 
influence  of the glass block, the optics block covers only a small portion of the trap's surface,
allowing for
trapping away from to the glass. 
The shape of the optics block is chosen, such that 
free space laser access to an ion that is trapped in the addressing zone is enabled,
as shown in 
Fig.~\ref{fig:introduction: trap render}. 

The waveguide  height inside the optics block is matched to the ion height of 
$x_0 = 170\,\mu\text{m}$. 
This height was chosen to be sufficiently large to prevent the waveguide output from clipping the trap surface but
low enough to keep the required trapping voltages below $40~\si{\volt}$, as this is the limit
of the experiment control system (see appendix~\ref{app: electrode layout}). 

\subsubsection{Electrode geometry}

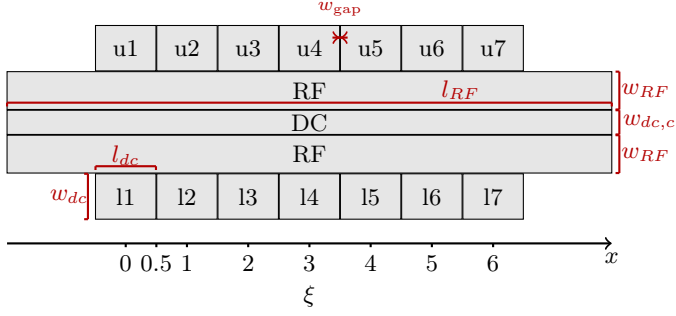
\begin{figure}[ht!]
  \centering
  \begin{tikzpicture}[scale=1]

  \draw[fill=gray!20, draw=black] (0.400,0.680) -- (0.400,1.280) -- (-0.400,1.280) -- (-0.400,0.680) -- cycle;
  \node at (0.000,0.980) {\small u4};
  \draw[fill=gray!20, draw=black] (1.210,0.680) -- (1.210,1.280) -- (0.410,1.280) -- (0.410,0.680) -- cycle;
  \node at (0.810,0.980) {\small u5};
  \draw[fill=gray!20, draw=black] (2.020,0.680) -- (2.020,1.280) -- (1.220,1.280) -- (1.220,0.680) -- cycle;
  \node at (1.620,0.980) {\small u6};
  \draw[fill=gray!20, draw=black] (2.830,0.680) -- (2.830,1.280) -- (2.030,1.280) -- (2.030,0.680) -- cycle;
  \node at (2.430,0.980) {\small u7};
  \draw[fill=gray!20, draw=black] (-0.410,0.680) -- (-0.410,1.280) -- (-1.210,1.280) -- (-1.210,0.680) -- cycle;
  \node at (-0.810,0.980) {\small u3};
  \draw[fill=gray!20, draw=black] (-1.220,0.680) -- (-1.220,1.280) -- (-2.020,1.280) -- (-2.020,0.680) -- cycle;
  \node at (-1.620,0.980) {\small u2};
  \draw[fill=gray!20, draw=black] (-2.030,0.680) -- (-2.030,1.280) -- (-2.830,1.280) -- (-2.830,0.680) -- cycle;
  \node at (-2.430,0.980) {\small u1};
  \draw[fill=gray!20, draw=black] (0.400,-1.280) -- (0.400,-0.680) -- (-0.400,-0.680) -- (-0.400,-1.280) -- cycle;
  \node at (0.000,-0.980) {\small l4};
  \draw[fill=gray!20, draw=black] (1.210,-1.280) -- (1.210,-0.680) -- (0.410,-0.680) -- (0.410,-1.280) -- cycle;
  \node at (0.810,-0.980) {\small l5};
  \draw[fill=gray!20, draw=black] (2.020,-1.280) -- (2.020,-0.680) -- (1.220,-0.680) -- (1.220,-1.280) -- cycle;
  \node at (1.620,-0.980) {\small l6};
  \draw[fill=gray!20, draw=black] (2.830,-1.280) -- (2.830,-0.680) -- (2.030,-0.680) -- (2.030,-1.280) -- cycle;
  \node at (2.430,-0.980) {\small l7};
  \draw[fill=gray!20, draw=black] (-0.410,-1.280) -- (-0.410,-0.680) -- (-1.210,-0.680) -- (-1.210,-1.280) -- cycle;
  \node at (-0.810,-0.980) {\small l3};
  \draw[fill=gray!20, draw=black] (-1.220,-1.280) -- (-1.220,-0.680) -- (-2.020,-0.680) -- (-2.020,-1.280) -- cycle;
  \node at (-1.620,-0.980) {\small l2};
  \draw[fill=gray!20, draw=black] (-2.030,-1.280) -- (-2.030,-0.680) -- (-2.830,-0.680) -- (-2.830,-1.280) -- cycle;
  \node at (-2.430,-0.980) {\small l1};
  \draw[fill=gray!20, draw=black] (4.000,-0.160) -- (4.000,0.160) -- (-4.000,0.160) -- (-4.000,-0.160) -- cycle;
  \node at (0.000,0.000) {\small DC};
  \draw[fill=gray!20, draw=black] (4.000,-0.670) -- (4.000,-0.170) -- (-4.000,-0.170) -- (-4.000,-0.670) -- cycle;
  \node at (0.000,-0.420) {\small RF};
  \draw[fill=gray!20, draw=black] (4.000,0.170) -- (4.000,0.670) -- (-4.000,0.670) -- (-4.000,0.170) -- cycle;
  \node at (0.000,0.420) {\small RF};
 
  \draw[red!70!black, thick] (-2.930,-1.280) -- (-2.930,-0.680);
  \draw[red!70!black, thick] (-2.930,-1.280) -- +(-0.050,0.000);
  \draw[red!70!black, thick] (-2.930,-0.680) -- +(-0.050,0.000);
  \node[red!70!black] at (-3.180,-0.980) {$w_{dc}$};
  \draw[red!70!black, thick] (-2.030,-0.580) -- (-2.830,-0.580);
  \draw[red!70!black, thick] (-2.030,-0.580) -- +(-0.000,-0.050);
  \draw[red!70!black, thick] (-2.830,-0.580) -- +(-0.000,-0.050);
  \node[red!70!black] at (-2.430,-0.430) { $l_{dc}$};
  \draw[red!70!black, thick] (4.100,-0.160) -- (4.100,0.160);
  \draw[red!70!black, thick] (4.100,-0.160) -- +(-0.050,0.000);
  \draw[red!70!black, thick] (4.100,0.160) -- +(-0.050,0.000);
  \node[red!70!black] at (4.50,0.000) {$w_{dc,c}$};
  \draw[red!70!black, thick] (4.000,0.260) -- (-4.000,0.260);
  \draw[red!70!black, thick] (4.000,0.260) -- +(-0.000,-0.050);
  \draw[red!70!black, thick] (-4.000,0.260) -- +(-0.000,-0.050);
  \node[red!70!black] at (2.000,0.430) {$l_{RF}$};
  \draw[red!70!black, thick] (4.100,0.170) -- (4.100,0.670);
  \draw[red!70!black, thick] (4.100,0.170) -- +(-0.050,0.000);
  \draw[red!70!black, thick] (4.100,0.670) -- +(-0.050,0.000);
  \node[red!70!black] at (4.450,0.420) {$w_{RF}$};
  \draw[red!70!black, thick] (4.100,-0.170) -- (4.100,  -0.670);
  \draw[red!70!black, thick] (4.100,-0.170) -- +(-0.050,-0.000);
  \draw[red!70!black, thick] (4.100,-0.670) -- +(-0.050,-0.000);
  \node[red!70!black] at (4.450,-0.420) {$w_{RF}$};
  \draw[>-<, red!70!black, thick]
    (.305,1.12) -- (0.505,1.12);
  \draw[red!70!black, thick,]
    (.305,1.12) -- (0.505,1.12);
  \node[red!70!black, above] at (0.405,1.3) {\scriptsize $w_{\mathrm{gap}}$};
  
  \draw[->, thick] (-4.0,-1.6) -- (4.0,-1.6) node[below] {$x$};
  \node at (0.000,-2.3) {$\xi$};
  \foreach \x/\label in {-2.43/0, -2.025/0.5, -1.62/1, -0.81/2, 0/3, 0.81/4, 1.62/5, 2.43/6} {
    \draw[thick] (\x,-1.6) -- (\x,-1.65);
    \node[below] at (\x,-1.65) {\label};
  }
\end{tikzpicture}
	\caption{
    Schematic of our trap geometry.
    The dimensions used to decribe the layout of the trap are shown in red, 
    which are
    the length 
    $l_{\mathrm{dc}}$ 
    and width
    $w_{\mathrm{dc}}$ 
    of the DC segmented electrodes, the width of the center DC electrode
    $w_{\mathrm{dc}}$ 
    and the RF electrodes
    $w_{\mathrm{RF}}$,
    as well as the length of the entire trap
    $l_{\mathrm{RF}}$.
    $w_{\mathrm{gap}}$ 
    is the width of the gap between the metal electrodes and 
    is the same across the entire trap. 
    We use the dimensionless parameter $\xi$ to describe 
    the ion position relative to the DC segmented electrodes.
	}
  \label{fig:fabrication: trap geom}
\end{figure}

The electronic layout of the trap features segmented electrodes for static potentials 
as well as RF electrodes to generate the 
ponderomotive pseudopotential. 
All metal structures on this device
(electrodes, bond pads, and leads)
are patterned within a single layer to reduce fabrication complexity.
The segmented DC electrodes are positioned outside the symmetric RF rails. 
to facilitate signal routing on a single layer, 
The resulting trap layout is illustrated in Fig.~\ref{fig:fabrication: trap geom}.

The finalized layout features RF electrodes with a width of 
$w_{\mathrm{RF}}=250~\si{\micro\meter}$, 
separated by a central DC electrode of width 
$w_{\mathrm{dc,c}}=160~\si{\micro\meter}$. 
The outer segmented DC electrodes have a width of 
$w_{\mathrm{dc}}=300~\si{\micro\meter}$ 
and a length of $l_{\mathrm{dc}}=400~\si{\micro\meter}$. 
All electrodes are electrically isolated by trenches with a uniform gap width of 
$w_{\mathrm{gap}}=5~\si{\micro\meter}$.
The choice of these parameters is elaborated in App.~\ref{app: electrode layout}.

\subsubsection{Influence of the dielectric on the trapping potential}

\begin{figure}[ht!]
  \centering
	\includegraphics[width=0.5\textwidth]{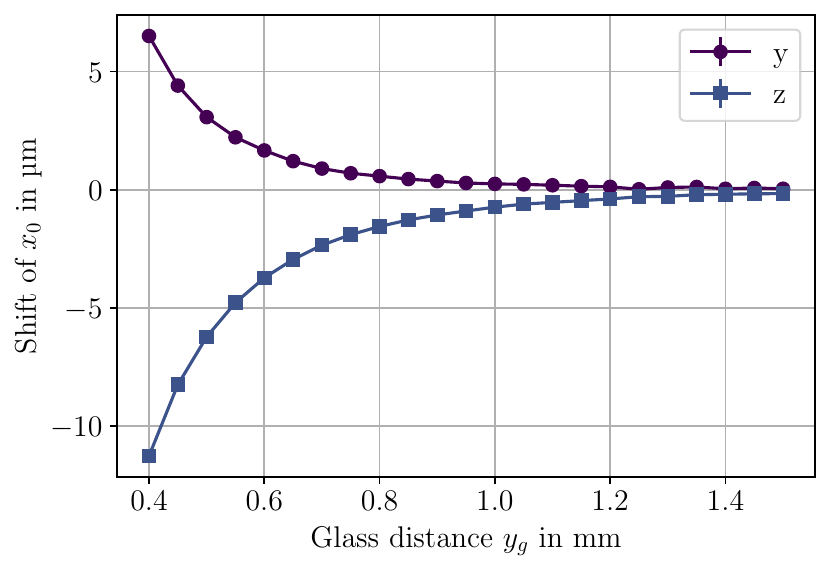}
	\caption{FEM simulation of the position of the RF null being shifted by the presence of a  glass. 
    The glass is modeled as a glass brick that that is sitting on the chips surface, 
    is $400\si{\micro\metre}$ thick, 
    is $3\si{\milli\meter}$ wide, and emerges from the chips edge 
    to a distance to the ion as specified by the x-axis of the plot.
    Moving the glass closer to the ion shifts the RF null towards the traps surface and closer to the glass. 
	}
  \label{fig:fabrication: shift of rf null vs glass distance}
\end{figure}

\begin{figure}[ht!]
  \centering
	\includegraphics[width=0.5\textwidth]{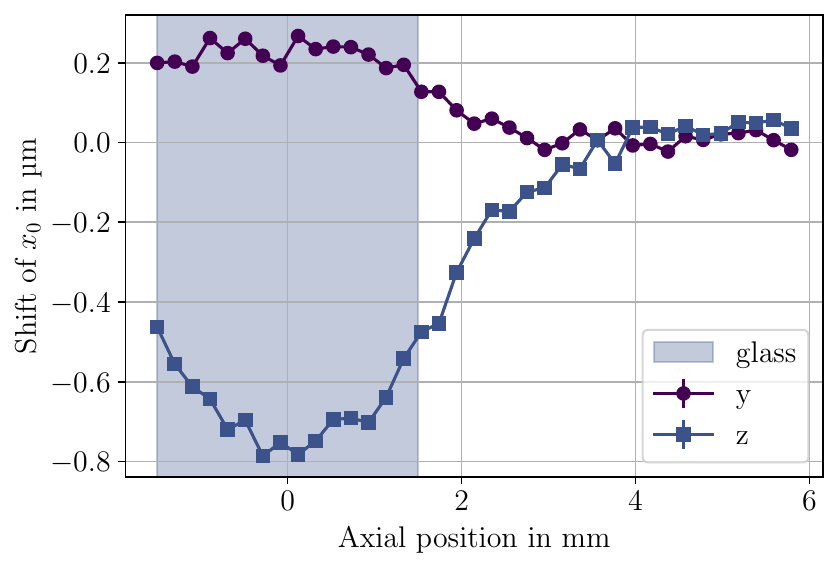}
	\caption{For $1\si{mm}$ distance of the glass block to the RF null, 
    the position of the RF null simulated along the trap axis using a FEM approach. 
    The position is given as offset from the position calculated without the glass block. 
    The blue area indicates axial positions where the ion is in front of the glass. 
	}
  \label{fig:fabrication: shift of rf null vs axial_position}
\end{figure}

The electrode structure of the trap was designed without the borosilicate glass block containing the waveguide. 
Now, we analyzed the effect of the optics block onto the RF trapping field with a Finite-Element Method (FEM).  
We model the trap substrate as 
borosilicate glass 
with a thickness of
$525~\si{\micro\meter}$ 
and the RF electrodes as metal on top
with a thickness of $2~\si{\micro\meter}$.
The rest of the trap surface is covered with a grounded metal layer with the same thickness, 
leaving  a $5~\si{\micro\meter}$ gap to the RF electrodes.
The optics block is modeled as a glass block that
 is $400~\si{\micro\meter}$ thick and $3~\si{mm}$ long in x-direction.
In the radial direction ($y$), the glass block extends from the chip's edge  
to a distance $y_g$ to 
the trapping position
calculated without the optics block.

We extract the trapping position by applying a potential to the RF electrodes and estimating 
the position of the potential minimum in the y, z plane while the axial position is centered around the glass block.
The effect of the dielectric block on the trapping potential is given by the deviation from the undistorted trapping position 
as a function of $y_g$ as shown in 
Fig.~\ref{fig:fabrication: shift of rf null vs glass distance}.

For the final trap design, 
we choose a distance of $y_g=1~\si{mm}$ where  the trapping positions shifts around 
$0.8~\si{\micro\meter}$ in the z-direction 
and around
$0.2~\si{\micro\meter}$ in the y-direction. 
For the final geometry of the glass block on the trap  
the shift of the ion position along the trap axis is shown in 
Fig.~\ref{fig:fabrication: shift of rf null vs axial_position}.

\subsection{Ion trap fabrication}

\tikzset{
    squarednode/.style={rectangle, draw=white, fill=lightgray!100, very thick, minimum size=5mm},
    squarednode_black/.style={rectangle, draw=black, fill=lightgray!100, very thick, minimum size=5mm},
    roundnode/.style={circle, draw=black, fill=white, very thick, minimum size=5mm},
    scalebar/.style={rectangle, draw=white, fill=lightgray!100, very thick},
    arrow_white/.style={color=white, very thick, ->},
    arrow_black/.style={color=black, very thick, ->},
    thickwhite/.style={white, line width=1.5 pt, dash pattern=on 0.1pt off 6pt, line cap=round},
    thickred_laser/.style={red!70!white, line width=1.5 pt, dash pattern=on 6pt off 3pt, line cap=round},
    thickblue_laser/.style={blue!70!white, line width=1.5 pt, dash pattern=on 6pt off 3pt, line cap=round},
    thickred/.style={red, line width=1.5 pt, dash pattern=on 3pt off 3pt, line cap=round},
    pluscircle/.pic={
        \filldraw[draw=black, fill=blue!40!white, thick] (0,0) circle (0.15cm); 
        \draw[thick] (-0.08,0) -- (0.08,0);  
        \draw[thick] (0,-0.08) -- (0,0.08); 
    },
    laser_beam/.pic={
        \shade[left color=red!5!white, right color=red!80!white, opacity=0.5] 
            (0,0) -- (4.65, 0.4) -- (4.65, -0.4) -- cycle;
    },
    cloud/.pic={
        \shade[shading=radial, inner color=green, outer color=green!50!white, opacity=0.5, yscale=0.5] 
            (0,0) circle (0.5cm);
    },
}

In this section, we detail the fabrication process for the ion trap device. 
The majority of the wafer-level fabrication was performed at the semiconductor manufacturing facilities of Infineon Technologies Austria AG in Villach on wafers with $200~\si{mm}$ diameter. 
A key feature of this architecture is the physical separation of the optical waveguide layer from the trap electrode substrate, 
which allows for the independent and parallel fabrication of both components. 
During this parallel flow, the optics wafer was processed at Joanneum Research in Weiz for the inscription of the femtosecond-laser-written (FLW) waveguides and ablation trenches. The used writing parameters are denoted in Table~\ref{tab:laser_parameters}. Specifics about the laser writing process are detailed in Sec.~\ref{sec: waveguides}.

\begin{table}[ht!]
    \centering
    \begin{tabular}{c|c|c}
        & FLW waveguide & ablation trench \\
        \hline
        wavelength & 520 nm & 520 nm \\
        laser power & 126 mW & 650 mW \\
        pulse energy & 126 nJ & 5 µJ \\
        pulse length & 209 fs & 209 fs \\
        repetition rate & 1 MHz & 125 kHz \\
        polarization & circular & circular \\
        writing speed & 14 mm/s & 3 mm/s \\
        overscans & 6 & 4 \\
        objective & \makecell{Zeiss Epiplan- \\ Neofluar 20x 0.5} & \makecell{Zeiss Epiplan- \\ Neofluar 20x 0.5}
    \end{tabular}
    \caption{Writing parameters for FLW waveguides and ablation trenches. 
    Laser parameters were adjusted for a mode-field radius of 
    $\approx5~\si{\micro\meter}$, 
    to minimize the beam waist at the ion's location in the absence of focusing optics. Note that the writing parameters are different to the ones in Sec.~\ref{sec: waveguides}.
    }
    \label{tab:laser_parameters}
\end{table}

\begin{figure}[ht!]
    \centering
    \begin{tikzpicture}
        \coordinate (xyz) at (3.6,-3);
        \node[inner sep=0pt] (image) at (0,0)
            {\includegraphics[width=.5\textwidth]{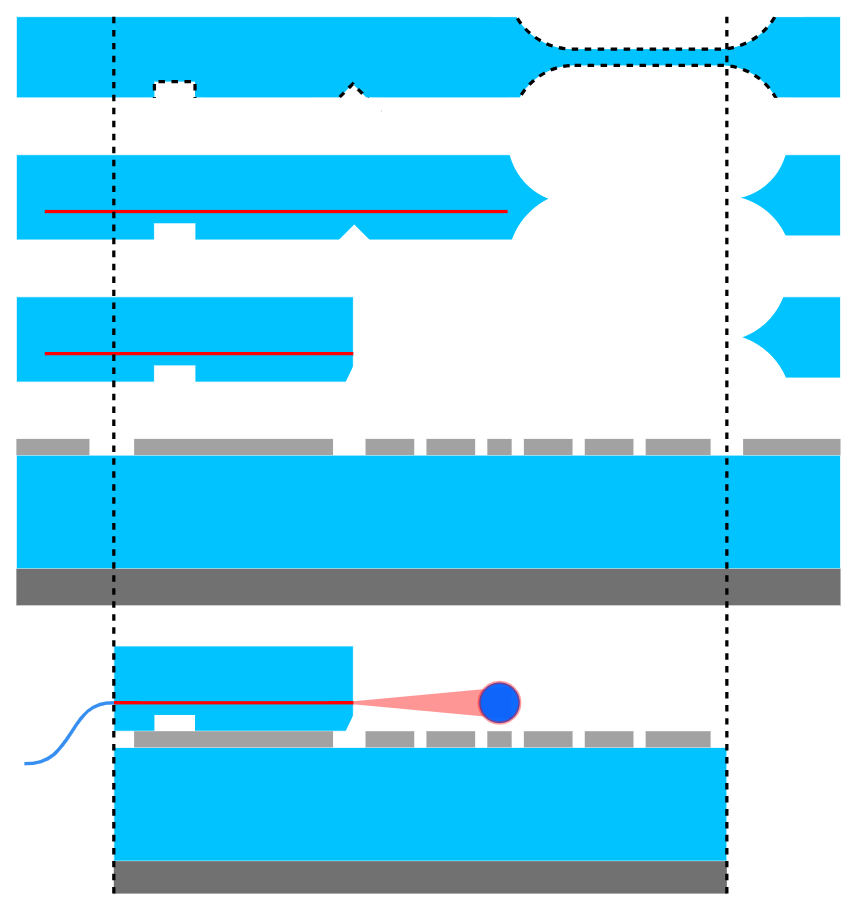}};
        \node[squarednode_black] at (-2.65,4.55) (a) {a};
        \node[squarednode_black] at (2.4,4.20) (b) {b};
        \node[squarednode_black] at (-0.8,4.55) (c) {c};
        \node[squarednode_black] at (-1.6,3.2) (d) {d};
        \node[squarednode_black] at (2,2.7) (e) {e};
        \node[squarednode_black] at (-0.1,1.2) (f) {f};
        \node[squarednode_black] at (1.2,0.9) (g) {g};
        \node[squarednode_black] at (0.3,-1.35) (h) {h};
        \node[squarednode_black] at (-3.8,-3.8) (i) {i};
        \node[text=white] at (0.75,-2.55) {\large \textbf +};
        \node at (0,5.1) {productive space};
        \node[align=center] at (-3.8,5) {dummy \\ space};
        \node[align=center] at (3.8,5) {dummy \\ space};
        \node[roundnode] at (-4.2,4.17) {1};
        \node[roundnode] at (-4.2,2.7) {2};
        \node[roundnode] at (-4.2,1.2) {3};
        \node[roundnode] at (-4.2,-0.5) {4};
        \node[roundnode] at (-4.2,-2.3) {5};
        \draw[arrow_black] (a.south) -- +(0,-0.3);
        \draw[arrow_black] (c.south) -- +(0,-0.3);
        \draw[arrow_black] (d.south) -- +(0,-0.3);
        \draw[arrow_black] (e.west) -- +(-0.3,0);
        \draw[arrow_black] (f.west) -- +(-0.3,0);
        \draw[arrow_black] (g.south) -- +(0,-0.3);
        \draw[arrow_black] (h.east) -- +(0.3,0);
        \draw[arrow_black] (i.north) -- +(0,0.3);
        \draw[arrow_black] (xyz) -- +(0,0.5) node[near end, left] {z};
        \draw[arrow_black] (xyz) -- +(0.5,0) node[near end, below] {y};
    \end{tikzpicture}
	\caption{Fabrication scheme of an ion trap device. 
    In steps 1-3 the optics wafer is formed by introducing alignment marks (a) which are used to align the subsequent fabrication steps to, such as
    the through-etch from both sides of the substrate (b), the ablation trench (c) and FLW waveguide scribing (d).
    Since the through-etch produces a curved sidewall with a peak (e), a cleaving process provides a smooth and flat front end of the waveguide (f). In step 4 a metal layer (g) is structured onto a wafer stack of borosilicate glass and crystalline silicon (h) to form a trap wafer. Subsequent wafer bonding, dicing and fiber (i) assembly to the rear end of the waveguide results in functional ion trap devices (step 5).}
    \label{fig:fabrication_steps}
\end{figure}
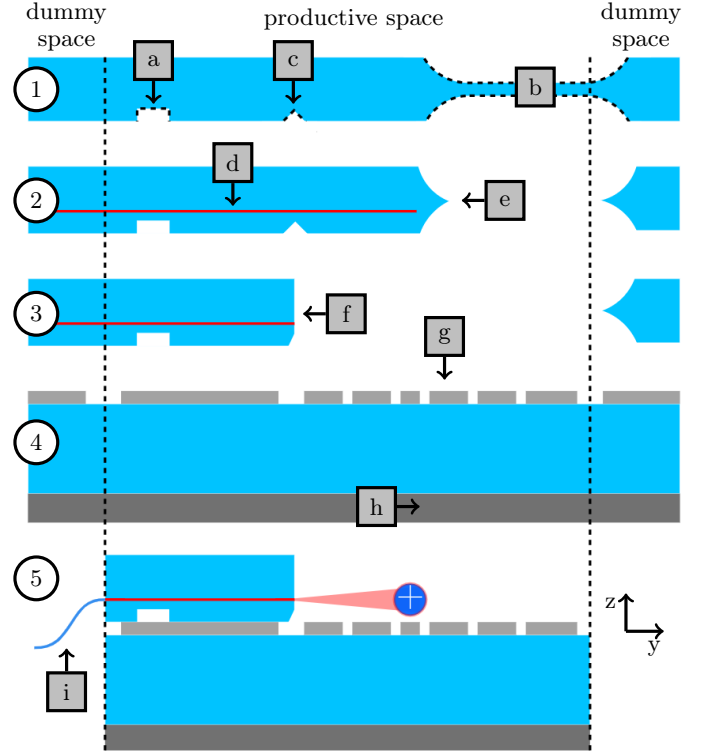

The fabrication of the glass block starts with 
a $400~\si{\micro\meter}$ thick borosilicate substrate. For the following description please refer to Fig.~\ref{fig:fabrication_steps}, steps 1 and 2. First, a shallow etch of alignment structures (a) into the substrate is performed. These alignment marks act as a reference for the subsequent through-etch (b) with hot 50\% hydrofluoric acid from both sides. After the etch, the optics wafer is sent to Joanneum Research in Weiz for waveguide inscription (d) and ablation trench (c) writing. At the ablation trench the tips of the optics block is cleaved off via manual application of a downward force onto the glass with plastic tweezers. This process results in a smooth and flat end of the inscribed waveguide (f). A thorough analysis of the end facets around the waveguide end is presented in Sec.~\ref{subsec: nosetip characterization}.

The trap wafer is fabricated (Fig.~\ref{fig:fabrication_steps}, 
step 4) on a $525~\si{\micro\meter}$ thick borosilicate wafer that has been bonded onto a $200~\si{\micro\meter}$ silicon wafer (h) for ease of handling. An aluminum layer with a thickness of $2~\si{\micro\meter}$ is deposited via magnetron sputtering.
A subsequent optical lithography step provides a resist mask with a minimal gap size of $5~\si{\micro\meter}$, followed by a reactive-ion etch process to structure the metal layer to form the electrodes and their wiring of a surface ion trap.

Finally, as in Fig.~\ref{fig:fabrication_steps}, step 5, the optics wafer is anodically bonded onto the trap wafer. The silicon carrier wafer might be removed from the glass substrate. Afterwards, the wafer is mechanically diced to form single ion trap devices. 

\subsection{Ion trap module assembly}
\label{subsec: fabrication assembly}

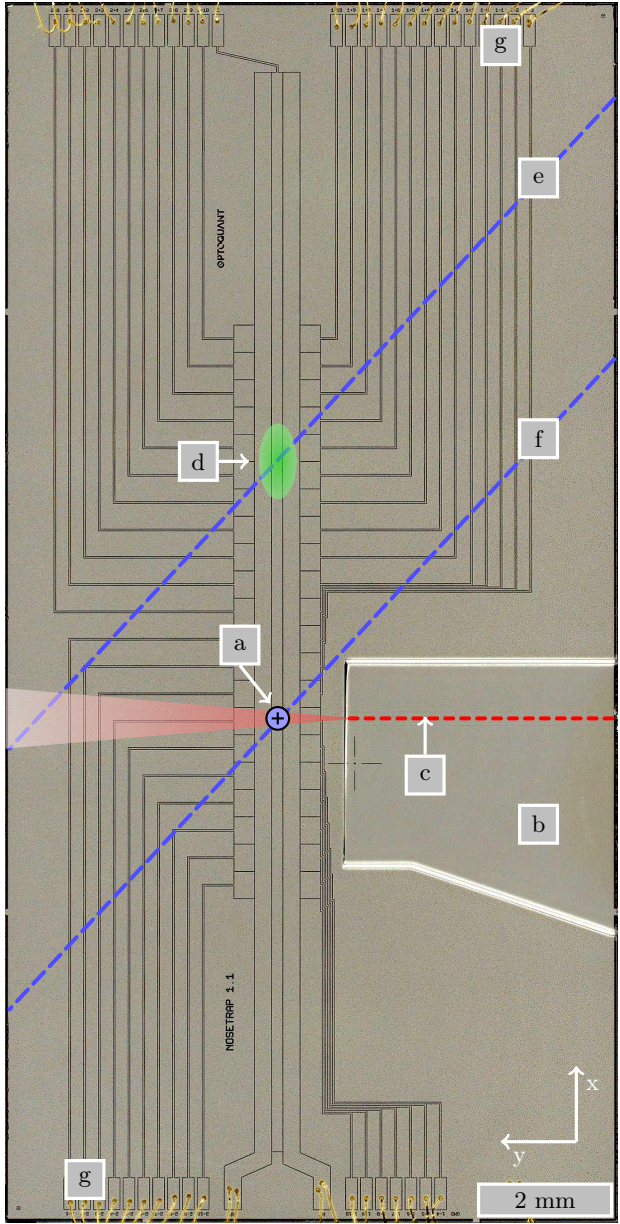
\begin{figure}[ht!]
    \centering
    \begin{tikzpicture}
        \coordinate (xyz) at (3.5,-7);
        \node[inner sep=0pt] (image) at (0,0)
            {\includegraphics[width=.9\textwidth, angle=90]{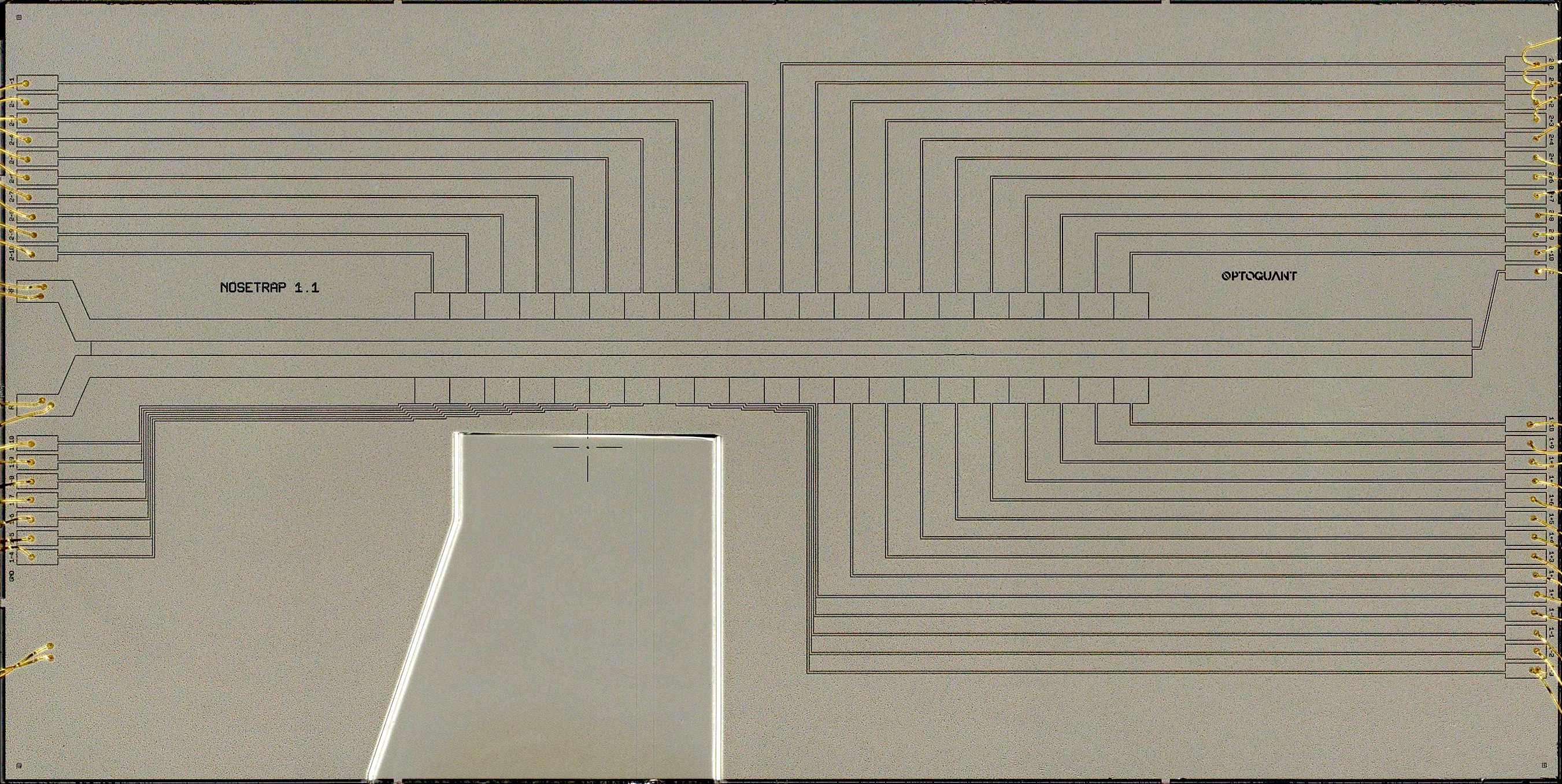}};
        \node[scalebar, minimum width=17.6mm] at (image.south east) [above left=2pt and 2 pt] (scalebar) {2 mm};
        \draw[thickblue_laser] (4,3.35) -- (-4,-5.25);
        \draw[thickblue_laser] (4,6.80) -- (-4,-1.8);
        \draw[thickred] (0.5,-1.4) -- (4,-1.4);
        \pic[rotate=180] at (0.6,-1.4) {laser_beam};
        \pic at (-0.45,-1.4) {pluscircle};
        \draw[arrow_white] (xyz) -- +(-1,0) node[near end, below] {y};
        \draw[arrow_white] (xyz) -- +(0,1) node[near end, right] {x};
        \node[squarednode] at (-0.95,-0.43) (a) {a};
        \node[squarednode] at (1.5,-2.15) (c) {c};
        \node[squarednode] at (3,-2.8) (b) {b};
        \node[squarednode] at (3,5.75) (e) {e};
        \node[squarednode] at (3,2.3) (f) {f};
        \node[squarednode] at (-1.5,2) (d) {d};
        \node[squarednode] at (-3,-7.5) (g) {g};
        \node[squarednode] at (2.5,7.5) (g) {g};
        \draw[arrow_white] (a.south) -- +(0.4,-0.5);
        \draw[arrow_white] (c.north) -- +(0,0.5);
        \draw[arrow_white] (d.east) -- +(0.4,0);
        \pic[rotate=90] at (-0.45,2) {cloud};
    \end{tikzpicture}
	\caption{Microscopy image of completed the ion trap chip after assembly (device B). The ion (a) sits $\sim 1~\si{\milli\meter}$ in front of the glass block (b) containing a FLW waveguide (c) for $729~\si{\nano\meter}$ laser-light. A dedicated loading zone (d) is situated a few millimeters away from the glass block. Additional free-space laser access at 45° angle (e and f)  are used for the other ionization, ion detection and re-pumping. On the top and bottom of the image wire bonds (g)  are visible, that are used for electrical contact to the carrier PCB.}
    \label{fig:microscopy_image_overview}
\end{figure}

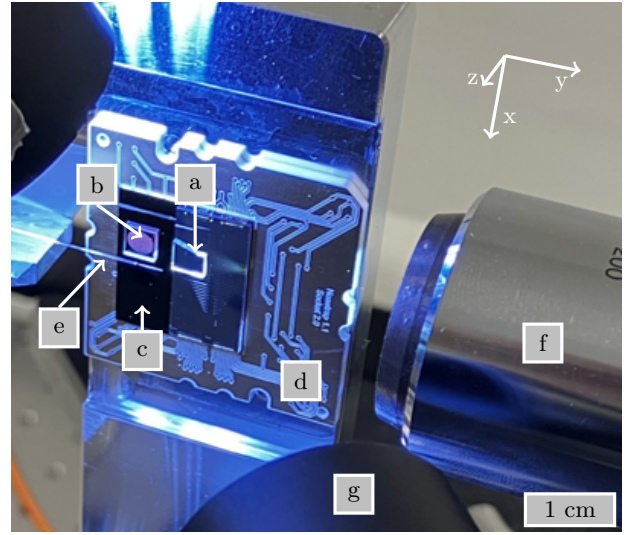
\begin{figure}[ht!]
    \centering
    \begin{tikzpicture}
        \coordinate (xyz) at (2.5,2.8);
        \node[inner sep=0pt] (image) at (0,0)
            {\includegraphics[width=.45\textwidth]{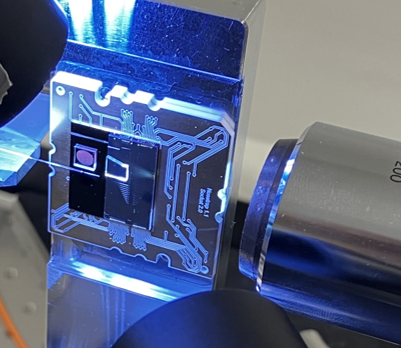}};
        \node[scalebar, minimum width=11.8mm] at (image.south east) [above left=2pt and 2 pt] (scalebar) {1 cm};
        \node[squarednode] at (-1.6,1.1) (A) {a};
        \node[squarednode] at (-2.9,1.1) (B) {b};
        \node[squarednode] at (-2.3,-1.1) (C) {c};
        \node[squarednode] at (-0.2,-1.5) (D) {d};
        \node[squarednode] at (-3.4,-0.8) (E) {e};
        \node[squarednode] at (3,-1) (F) {f};
        \node[squarednode] at (0.5,-3) (G) {g};
        \draw[arrow_white] (A.south) -- (-1.6,0.2);
        \draw[arrow_white] (B.south) -- (-2.3,0.4);
        \draw[arrow_white] (C.north) -- (-2.3,-0.5);
        \draw[arrow_white] (E.north) -- (-2.8,0.1);
        \draw[arrow_white] (xyz) -- +(-0.3,-0.4) node[near end, left] {z};
        \draw[arrow_white] (xyz) -- +(1,-0.2) node[near end, below] {y};
        \draw[arrow_white] (xyz) -- +(-0.2,-1.1) node[near end, right] {x};
    \end{tikzpicture}
	\caption{Photograph of assembly setup. The glass block (a) as part of the ion trap is mounted on a silicon interposer (c), which is glued on a carrier PCB (d). An optical fiber (e) is approached via a movable stage to align with the waveguide in the glass block. The fiber support chip (b) adds mechanical support to the fiber during handling. A microscope (f) opposite of the glass block tip allows imaging of the light field produced by the waveguide. The UV light source (g) enables curing of the UV curable glue.}
    \label{fig:ass_photo}
\end{figure}

The ion trap chip is assembled 
on a carrier PCB \cite{Anmasser2026}, as seen in Fig.~\ref{fig:ass_photo} D, 
to be able to mount and establish electrical contact in the experimental setup.
The trap is glued onto the PCB end electrical connections are made
with wire bonds. 
Fig.~\ref{fig:microscopy_image_overview} shows a microscopy image of a complete ion trap chip.

After the electrical assembly, an optical fiber (Thorlabs S630-HP, Fig.~\ref{fig:ass_photo}/e) is attached to the ion trap (a) with the aid of a UV curable glue (Norland 68). 
A glass support chip (b) is positioned in close vicinity of the fiber-chip interface, to improve the mechanical stability of the setup. 
Both ion trap chip and support chip are glued onto a $200~\si{\micro\meter}$ thick silicon interposer (c) to reduce mechanical stress during cooldown. 
This silicon interposer is assembled onto a 
PCB (d) which fans out the electrical connections. An optical microscope with a Mitutoyo Plan Apo 20x/0.42 objective lens (f) is used to measure laser light intensity exiting the front end of the waveguide. Additionally, the microscope is used to image the beam intensity profile a the ion's position. 

\begin{figure}[ht!]
    \centering
    \begin{tikzpicture}
        \coordinate (xyz) at (-3,-1);
        \node[inner sep=0pt] (image) at (0,0)
            {\includegraphics[width=.45\textwidth]{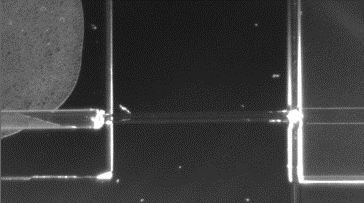}};
        \node[scalebar, minimum width=15.5mm] at (image.south east) [above left=2pt and 2 pt] (scalebar) {1 mm};
        \node[squarednode] at (3.3,0.6) (c) {c};
        \node[squarednode] at (-2.8,1.3) (b) {b};
        \node[squarednode] at (1.8,-1.3) (a) {a};
        \draw[arrow_white] (c.south) -- (3.2,-0.15);
        \draw[arrow_white] (c.south) -- (3.45,-0.5);
        \draw[arrow_white] (b.south) -- +(0,-0.5);
        \draw[arrow_white] (a.north) -- +(0,0.5);
        \draw[arrow_white] (xyz) -- +(0.5,0) node[near end, right] {y};
        \draw[arrow_white] (xyz) -- +(0,-0.5) node[near end, left] {x};
    \end{tikzpicture}
    \begin{tikzpicture}
        \coordinate (xyz) at (-3,-2.3);
        \node[inner sep=0pt] (image) at (0,0)
            {\includegraphics[width=.45\textwidth]{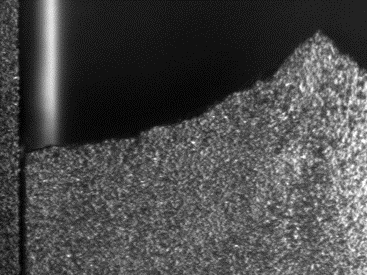}};
        \node[scalebar, minimum width=32.5mm] at (image.south east) [above left=2pt and 2 pt] (scalebar) {100 µm};
        \node[squarednode] at (2.95,1.2) (f) {f};
        \node[squarednode] at (-2.8,1.3) (e) {e};
        \node[squarednode] at (0.5,-0.5) (d) {d};
        \draw[arrow_white] (f.north) -- +(0,0.5);
        \draw[arrow_white] (e.west) -- +(-0.5,0);
        \draw[arrow_white] (d.south) -- +(0,-0.5);
        \draw[arrow_white] (xyz) -- +(0.5,0) node[near end, right] {z};
        \draw[arrow_white] (xyz) -- +(0,0.5) node[near end, left] {x};
    \end{tikzpicture}
	\caption{
    Microsopy images during assembly. Upper image: fiber tip (a) is glued to the glass block as well as the support chip (b). Orientation markings (c) in glass block are visible. Lower image: Microscopy of glass block sidewall (d) near fiber attach side, bonded to ion trap wafer (e). A peak (f) is visible due to the through-etch from both sides of the substrate (compare to Fig.~\ref{fig:fabrication_steps}/e). The rough surface is due to mechanical dicing. Scattering losses of the laser light injected into the waveguide via the fiber are minimized by the UV curing glue.
	}
    \label{fig:ass_micro}
\end{figure}
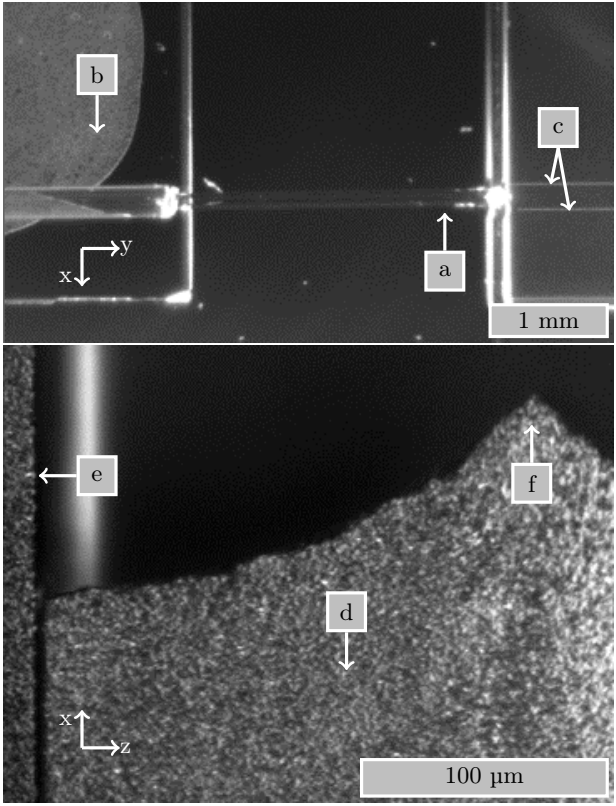

The optical assembly procedure involves a few dedicated steps. 
First, the coating of the fiber is removed for roughly $5~\si{\milli\meter}$ beyond the cleaved fiber tip (Fig.~\ref{fig:ass_micro}/a). 
The fiber is mounted onto a 6-axis hexapod stage positioning the fiber facet in between orientation markings (c), that were inscribed $100~\si{\micro\meter}$ to the left and right of the productive waveguide. A microscope is used to image the $730~\si{\nano\meter}$ laser-light intensity profile after the waveguide facet. The position of maximum fiber-to-waveguide coupling is marked and the waveguide retracted ~$3~\si{\centi\meter}$ away from the glass block.

In the second step, a small droplet of Norland 68 UV curable glue is manually applied to the retracted fiber tip. Using the marked coordinates the fiber is positioned in the prior optimum location, where a second optimization is performed - now with the glue filling the space between fiber tip and rear end of the waveguide. As soon as maximum transmission is found, UV light is applied to cure the glue.

During the third step, an additional larger droplet of glue is applied to the support chip (Fig.~\ref{fig:ass_micro}/b) and the coated part of the fiber directly positioned above the support chip followed by a curing step in order to mechanically fixate the fiber and prevent fiber tip detachment.

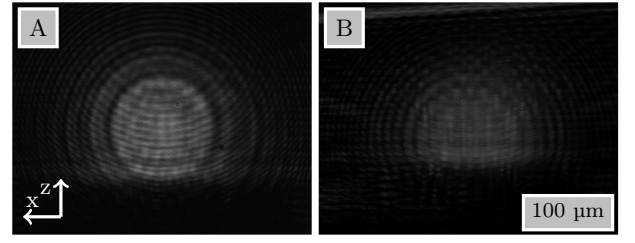
\begin{figure}[ht!]
    \centering
    \begin{tikzpicture}
        \coordinate (xyz) at (-3.35,-1.3);
        \node[inner sep=0pt] (image) at (0,0)
            {\includegraphics[width=.45\textwidth]{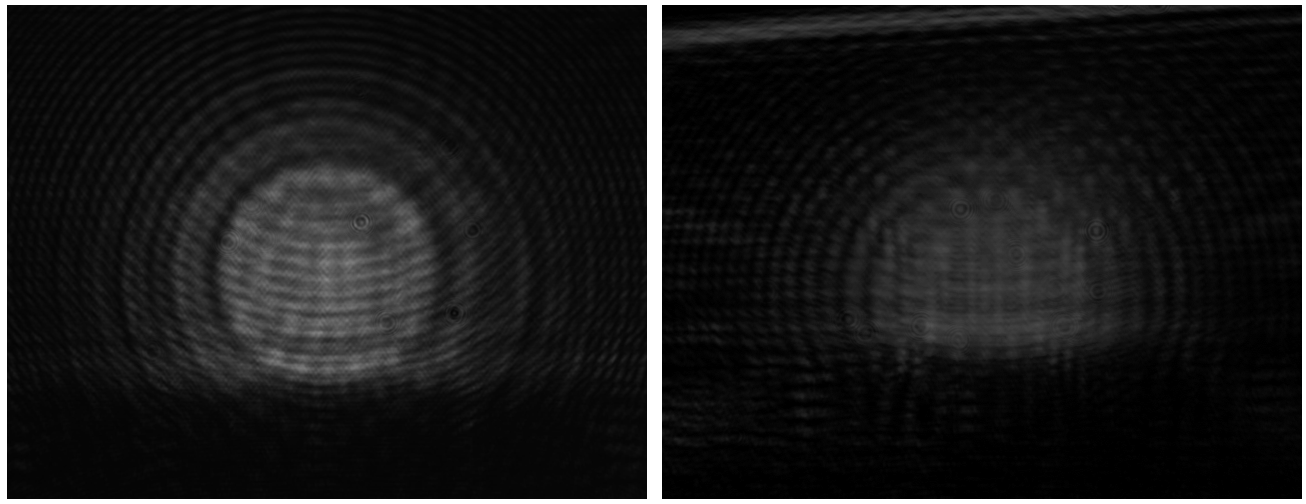}};
        \node[scalebar, minimum width=11.8mm] at (image.south east) [above left=2 pt and 2 pt] (scalebar) {\footnotesize 100 µm};
        \node[squarednode] at (-3.65,1.2) (A) {A};
        \node[squarednode] at (0.4,1.2) (B) {B};
        \draw[arrow_white] (xyz) -- +(0,0.5) node[near end, left] {z};
        \draw[arrow_white] (xyz) -- +(-0.5,0) node[near end, above] {x};
    \end{tikzpicture}
	\caption{Light intensity profile of device A and B after assembly at ion's position, 1 mm away from the nose tip. A loss of intensity in the lower part of the images is due to clipping of the beam by the ion trap chip as well as the carrier PCB in total extending $\sim1.9~\si{\centi\meter}$ from the glass block facet. Further information on the beam profile of device B is given in Fig.~\ref{fig:fabrication: modeprofile datahero device}.}
    \label{fig:chip1_light_profile2}
\end{figure}

\begin{figure}[ht!]
  \centering
  \includegraphics[width=0.45\textwidth]{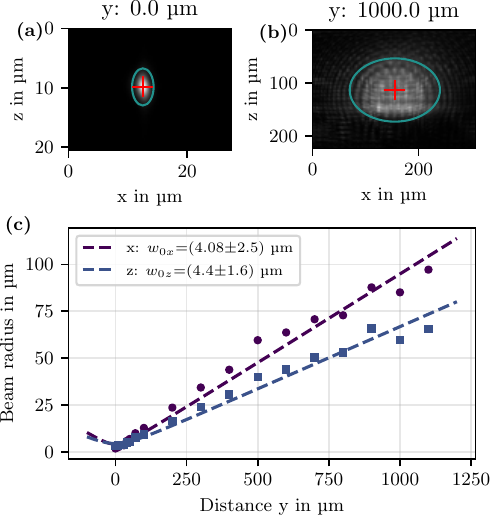}
	\caption{
    $D4\sigma$-radii in x- and z-direction of the beam profile of device B at several distances from the glass block facet. The higher width in the x-direction compared to the z-direction is in accordance with the assumption of the waveguide geometry pointed out in Sec.~\ref{subsec: waveguide formation}.
	}
  \label{fig:fabrication: modeprofile datahero device}
\end{figure}

Two functional ion trap devices were fabricated.
Device A has a total waveguide transmission efficiency, including reflection losses, of $0.47 \pm 0.03$, whereas device B has a lower transmission efficiency of $0.34 \pm 0.03$. The intensity profile of the beam at the ion location 1 mm away from the glass block tip is shown in Fig.~\ref{fig:chip1_light_profile2}. In device B the laser beam was imaged at various locations from the waveguide facet to beyond the ion's position. The $D4\sigma$ beam size~\cite{Siegman98} in x- and z-direction was measured as shown in Fig.~\ref{fig:fabrication: modeprofile datahero device}. Directly at the facet the beam diameter as extracted from the fit is $(4.1\pm2.5)~\si{\micro\meter}$ and $(4.4\pm1.7)~\si{\micro\meter}$ in x- and z-direction respectively. At the ion position the laser beam has diverged to a radius of $(94.9\pm2.4)~\si{\micro\meter}$ and $(66.8\pm1.8)~\si{\micro\meter}$ in x- and z-direction respectively.

\subsection{Characterization of the outcoupling facet}
\label{subsec: nosetip characterization}

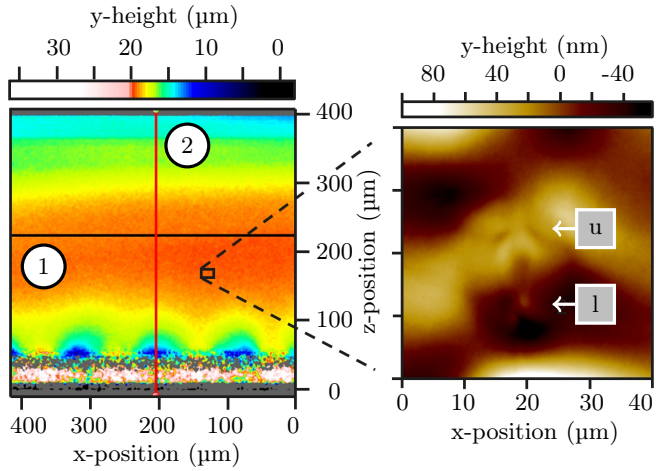
\begin{figure}[ht!]
    \centering
    \begin{tikzpicture}
        \node[inner sep=0pt] (image) at (0,0)
            {\includegraphics[width=.48\textwidth]{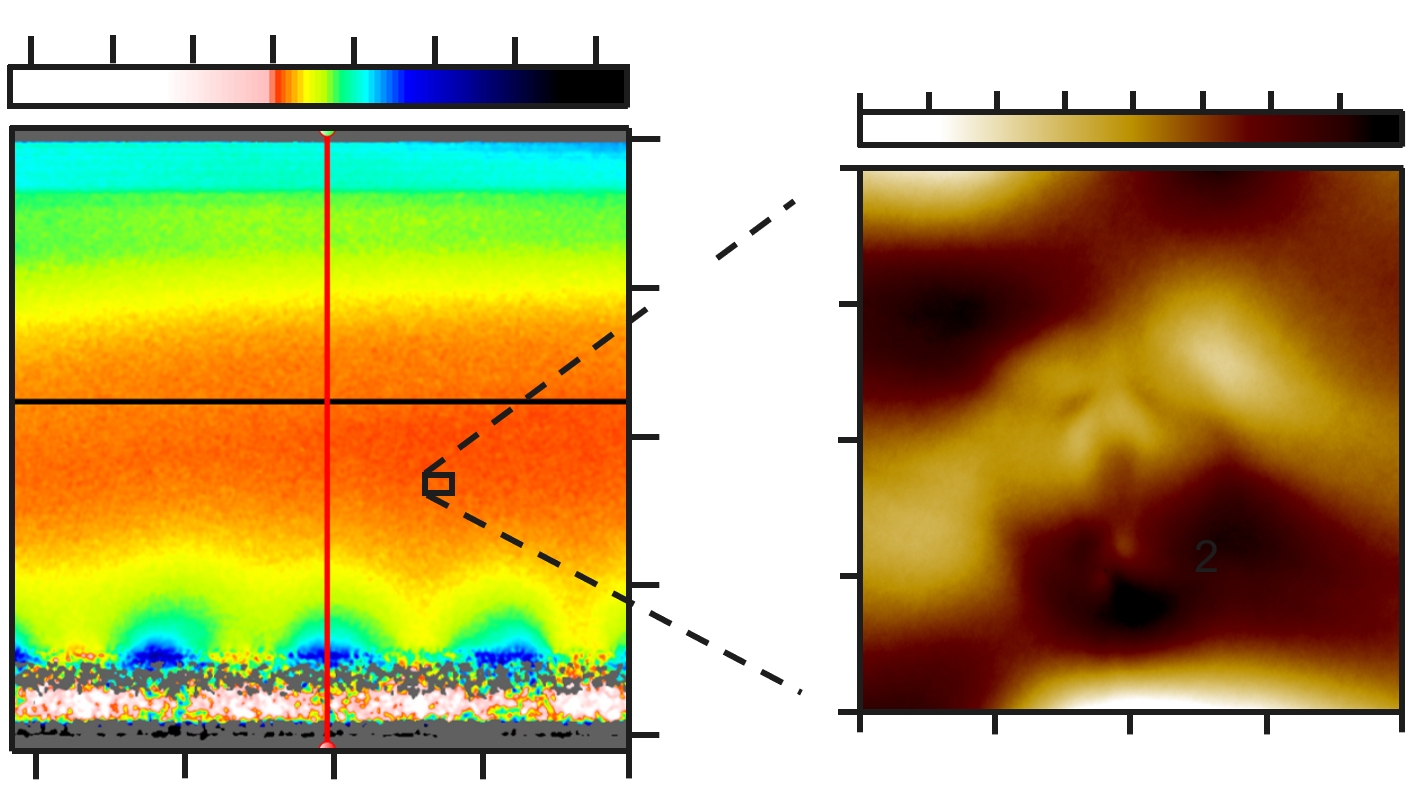}};
        \node[roundnode] at (-3.8,-0.4) (1) {1};
        \node[roundnode] at (-1.9,1.2) (2) {2};
        \node[align=center] at (-4.0,-2.5) {400};
        \node[align=center] at (-3.2,-2.5) {300};
        \node[align=center] at (-2.3,-2.5) {200};
        \node[align=center] at (-1.4,-2.5) {100};
        \node[align=center] at (-0.5,-2.5) {0};
        \node[align=center] at (-2.3,-2.9) {x-position (µm)};
        \node[align=center] at (-3.6,2.5) {30};
        \node[align=center] at (-2.65,2.5) {20};
        \node[align=center] at (-1.65,2.5) {10};
        \node[align=center] at (-0.7,2.5) {0};
        \node[align=center] at (-2.2,2.9) {y-height (µm)};
        \node[align=left] at (0.05,1.65) {400};
        \node[align=left] at (0.05,0.7) {300};
        \node[align=left] at (0.05,-0.2) {200};
        \node[align=left] at (0.05,-1.1) {100};
        \node[align=left] at (0.05,-2.05) {0};
        \node[align=center] at (0.95,-2.2) {0};
        \node[align=center] at (1.75,-2.2) {10};
        \node[align=center] at (2.6,-2.2) {20};
        \node[align=center] at (3.45,-2.2) {30};
        \node[align=center] at (4.1,-2.2) {40};
        \node[align=center] at (2.7,-2.6) {x-position (µm)};
        \node[align=center] at (3.8,2.15) {-40};
        \node[align=center] at (3.03,2.15) {0};
        \node[align=center] at (2.20,2.15) {40};
        \node[align=center] at (1.35,2.15) {80};
        \node[align=center] at (2.7,2.5) {y-height (nm)};
        \node[align=center, rotate=90] at (0.5,-0.2) {z-position (µm)};
        \node[squarednode] at (3.5,0.1) (1) {u};
        \node[squarednode] at (3.5,-0.9) (2) {l};
        \draw[arrow_white] (1.west) -- +(-0.3,0);
        \draw[arrow_white] (2.west) -- +(-0.3,0);
    \end{tikzpicture}
	\caption{Left side: A typical height profile of a cleaved glass facet acquired via optical profilometry. Between 0 and $50 ~\si{\micro\meter}$ in z-position the ablation trench is visible. Please note this profile has been leveled. Right side: AFM scan of cleaved front end of the waveguide at waveguide location. There is a topographical feature which corresponds to the optically visible waveguide structure (Fig.~\ref{fig:waveguides:overview} a and b). The upper waveguide is at position u, whereas the lower waveguide sits at position l.}
    \label{fig:nose_tip_heatmap}
\end{figure}

The waveguide (Fig.~\ref{fig:fabrication_steps}/d) delivers the light to the ion. For correct beam placement and shape a smooth end facet is required. 
Our approach to generate a suitable facet is to cleave of the tip of the optics block. 
The cleave is initiated with help of the inscribed ablation trench.
The glass block is placed on top of a acrylic glass piece such that the ablation trench is aligned with an edge of the acrylic glass. 
A downward force is applied with plastic tweezers on the protruding glass to create the cleave.
This is a manual process with low repeatability due to the inhomogeneous force distribution during the cleave. 
This may result in varying and slanted or curved facets, impairing beam pointing accuracy and beam quality. 
Fig.~\ref{fig:nose_tip_heatmap} shows height-profile in a cleaved glass tip facet as obtained via optical profilometry. 

Along the cuts 1 and 2 of Fig.~\ref{fig:nose_tip_heatmap} we observe curvatures along the x- and z-axis with height differences of $\sim60~\si{\nano\meter}$ over a length of $650~\si{\micro\meter}$ and $\sim 5~\si{\micro\meter}$ over a length of $400~\si{\micro\meter}$, respectively. The waveguide is placed at $z = 170~\si{\micro\meter}$ above the electrode layer. The tilt of a plane fitted through the the height data inside a radius of $4 ~\si{\micro\meter}$ around the waveguide with respect to the optical axis of the profilometer is below 1° in 5 measured samples.

Waveguide writing changes the density of the glass locally at the waveguide position, resulting in a perturbation of the cleave and thus a perturbation of the facet surface topography around the waveguide position as can be seen in an AFM surface image (Fig.~\ref{fig:nose_tip_heatmap}). The local height deviation within a length scale of $2~\si{\micro\meter}$ is below $50~\si{\nano\meter}$. Similar samples without inscribed waveguides have shown a RMS surface roughness of $1.0 \pm 0.1~\si{\nano\meter}$.

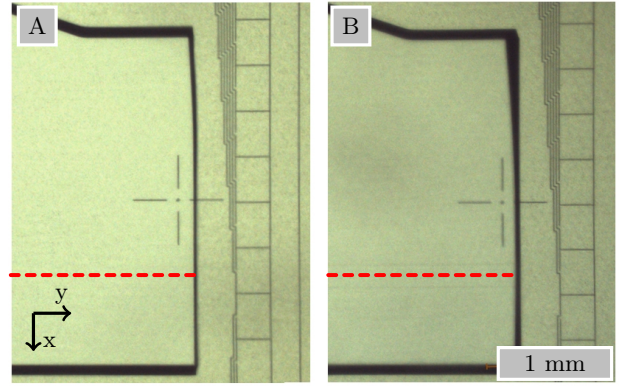
\begin{figure}[ht!]
    \centering
    \begin{tikzpicture}
        \coordinate (xyz) at (-3.7,-1.6);
        \node[inner sep=0pt] (image) at (0,0)
            {\includegraphics[width=.45\textwidth]{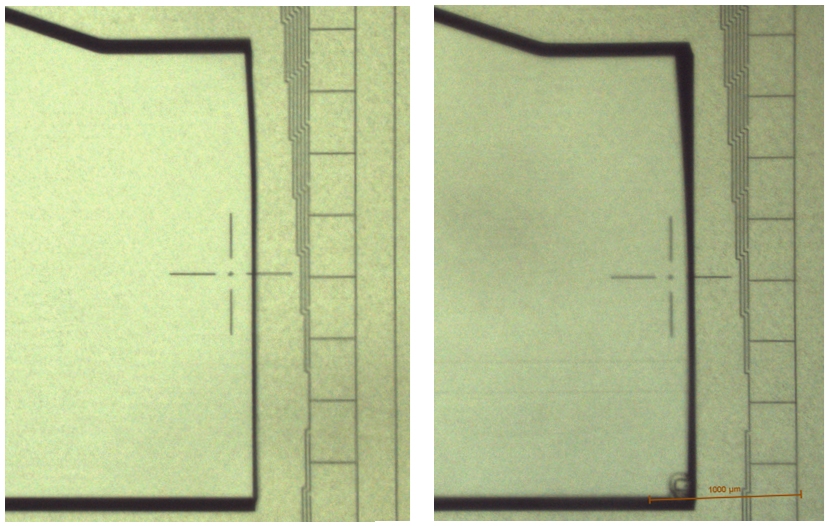}};
        \node[scalebar, minimum width=15mm] at (image.south east) [above left=2 pt and 2 pt] (scalebar) { 1 mm};
        \node[squarednode] at (-3.65,2.2) (A) {A};
        \node[squarednode] at (0.5,2.2) (B) {B};
        \draw[arrow_black] (xyz) -- +(0,-0.5) node[near end, right] {x};
        \draw[arrow_black] (xyz) -- +(0.5,0) node[near end, above] {y};
        \draw[thickred] (0.2,-1.095) -- (2.7,-1.095);
        \draw[thickred] (-4,-1.095) -- (-1.55,-1.095);
    \end{tikzpicture}
	\caption{Microscope image of glass blocks after ion trap assembly. Devices A and B correspond to images a and b. On image B the orientation markings $100 ~\si{\micro\meter}$ to the left and right of the waveguide are faintly visible. Device A seems to have a facet of the waveguide more orthogonal to the ion trap surface compared to device B, which results in the beam center better aligned to the ion height.}
    \label{fig:nose_micro}
\end{figure}

Optical profiling is not available for assembled devices, since mechanical and optical access to the outcoupling facet is restricted by the bulk of the chip. Therefore, the chip facet tilt angles in our devices under test have been inspected by optical microscopy. Fig.~\ref{fig:nose_micro} shows the glass block tips of device A and B. 
The thickness of the black shadow at the facet position indicates the magnitude of the surface topography, giving an upper bound for the facet angle with respect to the ion trap surface, assuming a surface profile like in \ref{fig:nose_tip_heatmap}. 
For device A the facet angle is assumed $<85.7\si{\degree} \pm 0.5\si{\degree}$, whereas device B shows a worse bound with an angle $< 83.3\si{\degree} \pm 0.5\si{\degree}$.

\section{Trapping performance}
\label{sec: measurements}
In this section we investigate the trapping performance.
First we show a general characterization of the trap
such as heating rate and stray fields. 
Next, we show measurements on the interaction of the ion with the integrated waveguide.

The measurements shown in this section
were taken on two experimental setups, one located at the university of Innsbruck and 
the other at Infineon Technologies Austria AG in Villach. 
Both systems are operated with $^{40}\mathrm{Ca}^+$ ions in a closed-cycle cryostat (Kiutra T-Type) at $T \approx 8~\si{K}$.
The $^{40}\mathrm{Ca}^+$ ions are produced from a neutral atom flux with a two-step ionization process,
using laser beams
with wavelengths of
$423~\si{nm}$
and 
$375~\si{nm}$.
For Doppler cooling and detection of the ions, 
we use 
a laser with a wavelength of 
$397~\si{nm}$.
For resolved sideband cooling and as qubit transition,
we use the quadrupole transition
$4^2\mathrm{S}_{1/2} \leftrightarrow 
3^2\mathrm{D}_{5/2}$,
addressed with laser with  a wavelength of $729~\si{nm}$.

The ponderomotive trapping potential is generated by applying
a RF electrical signal at $31.7~\si{MHz}$ with an amplitude  of 
$U_{RF} \approx 110~\si{V}$ 
to
$U_{RF} \approx 160~\si{V}$ 
resulting in radial secular frequencies of the ion of 
$\omega_r \approx 2\pi \cdot  2~\si{MHz}$ to 
$\omega_r \approx 2\pi \cdot  3~\si{MHz}$. 
Axial confinement is achieved by applying DC voltages as detailed in Sec.~\ref{sec: fabrication}.

\subsection{Heating rates}

This section shows measurements using trapped ions, to characterize the ion trap on borosilicate glass and the effects of the dielectric block. Heating rate measurements are done at the experimental setup at Infineon Technologies Austria AG in Villach. 
The static voltages for axial confinement are generated using 
a 32 channels DAC board ($\pm 10~\si{V}$), amplified to $\pm 40~\si{V}$. 
The voltages are filtered directly after the amplifiers outside the vacuum chamber with third-order lowpass filters with a 
cutoff frequency of $16~\si{Hz}$
and again inside the cryostat with first order lowpass filters 
with 
cutoff frequency $5~\si{\kilo Hz}$.
We apply static trapping voltages as derived in 
App.~\ref{app:voltage calculation}. 
A comparison between simulated and measured axial secular frequency is shown in App.~\ref{app: secular freq}.

\begin{figure}[ht!]
  \centering
  \includegraphics[width=0.5\textwidth]{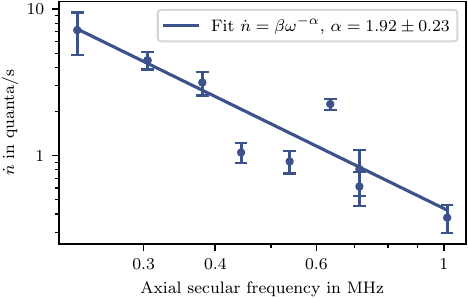}
	\caption{
    The heating rate measured on the 
    $\mathrm{S}_{1/2} \leftrightarrow 
    \mathrm{D}_{5/2}$,
    transition via sideband thermometry. 
    Each datapoint results from a weighted linear fit of ion temperature data, 
    where the weights are calculated from the quantum shot noise of the red and blue sideband data. 
	}
  \label{fig:measurements: heating rates vs axial freq}
\end{figure}

Heating rates along the trap axis  at an axial frequency of $1~\si{MHz}$ were measured using resolved sideband spectroscopy and range between 
$0.5~\si{ph/s}$
and
$1.5~\si{ph/s}$.
We did not measure an influence of dielectric optics block on the traps surface on the heating rate. This observation is consistent with simulations using the model of
\cite{Teller2021}  
which predicts negligible heating at a distance of $1~\si{mm}$.
The heating rate on the axial mode as function of the axial mode frequency is shown in 
Fig.~\ref{fig:measurements: heating rates vs axial freq}.
A powerlaw fit results in an exponent $1.92\pm0.23$, indicating that the resulting heating rate is most likely due to technical noise on the DC electrodes, 
which is filtered by first order RC filters located within the cryogenic chamber close to the trap.

\begin{figure}[ht!]
  \centering
  \includegraphics[width=0.5\textwidth]{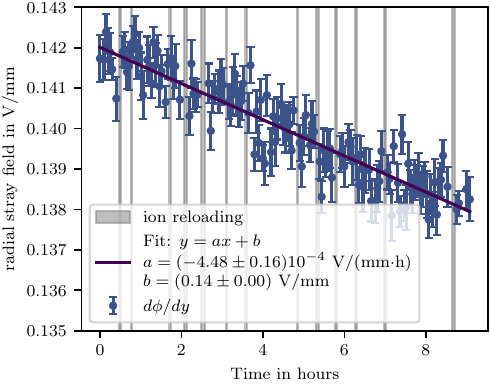}
	\caption{
    Stray charge measurement over 9 hours  in the addressing zone. 
    The stray charges is measured by scanning over y compensation shims and 
    measuring where the excitation of the 
    micromotion sideband of the
    $\mathrm{S}_{1/2} \leftrightarrow 
    \mathrm{D}_{5/2}$
    transition is minimal.
    The grey areas indicate time intervals where ions where reloaded 
    (i.e. photoionization lasers as well as the oven is switched on )
    The accumulated loading duration over the course of the four hours was $40$~minutes
    which corresponds to $\approx 7 ~\%$ of the time. 
	}
  \label{fig:measurements: x stray fields over time close to glass}
\end{figure}
\begin{figure}[ht!]
  \centering
  \includegraphics[width=0.5\textwidth]{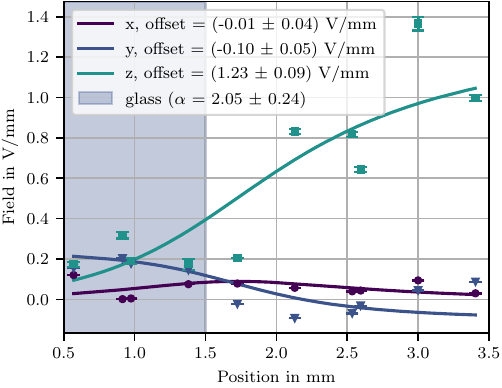}
	\caption{
    Measured stray fields along the trap axis. 
    The simulated data represents a uniform charge distribution on the glass, simulated using a FEM approach,
    The simulated stray fields are fitted to the measured stray field data with a scaling factor 
    and a individual offset in each spatial direction. 
    The fitted stray field is $2~\si{e \per \micro\metre^2}$. 
    }    
  \label{fig:measurements: stray fields radial glass}
\end{figure}

\subsection{Stray fields}
We also characterized the stray field as a function of the ion position and compare 
it to simulated values. 
To measure the radial stray field, 
we probe
the micromotion sideband of the 
$\mathrm{S}_{1/2} \leftrightarrow 
\mathrm{D}_{5/2}$,
transition while varying the voltages that compensate an electric field in $y$ or $z$ 
and 
identify the voltage 
configuration that minimizes micromotion.
The stray field is determined by taking 
the negative of the electric field simulated at these specific 
correcting voltages.

We measure the axial stray fields
we slowly 
modulating
the curvature of the axial confinement
and adjusting the stray field compensation  to a value where 
the position of the ion stays constant. 
The precision of this method is given by the resolution of the optical system used to determine the position of the ion 
as well as the span of the modulation of the axial confinement.
We have a resolution of the camera of 
($2~\si{\micro\metre \per pixel}$).
By modulating the axial secular frequency from $1~\si{MHz}$ to $100~\si{kHz}$, 
we can observe stray fields down to $10^{-3}~\si{V \per mm}$
as discussed in App.~\ref{sec:app: stray field measurement}.

The radial stray field of the glass block was monitored over time in the addressing zone at a distance of  $1~\si{mm}$ to the glass.
The measurement is carried out over night
over the course of 9 hours as shown in Fig.~\ref{fig:measurements: x stray fields over time close to glass}.
The drift of the stray field has an average of 
$(-4.48 \pm 0.16)10^{-4}\si{V \per mm \per hour}$.
During the measurement duration, ion reloading was necessary multiple times, as marked in the figure. 
Fig.~\ref{fig:measurements: x stray fields over time close to glass} 
shows that loading events do not yield fast stray field changes, 
indicating that 
neutral atom flux and photoionization lasers 
are not a major cause of stray fields for short ion loading 
periods
on the order of a few minutes. 

In addition to the temporal stability measurements,
we characterize the spatial dependence of the stray fields along the trap axis. 
The results, illustrated in Fig.~\ref{fig:measurements: stray fields radial glass}, 
are compared to a numerical model based on the Finite-Element Method (FEM) 
to identify the influence of the glass block.

In the simulation, we model the optics block with a uniform surface charge density 
of $1~\si{e\per\micro\meter^2}$ applied to all exposed surfaces. 
From this model, we extract the resulting electric field components 
at the ion's position as a function of its axial coordinate. 
We fit the simulated fields to the experimental measurements.
We employ a linear fit model consisting of a global scaling factor $\alpha$ and a vector of independent spatial offsets, 
to account for global stray field contributions, which may arise from technical noise, 
background charges, or inaccuracies in the calculated shimsets, 
The total stray field model is then 
\begin{equation}
\mathbf{E}_m (x,y,z) = \alpha \mathbf{E}_s (x,y,z) + \mathbf{E}_o  
\end{equation}
with spatially independent offsets $\mathbf{E}_o$ and simulated fields $\mathbf{E}_s$.

The best fit to the experimental data yields a scaling factor $\alpha = 2.05 \pm 0.24$,
corresponding to an inferred surface charge density on the glass block of $(2.05\pm0.24)~\si{e\per\micro\meter^2}$. 
The extracted offsets for each direction, representing the background field components, 
are 
$E_{x,o} = (-0.01\pm 0.04)~\si{V\per\milli\meter}$, 
$E_{y,o} = (-0.1\pm 0.05)~\si{V\per\milli\meter}$, 
and 
$E_{z,o} = (1.23\pm 0.09)~\si{V\per\milli\meter}$. 
The agreement between the spatial profile of the measured data and the scaled FEM simulation suggests 
that the position-dependent stray fields are primarily driven by the presence of the dielectric block.

\subsection{Integrated addressing of a trapped ion}

This section presents measurements that characterize the light intensity distribution emitted by the integrated femtosecond laser written waveguide on a trapped ion.
We measure the Rabi frequency of the 
$4^2\mathrm{S}_{1/2} \leftrightarrow 3^2\mathrm{D}_{5/2}$ transition, driven by the integrated waveguide, 
over a range of discrete ion positions in front of the waveguide, from which we can extract the 
light intensity profile emitted by the waveguide.

These measurements were done at the experimental setup at University of Innsbruck.
As qubit laser, we use a 
diode laser~\cite{Freund2024, Pogorelov2021}. 
The light is fed into the vacuum system using a 
'SQS Vacuum Pigtailed Fiber Optic Feedthrough',
custom made with a fiber for light with a wavelength of $729~\si{\nano\meter}$.
On the vacuum side of the feedthrough, the fiber is mated to the fiber that is glued to the 
ion trap chip using a mating sleeve. 
The fiber is guided through the 
$40\si{K}$
and
$4\si{K}$
cold shields.

\begin{figure}[ht!]
  \centering
  \includegraphics[width=0.5\textwidth]{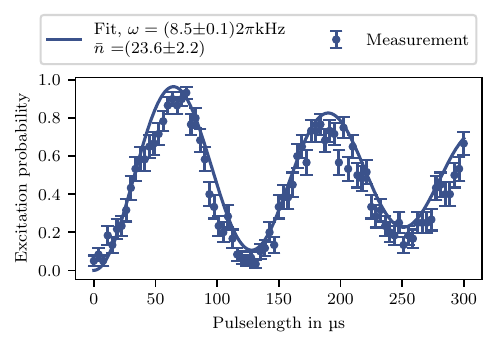}
	\caption{
  Rabiflop driven using the integrated FLW waveguide for the laserlight with a wavelength of $729~\si{\nano\meter}$. 
  Other wavelenght (for doppler cooling, repumping and state reset) were delivered via free-space optics. 
    The ion was
    trapped at a position $x=-10~\si{\micro\meter}$ with respect to waveguide.
	}
  \label{fig:measurements: rabiflop}
\end{figure}

A single ion is trapped in the addressing region, 
at an axial propagation distance of $1~\si{mm}$ from the emission point of the waveguide.
We measure the excitation probability as function of the interaction duration of the qubit laser which should yield Rabi oscillations.
The excited state excitation probability is governed by Rabi oscillations with 
multiple frequency components due to thermally distributed occupation probability
\cite{Leibfried2003}:
\begin{equation}
    P_{|1\rangle}(t) = \sum_{n=0}^{\infty} P_n(\bar{n}) \sin^2 \left( \frac{\Omega_{n,n} t}{2} \right)
\end{equation}
Where $P_n(\bar{n})$ is the phonon occupation probability of an ion with mean phonon number $\bar{n}$:
\begin{equation}
P_n(\bar{n}) = \frac{1}{\bar{n} + 1} \left( \frac{\bar{n}}{\bar{n} + 1} \right)^n.
\end{equation}
The Rabi frequency is given by $\Omega_{n,n} = \Omega(1 - \eta^2 n)$.
with $\eta$ the Lamb-Dicke parameter. 
A measured Rabi flop is shown in Fig.~\ref{fig:measurements: rabiflop}.

\begin{figure}[ht!]
  \centering
  \includegraphics[width=0.5\textwidth]{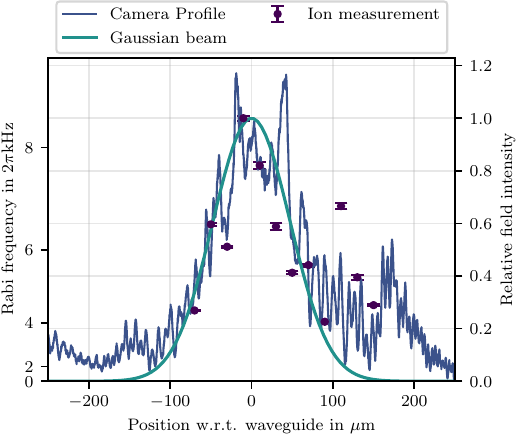}
	\caption{
    Rabi frequencies measured along to trap axis, 
    using the integrated FLW waveguide for the
    qubit laser at $729~\si{nm}$ wavelength.
    The blue curve shows the profile of the beam extracted from Fig.~\ref{fig:chip1_light_profile2}B
    and is scaled to fit the experimental data. 
    The turquoise line represents a gaussian mode with a radius of $94.9~\si{\micro\meter}$, 
    which is the modefield diameter at the ion measured in Fig.~\ref{fig:fabrication: modeprofile datahero device}.
	}
  \label{fig:measurements: position vs intensity}
\end{figure}

We measure the relative light intensity along the trap axis 
by shuttling the ion
in steps of 20 \si{\micro\meter}.
At each position, we measure the local Rabi frequency,
as shown by the purple data points in Fig.~\ref{fig:measurements: position vs intensity}. The measurement range in the negative x-direction is limited by geometrical constraints on the free-space access for the cooling and repumping lasers.
The intensity of the local field cannot be inferred from the Rabi frequency, because the polarization of the emitted light is not known.

Additionally, the intensity profile of the light was measured with a camera as described in Sec.~\ref{subsec: fabrication assembly} (blue curve). The right axis in Fig.~\ref{fig:measurements: position vs intensity} displays the relative laser intensity, normalized to its maximum measured value from these measurements and is rescaled to match the one estimated from the ion measurements. The camera profile reveals a main intensity peak centered around the waveguide position. Moreover,
an additional secondary peak is present near $x=170~\si{\micro\meter}$. 
For comparison, the intensity distribution of an ideal Gaussian beam with a radius of $94.9~\si{\micro\meter}$
(taken from Fig.~\ref{fig:fabrication: modeprofile datahero device}) is also shown. The overall shape and width of the main peak show good agreement between the camera profile and ion data.
A notable discrepancy is the secondary peak observed in the ion data around $x=110~\si{\micro\meter}$,
which is not fully understood, as it is absent in the camera profile.

Both the camera data and the ion-based measurements show significant local intensity variations.
These fluctuations likely originate from surface imperfections in the manually cleaved out-coupling facet of the waveguide.
Local defects on the glass surface can perturb the emitted wavefront,
leading to interference effects that may be the cause of the observed fluctuations in the intensity profile
as well as the secondary peak.

\section{Summary and conclusions}
\label{sec: summary}

In this work, we have demonstrated an ion surface trap platform, 
fabricated on a borosilicate glass substrate with an integrated femtosecond-laser-written waveguide. 
By inscribing the waveguide parallel to the trap's surface, this design delivers light directly to the ion, circumventing the need for out-of-plane coupling optics traditionally required in planar slab waveguides. 

We have shown that the waveguide mode field diameter can be reliably engineered and tuned by adjusting the laser writing parameters.
End-to-end testing with trapped \calciumforty ions in a cryogenic environment validated that the platform maintains excellent trapping conditions. 
Heating rates remained low, between 0.5 quanta/s and 1.5 quanta/s at an axial secular frequency of 
$1 \cdot 2\pi ~\si{MHz}$. 

Crucially, the millimeter-scale proximity of the bonded dielectric optics block on the trap surface did not induce any measurable dielectric heating. 
Furthermore, stray electric fields introduced by the glass block were found to be remarkably stable, 
exhibiting an average drift of
$(-4.48 \pm 0.16)10^{-4}\si{V \per mm \per hour}$.
Because of this stability, experimental recalibration of the micromotion compensation is only necessary every few hours. 
Using these integrated photonic elements, we successfully performed coherent qubit operations, 
by demonstrating Rabi oscillations.

Despite these successes, the current platform faces limitations that will inform future design iterations. 
A primary constraint of femtosecond-laser-written waveguides is their minimum radius of curvature, 
which is currently limited to the order of a few millimeters. 
Additionally, the straight, unlensed waveguide outputs a beam radius of roughly 100 µm at the ion's location, limiting the integrated addressing to a global beam. 
However, this platform is highly compatible with hybrid integration, 
the waveguides can be combined with optical elements added via a pick-and-place. 
For instance, adding a focusing borosilicate glass ball lens has been shown to achieve tight, 
nearly circular beam waists of approximately $5~\si{\micro\meter}$
\cite{Gruenberg2023}. These or similar lenses should be compatible with our approach.

\section*{Acknowledgments}
\label{sec: acknowledgments}

This work was supported by the 
 Austrian Research Promotion Agency
(FFG) through the "Optically Integrated Quantum Computing (OptoQuant)" project, Project number 884453. 
We gratefully acknowledge support by the European Union’s Horizon Europe research and innovation program under Grant Agreement Number 101114305 (“MILLENION-SGA1” EU Project), 
the European Union’s Horizon Europe research and innovation program under Grant Agreement Number 101046968 (BRISQ), 
the Austrian Science Fund (FWF Grant-DOI 10.55776/F71, 10.55776/COE1), 
by the European Union’s Horizon Chips Joint-Undertaking under Grant Agreement Number 101288915 (“CHAMP-ION-SGA1”), 
as well as the Intelligence Advanced Research Projects Activity (IARPA) and the Army Research Office, under the Entangled Logical Qubits program through Cooperative Agreement Number W911NF-23-2-0216. 

\section*{Author Contributions}
J.W., 
A.Z., 
P.H., 
M.S., 
M.V., 
T.M., 
B.L., 
K.S. 
and 
P.S. 
designed the experiments. 
J.W., 
A.Z., 
P.H., 
M.S., 
V.S., 
M.P. 
and 
M.V. 
carried out the measurements and analyzed the data. 
J.W., 
A.Z. 
K.S. 
and 
P.H. 
fabricated the devices. 
J.W., 
A.Z.,
P.H., 
K.S. 
and 
P.S. 
wrote the manuscript. 
All authors reviewed the manuscript. 
C.R., K.S., T.M., B.L. 
and 
P.S. 
supervised the project.

\clearpage
\appendix

\section{Electrode layout of the ion trap}
\label{app: electrode layout}

We find the optimal trap geometry by using the width of the RF electrodes,
$w_{\mathrm{RF}}$,
as our primary independent variable,
from which all other 
design parameters are derived: The separation between the RF electrodes is 
chosen so that the height of the pseudopotential minimum matches the desired 
ion height of $170~\si{\micro\meter}$, and the length of the DC electrode is chosen such that the 
maximum required DC voltage for  axial confinement is minimal. 
How the static voltages for trapping are calculated is shown in appendix
\ref{app:voltage calculation}.

\begin{figure}[ht!]
  \centering
  \includegraphics[width=0.5\textwidth]{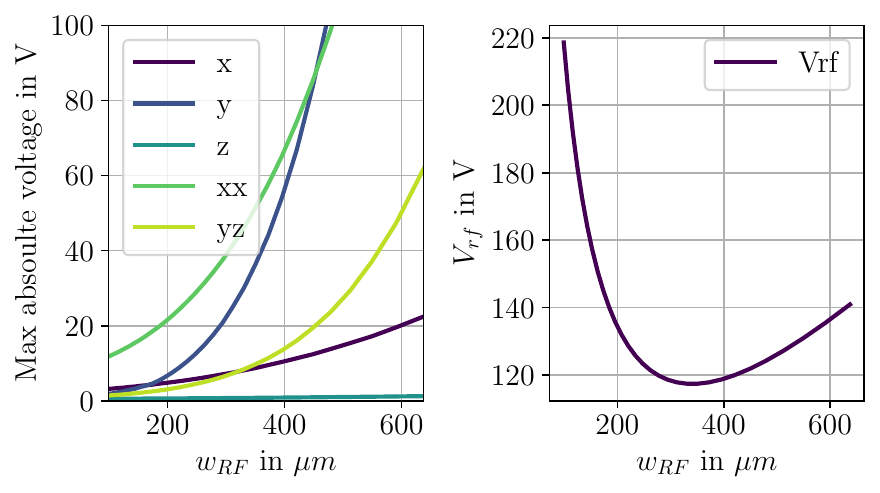}
	\caption{
  \label{fig:fabrication: voltages vs rf width} 
    Voltages required for ion trapping at an ion height of 
    $x_0=170~\si{\micro\meter}$, and secular frequencies of $1~\si{MHz}$ for 
    the axial and $2~\si{MHz}$ for the radial modes, depending on the width 
    of the RF electrodes. Left: Maximum DC voltage required to generate 
    $1\si{MHz}$ axial frequency. Right: Amplitude of the RF voltage required 
    to generate $2\si{MHz}$ radial secular frequencies.
	}
\end{figure}

The required RF amplitude and DC voltages as a function of $w_{\mathrm{RF}}$ 
are shown in Fig.~\ref{fig:fabrication: voltages vs rf width}.
The two graphs depict the design tradeoff: The required RF amplitude reaches 
a global minimum of $V_{\mathrm{RF}} \approx 117~\si{\volt}$ at an RF 
electrode width of $w_{\mathrm{RF}} \approx 350~\si{\micro\meter}$. 
However, wider RF rails push the outer DC segments further away from the ion. 
This increased distance inflates the DC voltages needed to maintain the same 
axial confinement. 

We balanced these competing requirements by selecting an
RF electrode width of 
$w_{\mathrm{RF}} = 250~\si{\micro\meter}$. At this dimension, the required RF 
amplitude remains moderate (below $140~\si{\volt}$), while the maximum 
required DC voltage is kept well below our hardware limit of $40~\si{\volt}$. 
Based on this RF width, the corresponding optimal length for the DC segments 
is $l_{\mathrm{dc}} = 400~\si{\micro\meter}$. Finally, we set the width of 
the DC segments to $w_{\mathrm{dc}} = 300~\si{\micro\meter}$; our simulations 
indicated that widening them beyond this point yields no relevant reduction 
in the required trapping voltages.

\section{Voltage sets for shuttling}
\label{app:voltage calculation}

This appendix describes the methodology used to calculate the voltage
waveforms required for ion shuttling. 
We extend the standard analytical calculation
of static shimsets~\cite{Allcock2010}, to generate continuous waveforms while dynamically
switching between subsets of available electrodes. 
Notably, this approach
achieves continuity of the voltages without the need for a
computational optimization routine. 
We first briefly outline the
prerequisites for calculating static trapping potentials before
detailing the construction of the dynamic shuttling waveforms.

\subsection{Stationary trapping}
\label{app:voltage theory}

The potential of a single electrode is calculated using a
Biot-Savart-like 
approach~\cite{Oliveira_2001}, assuming a gapless infinite plane. 
We extract \textit{unit voltage potentials} of each electrode by applying $1\si{\volt}$ to the target electrode while grounding all others. 
For RF electrodes, we use the pseudopotential approximation~\cite{Dehmelt1968}.

The total potential $\phi = \sum_i u_i \phi_i$ is a superposition of unit 
potentials $\phi_i$ scaled by the applied voltages
 $\mathbf{u}=[u_1, \dots, u_n]^T$.
 We characterize the effect of electrode $i$ on the ion by decomposing its
its unit voltage potential into spatial multipoles:
$\mathbf{B}_{-i} = \left[ 
\frac{\partial}{\partial x},
\frac{\partial}{\partial y},
\frac{\partial}{\partial z},
\frac{\partial^2}{\partial z^2} 
\right]^T
\phi_i(x,y,z)$.
And construct the multipole matrix 
$\mathbf{B} = [
\mathbf{B}_{-1}, 
\dots, 
\mathbf{B}_{-n}, ]$.
We relate 
the applied 
voltages $\mathbf{u}$ to the total multipole field $\mathbf{b}$ at the ion~\cite{Allcock2010}:

\begin{equation}
    \label{eq:fabrication: linear system of equations}
    \mathbf{b} = \mathbf{B}\mathbf{u}
\end{equation}

By solving this linear system of equations, specific shim sets can be calculated. 
E.g. solving 
    Eq.~\ref{eq:fabrication: linear system of equations} 
    with
$\mathbf{b}=[0,0,0,1]^T$ 
yields a voltage set that generates a harmonic trapping potential in the axial direction 
with a curvature of  
$1\si{\volt\per\metre^2}$.

The least-squares solution $\mathbf{u}_m$ is:
\begin{equation}
\label{eq:fabrication: single solution shim matrix}
    \mathbf{u}_m = \mathbf{B}^+ \mathbf{b}
\end{equation}
where $\mathbf{B}^+$ is the pseudo-inverse of $\mathbf{B}$.
The complete solution space $\mathbf{U}$ is defined by:
\begin{equation}
\label{eq:fabrication: shim full vector space}
    \mathbf{U} = \mathbf{u}_m + \mathrm{Span} \left\{ \mathbb{1}_n - \mathbf{B}^+ \mathbf{B} \right\}
\end{equation}

If 
Eq.~\ref{eq:fabrication: linear system of equations} has a solution, 
 $\mathbf{u}_m$ represents the solution with the smallest Euclidean norm. 
We can also find a specific voltage vector $\mathbf{u}_c \in \mathbf{U}$ that minimizes the Euclidean distance to an arbitrary target voltage vector $\mathbf{y}$ (which is not necessarily within $\mathbf{U}$) using:
\begin{equation}
    \label{eq:fabrication: closest solution}
    \mathbf{u}_c = \mathbf{u}_m +  \left( \mathbb{1}_n - \mathbf{B}^+ \mathbf{B} \right) \mathbf{y}
\end{equation}
This expression for $\mathbf{u}_c$ is used in the following subsection to calulate
smooth waveforms for shuttling.

\subsection{Waveforms for shuttling}

We express the  axial position of the ion during shuttling with the dimensionless parameter 
$\xi(x) = x / (w_{dc} + w_{gap})$ relative to the segmented DC electrodes. 
$\xi=0$ denotes the center of electrode pair 1 (see Fig.~\ref{fig:fabrication: trap geom}). 
Shuttling waveforms are generated by 
solving Eq.~\ref{eq:fabrication: linear system of equations} 
across $\xi$ and mapping these spatial potentials to time.
If the same set set of $n$ electrodes is used to calculate $\mathbf{B}$ 
for all positions, 
Eq.\ref{eq:fabrication: single solution shim matrix} 
can be used to calculate voltages for each position 
and the resulting voltages as function of position $\mathbf{u}_m(\xi)$ will be continuous.
However, this means that the voltages of all $n$ electrodes can be non-zero
thus 
one would need to able to change all of them simultanously in the experiment.

\begin{figure}[ht!]
  \centering
    \includegraphics[width=0.5\textwidth]{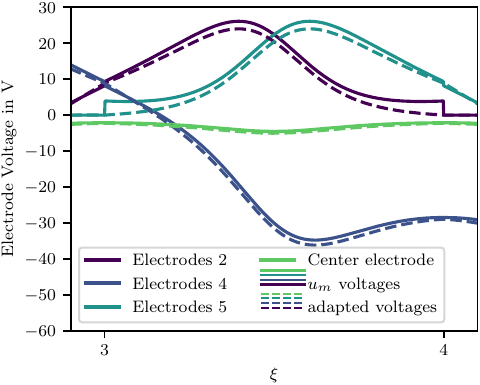}
    \caption{
    Shuttling waveforms applied to electrode pairs 2, 4, and 5 
    (following Fig.~\ref{fig:fabrication: trap geom}) to achieve an axial trapping potential curvature corresponding to a $1\si{\mega\hertz}$ secular frequency.
    \textbf{Solid lines:} Voltages calculated directly using Eq.~\ref{eq:fabrication: single solution shim matrix}. 
    Discontinuous jumps occur when the active subset of electrodes switches as the ion moves 
    (for $\xi<3$, 
    electrodes pairs 1,2,3,4 are used.
    For $3<\xi<4$, electrode pair 1 is swapped for pair 5.
    At $\xi=4$, pair 2 is swapped for pair 6.).
    \textbf{Dashed lines:} Continuous voltage waveforms achieved after the linear interpolation and projection smoothing process.
    }
  \label{fig:fabrication: waveform before and after smooting}
\end{figure}

We accommodate hardware limitations by restricting the active electrodes to a
sliding window of the four pairs closest to the ion's current position\footnote{For example, we use electrode pairs 1, 2, and 3 for trapping centered above pair 2. We use pairs 1, 2, 3, and 4 for shuttling from the center of pair 2 to the center of pair 3, and pairs 2, 3, 4, and 5 to shuttle to the center of pair 4, and so on.}. Directly applying Eq.~\ref{eq:fabrication: single solution shim matrix} while discretely switching this active electrode subset introduces undesirable discontinuous voltage jumps (solid lines in Fig.~\ref{fig:fabrication: waveform before and after smooting}).

We generate continuous waveforms by defining a set of 
\textit{nodes} $\mathcal{N}=[N_1, \dots, N_m]$ 
where electrode switching occurs. 
For each segment between $N_i$ and $N_{i+1}$, a specific electrode subset $s_i$ is active. 
Using Eq.~\ref{eq:fabrication: single solution shim matrix},
we first calculate valid shim sets $\mathbf{u}_{m, N_i}$ at each node $N_i$ 
using the intersection of adjacent electrode subsets ($s_{i-1} \cap s_i$). 

For intermediate positions ($N_i < \xi < N_{i+1}$), we linearly interpolate:
\begin{equation}
    \mathbf{u}_{\mathrm{interp}}(\xi) = \mathbf{u}_{m, N_i} + (\mathbf{u}_{m, N_{i+1}} - \mathbf{u}_{m, N_i}) \frac{\xi - N_i}{N_{i+1} - N_i}
\end{equation}
Because $\mathbf{u}_{\mathrm{interp}}(\xi)$ 
is not guaranteed to strictly satisfy the target potential constraints 
($\mathbf{u}_{\mathrm{interp}}(\xi) \notin \mathbf{U}$), 
we project it onto the valid solution space using 
Eq.~\ref{eq:fabrication: closest solution}
 to find the closest valid shim set $\mathbf{u}_{\mathrm{shim}}(\xi)$:
\begin{equation}
    \mathbf{u}_{\mathrm{shim}}(\xi) = \mathbf{u}_m(\xi) + \left( \mathbb{1}_n - \mathbf{B}^+ \mathbf{B} \right) \mathbf{u}_{\mathrm{interp}}(\xi)
\end{equation}
The resulting smoothed shim sets as a function of ion position are shown in Fig.~\ref{fig:fabrication: waveform before and after smooting}.

\section{Measured axial secular frequency}
\label{app: secular freq}

\begin{figure}[ht!]
  \centering
  \includegraphics[width=0.5\textwidth]{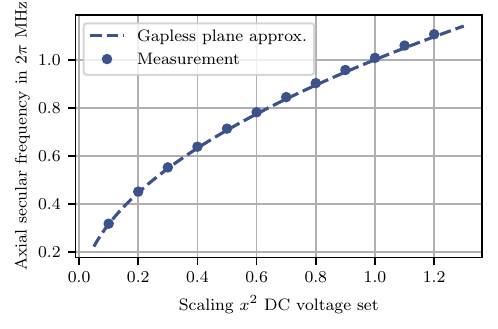}
	\caption{
    Measured axial secular frequency versus calculations derived from the gapless plane approximation method. 
    Frequencies were measured by recording ion fluorescence while sweeping the 
    RF parametric excitation. 
    The resonance frequency, i.e. the secular frequency of the ion was extracted by fitting a gaussian shaped curve to
    each data series and the fitted x-axis offset was used as secular frequency.
	}
  \label{fig:app: axial secular freqs}
\end{figure}

This section shows measurements of the axial secular frequency and compares it to values expected from the simulation.
Axial secular frequencies were determined via parametric excitation ("tickling") by applying an oscillating AC signal to a nearby segmented DC electrode. 
As the tickling frequency is swept, resonance with the ion's secular motion increases its motional energy, 
which is detected as a drop in fluorescence from the resonant cooling laser. 
Fig.~\ref{fig:app: axial secular freqs} compares these measured frequencies with simulated analytical values.

\section{Stray field measurement}
\label{sec:app: stray field measurement}

In order to accurately characterize and compensate for axial stray electric fields, we modulate the curvature of the axial confining potential. 
By adjusting the DC shim voltages until the ion's position remains strictly stationary on the imaging camera during this modulation, 
we can pinpoint the condition where the  stray field is compensated. 
The fundamental precision of this micromotion compensation technique is determined by the optical resolution of the 
camera used to track the ion and the modulation span of the axial secular frequency.

\begin{figure}[ht!]
  \centering
  \includegraphics[]{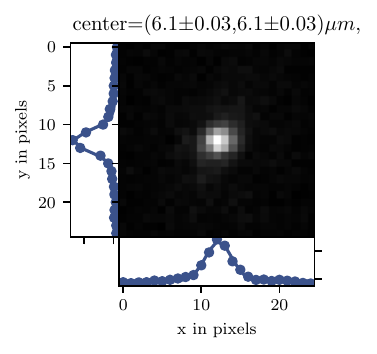}
  \includegraphics[]{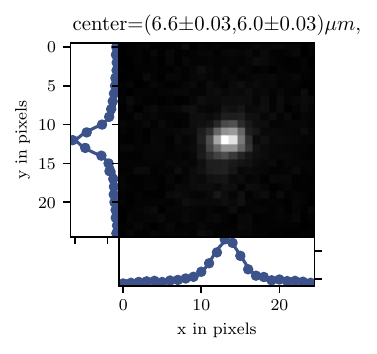}
	\caption{
    Camera images of the ion taken for two different x compensation values of 
    $-0.0214~\si{\volt\per\milli\meter}$ 
    (upper image)
     and 
    $-0.0204~\si{\volt\per\milli\meter}$ 
    (lower image).
    The plots on the side (bottom) of the images show the sum of counts of each row (columns) of pixels.
    The summed counts are fitted with gaussian fits to extract the $y$ ($x$) position of the ion. 
    The its show that the difference of the compensation value of
    $10^{-3}~\si{\volt\per\milli\meter}$  has shifted the ion by 
    $(0.5 \pm 0.042)~\si{\micro\meter}$.
    }
    \label{fig:app: ion image}
\end{figure}

\begin{figure}[ht!]
  \centering
  \includegraphics[width=0.5\textwidth]{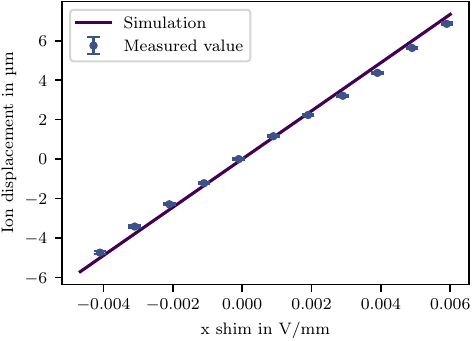}
	\caption{
    Ion displacement versus the applied $x$-shim compensation voltage, 
    measured at an axial secular frequency of $220\si{\kilo\hertz}$. 
    The origin on the $x$-axis represents the compensation setting where modulating the axial potential curvature 
    produced no visible ion displacement. 
    The data demonstrates a resolution limit of $\sim10^{-3}\si{\volt\per\milli\metre}$, 
    establishing the minimum resolvable field for micromotion compensation using this method.
    }
    \label{fig:app: position vs x shim}
\end{figure}

We quantify this resolution limit, by recording the ion's spatial displacement as a function of an applied $x$-shim electric field.
Example images for different compensation values are shown in Fig.~\ref{fig:app: ion image}.
The imaging system has a spatial resolution of $2\si{\micro\metre}$ per pixel. 
Fig.~\ref{fig:app: position vs x shim} displays the resulting displacement when tested at an axial secular frequency of $220\si{\kilo\hertz}$.
The zero point on the horizontal axis corresponds to the perfectly compensated state, where modulating the axial potential from $1\si{\mega\hertz}$ down to $100\si{\kilo\hertz}$ induced no observable change in the ion's position.

The measured ion position corresponds well with the simulated shift, as shown in  
Fig.~\ref{fig:app: position vs x shim}, 
As the slope of the ion shift is $(1.1 \cdot 10^3 \pm 9 )~\si{\volt \per \milli \meter}$
we can reliably resolve spatial shifts induced by $10^{-3}\si{\volt\per\milli\metre}$ field increments, as this already corresponds
to a shift of roughly one pixel on the camera. 
Consequently, we conclude that $10^{-3}\si{\volt\per\milli\metre}$ represents the resolution limit of our stray field measurement and micromotion compensation method using this optical technique.

\clearpage
\bibliography{main.bib}

\begin{thebibliography}{44}%
\makeatletter
\providecommand \@ifxundefined [1]{%
 \@ifx{#1\undefined}
}%
\providecommand \@ifnum [1]{%
 \ifnum #1\expandafter \@firstoftwo
 \else \expandafter \@secondoftwo
 \fi
}%
\providecommand \@ifx [1]{%
 \ifx #1\expandafter \@firstoftwo
 \else \expandafter \@secondoftwo
 \fi
}%
\providecommand \natexlab [1]{#1}%
\providecommand \enquote  [1]{``#1''}%
\providecommand \bibnamefont  [1]{#1}%
\providecommand \bibfnamefont [1]{#1}%
\providecommand \citenamefont [1]{#1}%
\providecommand \href@noop [0]{\@secondoftwo}%
\providecommand \href [0]{\begingroup \@sanitize@url \@href}%
\providecommand \@href[1]{\@@startlink{#1}\@@href}%
\providecommand \@@href[1]{\endgroup#1\@@endlink}%
\providecommand \@sanitize@url [0]{\catcode `\\12\catcode `\$12\catcode `\&12\catcode `\#12\catcode `\^12\catcode `\_12\catcode `\%12\relax}%
\providecommand \@@startlink[1]{}%
\providecommand \@@endlink[0]{}%
\providecommand \url  [0]{\begingroup\@sanitize@url \@url }%
\providecommand \@url [1]{\endgroup\@href {#1}{\urlprefix }}%
\providecommand \urlprefix  [0]{URL }%
\providecommand \Eprint [0]{\href }%
\providecommand \doibase [0]{https://doi.org/}%
\providecommand \selectlanguage [0]{\@gobble}%
\providecommand \bibinfo  [0]{\@secondoftwo}%
\providecommand \bibfield  [0]{\@secondoftwo}%
\providecommand \translation [1]{[#1]}%
\providecommand \BibitemOpen [0]{}%
\providecommand \bibitemStop [0]{}%
\providecommand \bibitemNoStop [0]{.\EOS\space}%
\providecommand \EOS [0]{\spacefactor3000\relax}%
\providecommand \BibitemShut  [1]{\csname bibitem#1\endcsname}%
\let\auto@bib@innerbib\@empty
\bibitem [{\citenamefont {Bruzewicz}\ \emph {et~al.}(2019)\citenamefont {Bruzewicz}, \citenamefont {Chiaverini}, \citenamefont {McConnell},\ and\ \citenamefont {Sage}}]{Bruzewicz2019}%
  \BibitemOpen
  \bibfield  {author} {\bibinfo {author} {\bibfnamefont {C.~D.}\ \bibnamefont {Bruzewicz}}, \bibinfo {author} {\bibfnamefont {J.}~\bibnamefont {Chiaverini}}, \bibinfo {author} {\bibfnamefont {R.}~\bibnamefont {McConnell}},\ and\ \bibinfo {author} {\bibfnamefont {J.~M.}\ \bibnamefont {Sage}},\ }\bibfield  {title} {\bibinfo {title} {Trapped-ion quantum computing: Progress and challenges},\ }\href {https://doi.org/10.1063/1.5088164} {\bibfield  {journal} {\bibinfo  {journal} {Applied Physics Reviews}\ }\textbf {\bibinfo {volume} {6}},\ \bibinfo {pages} {021314} (\bibinfo {year} {2019})}\BibitemShut {NoStop}%
\bibitem [{\citenamefont {Wang}\ \emph {et~al.}(2021)\citenamefont {Wang}, \citenamefont {Luan}, \citenamefont {Qiao}, \citenamefont {Um}, \citenamefont {Zhang}, \citenamefont {Wang}, \citenamefont {Yuan}, \citenamefont {Gu}, \citenamefont {Zhang},\ and\ \citenamefont {Kim}}]{Wang2021}%
  \BibitemOpen
  \bibfield  {author} {\bibinfo {author} {\bibfnamefont {P.}~\bibnamefont {Wang}}, \bibinfo {author} {\bibfnamefont {C.-Y.}\ \bibnamefont {Luan}}, \bibinfo {author} {\bibfnamefont {M.}~\bibnamefont {Qiao}}, \bibinfo {author} {\bibfnamefont {M.}~\bibnamefont {Um}}, \bibinfo {author} {\bibfnamefont {J.}~\bibnamefont {Zhang}}, \bibinfo {author} {\bibfnamefont {Y.}~\bibnamefont {Wang}}, \bibinfo {author} {\bibfnamefont {X.}~\bibnamefont {Yuan}}, \bibinfo {author} {\bibfnamefont {M.}~\bibnamefont {Gu}}, \bibinfo {author} {\bibfnamefont {J.}~\bibnamefont {Zhang}},\ and\ \bibinfo {author} {\bibfnamefont {K.}~\bibnamefont {Kim}},\ }\bibfield  {title} {\bibinfo {title} {Single ion qubit with estimated coherence time exceeding one hour},\ }\href {https://doi.org/10.1038/s41467-020-20330-w} {\bibfield  {journal} {\bibinfo  {journal} {Nature Communications}\ }\textbf {\bibinfo {volume} {12}},\ \bibinfo {pages} {233} (\bibinfo {year} {2021})}\BibitemShut {NoStop}%
\bibitem [{\citenamefont {Hughes}\ \emph {et~al.}(2025)\citenamefont {Hughes}, \citenamefont {Srinivas}, \citenamefont {Löschnauer}, \citenamefont {Knaack}, \citenamefont {Matt}, \citenamefont {Ballance}, \citenamefont {Malinowski}, \citenamefont {Harty},\ and\ \citenamefont {Sutherland}}]{ionq2025}%
  \BibitemOpen
  \bibfield  {author} {\bibinfo {author} {\bibfnamefont {A.~C.}\ \bibnamefont {Hughes}}, \bibinfo {author} {\bibfnamefont {R.}~\bibnamefont {Srinivas}}, \bibinfo {author} {\bibfnamefont {C.~M.}\ \bibnamefont {Löschnauer}}, \bibinfo {author} {\bibfnamefont {H.~M.}\ \bibnamefont {Knaack}}, \bibinfo {author} {\bibfnamefont {R.}~\bibnamefont {Matt}}, \bibinfo {author} {\bibfnamefont {C.~J.}\ \bibnamefont {Ballance}}, \bibinfo {author} {\bibfnamefont {M.}~\bibnamefont {Malinowski}}, \bibinfo {author} {\bibfnamefont {T.~P.}\ \bibnamefont {Harty}},\ and\ \bibinfo {author} {\bibfnamefont {R.~T.}\ \bibnamefont {Sutherland}},\ }\href {https://arxiv.org/abs/2510.17286} {\bibinfo {title} {Trapped-ion two-qubit gates with $>$99.99\% fidelity without ground-state cooling}} (\bibinfo {year} {2025}),\ \Eprint {https://arxiv.org/abs/2510.17286} {arXiv:2510.17286 [quant-ph]} \BibitemShut {NoStop}%
\bibitem [{\citenamefont {Smith}\ \emph {et~al.}(2025)\citenamefont {Smith}, \citenamefont {Leu}, \citenamefont {Miyanishi}, \citenamefont {Gely},\ and\ \citenamefont {Lucas}}]{smith2025}%
  \BibitemOpen
  \bibfield  {author} {\bibinfo {author} {\bibfnamefont {M.~C.}\ \bibnamefont {Smith}}, \bibinfo {author} {\bibfnamefont {A.~D.}\ \bibnamefont {Leu}}, \bibinfo {author} {\bibfnamefont {K.}~\bibnamefont {Miyanishi}}, \bibinfo {author} {\bibfnamefont {M.~F.}\ \bibnamefont {Gely}},\ and\ \bibinfo {author} {\bibfnamefont {D.~M.}\ \bibnamefont {Lucas}},\ }\bibfield  {title} {\bibinfo {title} {Single-qubit gates with errors at the ${10}^{\ensuremath{-}7}$ level},\ }\href {https://doi.org/10.1103/42w2-6ccy} {\bibfield  {journal} {\bibinfo  {journal} {Phys. Rev. Lett.}\ }\textbf {\bibinfo {volume} {134}},\ \bibinfo {pages} {230601} (\bibinfo {year} {2025})}\BibitemShut {NoStop}%
\bibitem [{\citenamefont {Ransford}\ \emph {et~al.}(2025)\citenamefont {Ransford}, \citenamefont {Allman}, \citenamefont {Arkinstall}, \citenamefont {III}, \citenamefont {Cooper}, \citenamefont {Delaney}, \citenamefont {Dreiling}, \citenamefont {Estey}, \citenamefont {Figgatt}, \citenamefont {Hall}, \citenamefont {Husain}, \citenamefont {Isanaka}, \citenamefont {Kennedy}, \citenamefont {Kotibhaskar}, \citenamefont {Madjarov}, \citenamefont {Mayer}, \citenamefont {Milne}, \citenamefont {Park}, \citenamefont {Reed}, \citenamefont {Ancona}, \citenamefont {Andersen}, \citenamefont {Andres-Martinez}, \citenamefont {Angenent}, \citenamefont {Argueta}, \citenamefont {Arkin}, \citenamefont {Ascarrunz}, \citenamefont {Baker}, \citenamefont {Barnes}, \citenamefont {Bartolotta}, \citenamefont {Berg}, \citenamefont {Besand}, \citenamefont {Bjork}, \citenamefont {Blain}, \citenamefont {Blanchard}, \citenamefont {Blume-Kohout}, \citenamefont {Bohn}, \citenamefont {Borgna}, \citenamefont {Botamanenko}, \citenamefont
  {Boutelle}, \citenamefont {Brown}, \citenamefont {Buckingham}, \citenamefont {Burdick}, \citenamefont {Burton}, \citenamefont {Carey}, \citenamefont {Carron}, \citenamefont {Chambers}, \citenamefont {Children}, \citenamefont {Colussi}, \citenamefont {Crepinsek}, \citenamefont {Cureton}, \citenamefont {Davies}, \citenamefont {Davis}, \citenamefont {DeCross}, \citenamefont {Deen}, \citenamefont {Delaney}, \citenamefont {DelVento}, \citenamefont {DeSalvo}, \citenamefont {Dominy}, \citenamefont {Duncan}, \citenamefont {Eccles}, \citenamefont {Edgington}, \citenamefont {Erickson}, \citenamefont {Erickson}, \citenamefont {Ertsgaard}, \citenamefont {Evans}, \citenamefont {Evans}, \citenamefont {Fabrikant}, \citenamefont {Fischer}, \citenamefont {Foltz}, \citenamefont {Foss-Feig}, \citenamefont {Francois}, \citenamefont {Freyberg}, \citenamefont {Gao}, \citenamefont {Garay}, \citenamefont {Garvin}, \citenamefont {Gaudiosi}, \citenamefont {Gilbreth}, \citenamefont {Giles}, \citenamefont {Glynn}, \citenamefont
  {Graves}, \citenamefont {Hansen}, \citenamefont {Hayes}, \citenamefont {Heidemann}, \citenamefont {Higashi}, \citenamefont {Hilbun}, \citenamefont {Hines}, \citenamefont {Hlavaty}, \citenamefont {Hoffman}, \citenamefont {Hoffman}, \citenamefont {Holliman}, \citenamefont {Hooper}, \citenamefont {Horning}, \citenamefont {Hostetter}, \citenamefont {Hothem}, \citenamefont {Houlton}, \citenamefont {Hout}, \citenamefont {Hutson}, \citenamefont {Jacobs}, \citenamefont {Jacobs}, \citenamefont {Johannsen}, \citenamefont {Johansen}, \citenamefont {Jones}, \citenamefont {Julian}, \citenamefont {Jung}, \citenamefont {Keay}, \citenamefont {Klein}, \citenamefont {Koch}, \citenamefont {Kondo}, \citenamefont {Kong}, \citenamefont {Kosto}, \citenamefont {Lawrence}, \citenamefont {Liefer}, \citenamefont {Lollie}, \citenamefont {Lucchetti}, \citenamefont {Lysne}, \citenamefont {Lytle}, \citenamefont {MacPherson}, \citenamefont {Malm}, \citenamefont {Mather}, \citenamefont {Mathewson}, \citenamefont {Maxwell}, \citenamefont
  {McCaffrey}, \citenamefont {McDougall}, \citenamefont {Mendoza}, \citenamefont {Mills}, \citenamefont {Morrison}, \citenamefont {Narmour}, \citenamefont {Nguyen}, \citenamefont {Nugent}, \citenamefont {Olson}, \citenamefont {Ouellette}, \citenamefont {Parks}, \citenamefont {Peters}, \citenamefont {Petricka}, \citenamefont {Pino}, \citenamefont {Polito}, \citenamefont {Preidl}, \citenamefont {Price}, \citenamefont {Proctor}, \citenamefont {Pugh}, \citenamefont {Ratcliff}, \citenamefont {Raymondson}, \citenamefont {Rhodes}, \citenamefont {Roman}, \citenamefont {Roy}, \citenamefont {Ryan-Anderson}, \citenamefont {Sanchez}, \citenamefont {Sangiolo}, \citenamefont {Sawadski}, \citenamefont {Schaffer}, \citenamefont {Schow}, \citenamefont {Sedlacek}, \citenamefont {Semenenko}, \citenamefont {Shevchuk}, \citenamefont {Shore}, \citenamefont {Siegfried}, \citenamefont {Singhal}, \citenamefont {Sivarajah}, \citenamefont {Skripka}, \citenamefont {Sletten}, \citenamefont {Spaun}, \citenamefont {Sprenkle}, \citenamefont
  {Stoufer}, \citenamefont {Tader}, \citenamefont {Taylor}, \citenamefont {Thompson}, \citenamefont {Tobey}, \citenamefont {Tran}, \citenamefont {Tran}, \citenamefont {Vittorini}, \citenamefont {Volin}, \citenamefont {Walker}, \citenamefont {White}, \citenamefont {Wilson}, \citenamefont {Wolf}, \citenamefont {Wringe}, \citenamefont {Young}, \citenamefont {Zheng}, \citenamefont {Zuraski}, \citenamefont {Baldwin}, \citenamefont {Chernoguzov}, \citenamefont {Gaebler}, \citenamefont {Sanders}, \citenamefont {Neyenhuis}, \citenamefont {Stutz},\ and\ \citenamefont {Bohnet}}]{quantinuum2025}%
  \BibitemOpen
  \bibfield  {author} {\bibinfo {author} {\bibfnamefont {A.}~\bibnamefont {Ransford}}, \bibinfo {author} {\bibfnamefont {M.~S.}\ \bibnamefont {Allman}}, \bibinfo {author} {\bibfnamefont {J.}~\bibnamefont {Arkinstall}}, \bibinfo {author} {\bibfnamefont {J.~P.~C.}\ \bibnamefont {III}}, \bibinfo {author} {\bibfnamefont {S.~F.}\ \bibnamefont {Cooper}}, \bibinfo {author} {\bibfnamefont {R.~D.}\ \bibnamefont {Delaney}}, \bibinfo {author} {\bibfnamefont {J.~M.}\ \bibnamefont {Dreiling}}, \bibinfo {author} {\bibfnamefont {B.}~\bibnamefont {Estey}}, \bibinfo {author} {\bibfnamefont {C.}~\bibnamefont {Figgatt}}, \bibinfo {author} {\bibfnamefont {A.}~\bibnamefont {Hall}}, \bibinfo {author} {\bibfnamefont {A.~A.}\ \bibnamefont {Husain}}, \bibinfo {author} {\bibfnamefont {A.}~\bibnamefont {Isanaka}}, \bibinfo {author} {\bibfnamefont {C.~J.}\ \bibnamefont {Kennedy}}, \bibinfo {author} {\bibfnamefont {N.}~\bibnamefont {Kotibhaskar}}, \bibinfo {author} {\bibfnamefont {I.~S.}\ \bibnamefont {Madjarov}}, \bibinfo {author}
  {\bibfnamefont {K.}~\bibnamefont {Mayer}}, \bibinfo {author} {\bibfnamefont {A.~R.}\ \bibnamefont {Milne}}, \bibinfo {author} {\bibfnamefont {A.~J.}\ \bibnamefont {Park}}, \bibinfo {author} {\bibfnamefont {A.~P.}\ \bibnamefont {Reed}}, \bibinfo {author} {\bibfnamefont {R.}~\bibnamefont {Ancona}}, \bibinfo {author} {\bibfnamefont {M.~P.}\ \bibnamefont {Andersen}}, \bibinfo {author} {\bibfnamefont {P.}~\bibnamefont {Andres-Martinez}}, \bibinfo {author} {\bibfnamefont {W.}~\bibnamefont {Angenent}}, \bibinfo {author} {\bibfnamefont {L.}~\bibnamefont {Argueta}}, \bibinfo {author} {\bibfnamefont {B.}~\bibnamefont {Arkin}}, \bibinfo {author} {\bibfnamefont {L.}~\bibnamefont {Ascarrunz}}, \bibinfo {author} {\bibfnamefont {W.}~\bibnamefont {Baker}}, \bibinfo {author} {\bibfnamefont {C.}~\bibnamefont {Barnes}}, \bibinfo {author} {\bibfnamefont {J.}~\bibnamefont {Bartolotta}}, \bibinfo {author} {\bibfnamefont {J.}~\bibnamefont {Berg}}, \bibinfo {author} {\bibfnamefont {R.}~\bibnamefont {Besand}}, \bibinfo {author}
  {\bibfnamefont {B.}~\bibnamefont {Bjork}}, \bibinfo {author} {\bibfnamefont {M.}~\bibnamefont {Blain}}, \bibinfo {author} {\bibfnamefont {P.}~\bibnamefont {Blanchard}}, \bibinfo {author} {\bibfnamefont {R.}~\bibnamefont {Blume-Kohout}}, \bibinfo {author} {\bibfnamefont {M.}~\bibnamefont {Bohn}}, \bibinfo {author} {\bibfnamefont {A.}~\bibnamefont {Borgna}}, \bibinfo {author} {\bibfnamefont {D.~Y.}\ \bibnamefont {Botamanenko}}, \bibinfo {author} {\bibfnamefont {R.}~\bibnamefont {Boutelle}}, \bibinfo {author} {\bibfnamefont {N.}~\bibnamefont {Brown}}, \bibinfo {author} {\bibfnamefont {G.~T.}\ \bibnamefont {Buckingham}}, \bibinfo {author} {\bibfnamefont {N.~Q.}\ \bibnamefont {Burdick}}, \bibinfo {author} {\bibfnamefont {W.~C.}\ \bibnamefont {Burton}}, \bibinfo {author} {\bibfnamefont {V.}~\bibnamefont {Carey}}, \bibinfo {author} {\bibfnamefont {C.~J.}\ \bibnamefont {Carron}}, \bibinfo {author} {\bibfnamefont {J.}~\bibnamefont {Chambers}}, \bibinfo {author} {\bibfnamefont {J.}~\bibnamefont {Children}}, \bibinfo
  {author} {\bibfnamefont {V.~E.}\ \bibnamefont {Colussi}}, \bibinfo {author} {\bibfnamefont {S.}~\bibnamefont {Crepinsek}}, \bibinfo {author} {\bibfnamefont {A.}~\bibnamefont {Cureton}}, \bibinfo {author} {\bibfnamefont {J.}~\bibnamefont {Davies}}, \bibinfo {author} {\bibfnamefont {D.}~\bibnamefont {Davis}}, \bibinfo {author} {\bibfnamefont {M.}~\bibnamefont {DeCross}}, \bibinfo {author} {\bibfnamefont {D.}~\bibnamefont {Deen}}, \bibinfo {author} {\bibfnamefont {C.}~\bibnamefont {Delaney}}, \bibinfo {author} {\bibfnamefont {D.}~\bibnamefont {DelVento}}, \bibinfo {author} {\bibfnamefont {B.~J.}\ \bibnamefont {DeSalvo}}, \bibinfo {author} {\bibfnamefont {J.}~\bibnamefont {Dominy}}, \bibinfo {author} {\bibfnamefont {R.}~\bibnamefont {Duncan}}, \bibinfo {author} {\bibfnamefont {V.}~\bibnamefont {Eccles}}, \bibinfo {author} {\bibfnamefont {A.}~\bibnamefont {Edgington}}, \bibinfo {author} {\bibfnamefont {N.}~\bibnamefont {Erickson}}, \bibinfo {author} {\bibfnamefont {S.}~\bibnamefont {Erickson}}, \bibinfo {author}
  {\bibfnamefont {C.~T.}\ \bibnamefont {Ertsgaard}}, \bibinfo {author} {\bibfnamefont {B.}~\bibnamefont {Evans}}, \bibinfo {author} {\bibfnamefont {T.}~\bibnamefont {Evans}}, \bibinfo {author} {\bibfnamefont {M.~I.}\ \bibnamefont {Fabrikant}}, \bibinfo {author} {\bibfnamefont {A.}~\bibnamefont {Fischer}}, \bibinfo {author} {\bibfnamefont {C.}~\bibnamefont {Foltz}}, \bibinfo {author} {\bibfnamefont {M.}~\bibnamefont {Foss-Feig}}, \bibinfo {author} {\bibfnamefont {D.}~\bibnamefont {Francois}}, \bibinfo {author} {\bibfnamefont {B.}~\bibnamefont {Freyberg}}, \bibinfo {author} {\bibfnamefont {C.}~\bibnamefont {Gao}}, \bibinfo {author} {\bibfnamefont {R.}~\bibnamefont {Garay}}, \bibinfo {author} {\bibfnamefont {J.}~\bibnamefont {Garvin}}, \bibinfo {author} {\bibfnamefont {D.~M.}\ \bibnamefont {Gaudiosi}}, \bibinfo {author} {\bibfnamefont {C.~N.}\ \bibnamefont {Gilbreth}}, \bibinfo {author} {\bibfnamefont {J.}~\bibnamefont {Giles}}, \bibinfo {author} {\bibfnamefont {E.}~\bibnamefont {Glynn}}, \bibinfo {author}
  {\bibfnamefont {J.}~\bibnamefont {Graves}}, \bibinfo {author} {\bibfnamefont {A.}~\bibnamefont {Hansen}}, \bibinfo {author} {\bibfnamefont {D.}~\bibnamefont {Hayes}}, \bibinfo {author} {\bibfnamefont {L.}~\bibnamefont {Heidemann}}, \bibinfo {author} {\bibfnamefont {B.}~\bibnamefont {Higashi}}, \bibinfo {author} {\bibfnamefont {T.}~\bibnamefont {Hilbun}}, \bibinfo {author} {\bibfnamefont {J.}~\bibnamefont {Hines}}, \bibinfo {author} {\bibfnamefont {A.}~\bibnamefont {Hlavaty}}, \bibinfo {author} {\bibfnamefont {K.}~\bibnamefont {Hoffman}}, \bibinfo {author} {\bibfnamefont {I.~M.}\ \bibnamefont {Hoffman}}, \bibinfo {author} {\bibfnamefont {C.}~\bibnamefont {Holliman}}, \bibinfo {author} {\bibfnamefont {I.}~\bibnamefont {Hooper}}, \bibinfo {author} {\bibfnamefont {B.}~\bibnamefont {Horning}}, \bibinfo {author} {\bibfnamefont {J.}~\bibnamefont {Hostetter}}, \bibinfo {author} {\bibfnamefont {D.}~\bibnamefont {Hothem}}, \bibinfo {author} {\bibfnamefont {J.}~\bibnamefont {Houlton}}, \bibinfo {author} {\bibfnamefont
  {J.}~\bibnamefont {Hout}}, \bibinfo {author} {\bibfnamefont {R.}~\bibnamefont {Hutson}}, \bibinfo {author} {\bibfnamefont {R.~T.}\ \bibnamefont {Jacobs}}, \bibinfo {author} {\bibfnamefont {T.}~\bibnamefont {Jacobs}}, \bibinfo {author} {\bibfnamefont {M.}~\bibnamefont {Johannsen}}, \bibinfo {author} {\bibfnamefont {J.}~\bibnamefont {Johansen}}, \bibinfo {author} {\bibfnamefont {L.}~\bibnamefont {Jones}}, \bibinfo {author} {\bibfnamefont {S.}~\bibnamefont {Julian}}, \bibinfo {author} {\bibfnamefont {R.}~\bibnamefont {Jung}}, \bibinfo {author} {\bibfnamefont {A.}~\bibnamefont {Keay}}, \bibinfo {author} {\bibfnamefont {T.}~\bibnamefont {Klein}}, \bibinfo {author} {\bibfnamefont {M.}~\bibnamefont {Koch}}, \bibinfo {author} {\bibfnamefont {R.}~\bibnamefont {Kondo}}, \bibinfo {author} {\bibfnamefont {C.}~\bibnamefont {Kong}}, \bibinfo {author} {\bibfnamefont {A.}~\bibnamefont {Kosto}}, \bibinfo {author} {\bibfnamefont {A.}~\bibnamefont {Lawrence}}, \bibinfo {author} {\bibfnamefont {D.}~\bibnamefont {Liefer}},
  \bibinfo {author} {\bibfnamefont {M.}~\bibnamefont {Lollie}}, \bibinfo {author} {\bibfnamefont {D.}~\bibnamefont {Lucchetti}}, \bibinfo {author} {\bibfnamefont {N.~K.}\ \bibnamefont {Lysne}}, \bibinfo {author} {\bibfnamefont {C.}~\bibnamefont {Lytle}}, \bibinfo {author} {\bibfnamefont {C.}~\bibnamefont {MacPherson}}, \bibinfo {author} {\bibfnamefont {A.}~\bibnamefont {Malm}}, \bibinfo {author} {\bibfnamefont {S.}~\bibnamefont {Mather}}, \bibinfo {author} {\bibfnamefont {B.}~\bibnamefont {Mathewson}}, \bibinfo {author} {\bibfnamefont {D.}~\bibnamefont {Maxwell}}, \bibinfo {author} {\bibfnamefont {L.}~\bibnamefont {McCaffrey}}, \bibinfo {author} {\bibfnamefont {H.}~\bibnamefont {McDougall}}, \bibinfo {author} {\bibfnamefont {R.}~\bibnamefont {Mendoza}}, \bibinfo {author} {\bibfnamefont {M.}~\bibnamefont {Mills}}, \bibinfo {author} {\bibfnamefont {R.}~\bibnamefont {Morrison}}, \bibinfo {author} {\bibfnamefont {L.}~\bibnamefont {Narmour}}, \bibinfo {author} {\bibfnamefont {N.}~\bibnamefont {Nguyen}}, \bibinfo
  {author} {\bibfnamefont {L.}~\bibnamefont {Nugent}}, \bibinfo {author} {\bibfnamefont {S.}~\bibnamefont {Olson}}, \bibinfo {author} {\bibfnamefont {D.}~\bibnamefont {Ouellette}}, \bibinfo {author} {\bibfnamefont {J.}~\bibnamefont {Parks}}, \bibinfo {author} {\bibfnamefont {Z.}~\bibnamefont {Peters}}, \bibinfo {author} {\bibfnamefont {J.}~\bibnamefont {Petricka}}, \bibinfo {author} {\bibfnamefont {J.~M.}\ \bibnamefont {Pino}}, \bibinfo {author} {\bibfnamefont {F.}~\bibnamefont {Polito}}, \bibinfo {author} {\bibfnamefont {M.}~\bibnamefont {Preidl}}, \bibinfo {author} {\bibfnamefont {G.}~\bibnamefont {Price}}, \bibinfo {author} {\bibfnamefont {T.}~\bibnamefont {Proctor}}, \bibinfo {author} {\bibfnamefont {M.}~\bibnamefont {Pugh}}, \bibinfo {author} {\bibfnamefont {N.}~\bibnamefont {Ratcliff}}, \bibinfo {author} {\bibfnamefont {D.}~\bibnamefont {Raymondson}}, \bibinfo {author} {\bibfnamefont {P.}~\bibnamefont {Rhodes}}, \bibinfo {author} {\bibfnamefont {C.}~\bibnamefont {Roman}}, \bibinfo {author}
  {\bibfnamefont {C.}~\bibnamefont {Roy}}, \bibinfo {author} {\bibfnamefont {C.}~\bibnamefont {Ryan-Anderson}}, \bibinfo {author} {\bibfnamefont {F.~B.}\ \bibnamefont {Sanchez}}, \bibinfo {author} {\bibfnamefont {G.}~\bibnamefont {Sangiolo}}, \bibinfo {author} {\bibfnamefont {T.}~\bibnamefont {Sawadski}}, \bibinfo {author} {\bibfnamefont {A.}~\bibnamefont {Schaffer}}, \bibinfo {author} {\bibfnamefont {P.}~\bibnamefont {Schow}}, \bibinfo {author} {\bibfnamefont {J.}~\bibnamefont {Sedlacek}}, \bibinfo {author} {\bibfnamefont {H.}~\bibnamefont {Semenenko}}, \bibinfo {author} {\bibfnamefont {P.}~\bibnamefont {Shevchuk}}, \bibinfo {author} {\bibfnamefont {S.}~\bibnamefont {Shore}}, \bibinfo {author} {\bibfnamefont {P.}~\bibnamefont {Siegfried}}, \bibinfo {author} {\bibfnamefont {K.}~\bibnamefont {Singhal}}, \bibinfo {author} {\bibfnamefont {S.}~\bibnamefont {Sivarajah}}, \bibinfo {author} {\bibfnamefont {T.}~\bibnamefont {Skripka}}, \bibinfo {author} {\bibfnamefont {L.}~\bibnamefont {Sletten}}, \bibinfo {author}
  {\bibfnamefont {B.}~\bibnamefont {Spaun}}, \bibinfo {author} {\bibfnamefont {R.~T.}\ \bibnamefont {Sprenkle}}, \bibinfo {author} {\bibfnamefont {P.}~\bibnamefont {Stoufer}}, \bibinfo {author} {\bibfnamefont {M.}~\bibnamefont {Tader}}, \bibinfo {author} {\bibfnamefont {S.~F.}\ \bibnamefont {Taylor}}, \bibinfo {author} {\bibfnamefont {T.~H.}\ \bibnamefont {Thompson}}, \bibinfo {author} {\bibfnamefont {R.}~\bibnamefont {Tobey}}, \bibinfo {author} {\bibfnamefont {A.}~\bibnamefont {Tran}}, \bibinfo {author} {\bibfnamefont {T.}~\bibnamefont {Tran}}, \bibinfo {author} {\bibfnamefont {G.}~\bibnamefont {Vittorini}}, \bibinfo {author} {\bibfnamefont {C.}~\bibnamefont {Volin}}, \bibinfo {author} {\bibfnamefont {J.}~\bibnamefont {Walker}}, \bibinfo {author} {\bibfnamefont {S.}~\bibnamefont {White}}, \bibinfo {author} {\bibfnamefont {D.}~\bibnamefont {Wilson}}, \bibinfo {author} {\bibfnamefont {Q.}~\bibnamefont {Wolf}}, \bibinfo {author} {\bibfnamefont {C.}~\bibnamefont {Wringe}}, \bibinfo {author} {\bibfnamefont
  {K.}~\bibnamefont {Young}}, \bibinfo {author} {\bibfnamefont {J.}~\bibnamefont {Zheng}}, \bibinfo {author} {\bibfnamefont {K.}~\bibnamefont {Zuraski}}, \bibinfo {author} {\bibfnamefont {C.~H.}\ \bibnamefont {Baldwin}}, \bibinfo {author} {\bibfnamefont {A.}~\bibnamefont {Chernoguzov}}, \bibinfo {author} {\bibfnamefont {J.~P.}\ \bibnamefont {Gaebler}}, \bibinfo {author} {\bibfnamefont {S.~J.}\ \bibnamefont {Sanders}}, \bibinfo {author} {\bibfnamefont {B.}~\bibnamefont {Neyenhuis}}, \bibinfo {author} {\bibfnamefont {R.}~\bibnamefont {Stutz}},\ and\ \bibinfo {author} {\bibfnamefont {J.~G.}\ \bibnamefont {Bohnet}},\ }\href {https://arxiv.org/abs/2511.05465} {\bibinfo {title} {Helios: A 98-qubit trapped-ion quantum computer}} (\bibinfo {year} {2025}),\ \Eprint {https://arxiv.org/abs/2511.05465} {arXiv:2511.05465 [quant-ph]} \BibitemShut {NoStop}%
\bibitem [{\citenamefont {Pogorelov}\ \emph {et~al.}(2021)\citenamefont {Pogorelov}, \citenamefont {Feldker}, \citenamefont {Marciniak}, \citenamefont {Postler}, \citenamefont {Jacob}, \citenamefont {Krieglsteiner}, \citenamefont {Podlesnic}, \citenamefont {Meth}, \citenamefont {Negnevitsky}, \citenamefont {Stadler}, \citenamefont {H\"ofer}, \citenamefont {W\"achter}, \citenamefont {Lakhmanskiy}, \citenamefont {Blatt}, \citenamefont {Schindler},\ and\ \citenamefont {Monz}}]{Pogorelov2021}%
  \BibitemOpen
  \bibfield  {author} {\bibinfo {author} {\bibfnamefont {I.}~\bibnamefont {Pogorelov}}, \bibinfo {author} {\bibfnamefont {T.}~\bibnamefont {Feldker}}, \bibinfo {author} {\bibfnamefont {C.~D.}\ \bibnamefont {Marciniak}}, \bibinfo {author} {\bibfnamefont {L.}~\bibnamefont {Postler}}, \bibinfo {author} {\bibfnamefont {G.}~\bibnamefont {Jacob}}, \bibinfo {author} {\bibfnamefont {O.}~\bibnamefont {Krieglsteiner}}, \bibinfo {author} {\bibfnamefont {V.}~\bibnamefont {Podlesnic}}, \bibinfo {author} {\bibfnamefont {M.}~\bibnamefont {Meth}}, \bibinfo {author} {\bibfnamefont {V.}~\bibnamefont {Negnevitsky}}, \bibinfo {author} {\bibfnamefont {M.}~\bibnamefont {Stadler}}, \bibinfo {author} {\bibfnamefont {B.}~\bibnamefont {H\"ofer}}, \bibinfo {author} {\bibfnamefont {C.}~\bibnamefont {W\"achter}}, \bibinfo {author} {\bibfnamefont {K.}~\bibnamefont {Lakhmanskiy}}, \bibinfo {author} {\bibfnamefont {R.}~\bibnamefont {Blatt}}, \bibinfo {author} {\bibfnamefont {P.}~\bibnamefont {Schindler}},\ and\ \bibinfo {author}
  {\bibfnamefont {T.}~\bibnamefont {Monz}},\ }\bibfield  {title} {\bibinfo {title} {Compact ion-trap quantum computing demonstrator},\ }\href {https://doi.org/10.1103/PRXQuantum.2.020343} {\bibfield  {journal} {\bibinfo  {journal} {PRX Quantum}\ }\textbf {\bibinfo {volume} {2}},\ \bibinfo {pages} {020343} (\bibinfo {year} {2021})}\BibitemShut {NoStop}%
\bibitem [{\citenamefont {Proctor}\ \emph {et~al.}(2025)\citenamefont {Proctor}, \citenamefont {Young}, \citenamefont {Baczewski},\ and\ \citenamefont {Blume-Kohout}}]{Proctor2025}%
  \BibitemOpen
  \bibfield  {author} {\bibinfo {author} {\bibfnamefont {T.}~\bibnamefont {Proctor}}, \bibinfo {author} {\bibfnamefont {K.}~\bibnamefont {Young}}, \bibinfo {author} {\bibfnamefont {A.~D.}\ \bibnamefont {Baczewski}},\ and\ \bibinfo {author} {\bibfnamefont {R.}~\bibnamefont {Blume-Kohout}},\ }\bibfield  {title} {\bibinfo {title} {Benchmarking quantum computers},\ }\href {https://doi.org/10.1038/s42254-024-00796-z} {\bibfield  {journal} {\bibinfo  {journal} {Nature Reviews Physics}\ }\textbf {\bibinfo {volume} {7}},\ \bibinfo {pages} {105} (\bibinfo {year} {2025})}\BibitemShut {NoStop}%
\bibitem [{\citenamefont {Kielpinski}\ \emph {et~al.}(2002)\citenamefont {Kielpinski}, \citenamefont {Monroe},\ and\ \citenamefont {Wineland}}]{kielpinski2002}%
  \BibitemOpen
  \bibfield  {author} {\bibinfo {author} {\bibfnamefont {D.}~\bibnamefont {Kielpinski}}, \bibinfo {author} {\bibfnamefont {C.}~\bibnamefont {Monroe}},\ and\ \bibinfo {author} {\bibfnamefont {D.~J.}\ \bibnamefont {Wineland}},\ }\bibfield  {title} {\bibinfo {title} {Architecture for a large-scale ion-trap quantum computer},\ }\href {https://doi.org/10.1038/nature00784} {\bibfield  {journal} {\bibinfo  {journal} {Nature}\ }\textbf {\bibinfo {volume} {417}},\ \bibinfo {pages} {709} (\bibinfo {year} {2002})}\BibitemShut {NoStop}%
\bibitem [{\citenamefont {Pino}\ \emph {et~al.}(2021)\citenamefont {Pino}, \citenamefont {Dreiling}, \citenamefont {Figgatt}, \citenamefont {Gaebler}, \citenamefont {Moses}, \citenamefont {Allman}, \citenamefont {Baldwin}, \citenamefont {Foss-Feig}, \citenamefont {Hayes}, \citenamefont {Mayer}, \citenamefont {Ryan-Anderson},\ and\ \citenamefont {Neyenhuis}}]{pino2021}%
  \BibitemOpen
  \bibfield  {author} {\bibinfo {author} {\bibfnamefont {J.~M.}\ \bibnamefont {Pino}}, \bibinfo {author} {\bibfnamefont {J.~M.}\ \bibnamefont {Dreiling}}, \bibinfo {author} {\bibfnamefont {C.}~\bibnamefont {Figgatt}}, \bibinfo {author} {\bibfnamefont {J.~P.}\ \bibnamefont {Gaebler}}, \bibinfo {author} {\bibfnamefont {S.~A.}\ \bibnamefont {Moses}}, \bibinfo {author} {\bibfnamefont {M.~S.}\ \bibnamefont {Allman}}, \bibinfo {author} {\bibfnamefont {C.~H.}\ \bibnamefont {Baldwin}}, \bibinfo {author} {\bibfnamefont {M.}~\bibnamefont {Foss-Feig}}, \bibinfo {author} {\bibfnamefont {D.}~\bibnamefont {Hayes}}, \bibinfo {author} {\bibfnamefont {K.}~\bibnamefont {Mayer}}, \bibinfo {author} {\bibfnamefont {C.}~\bibnamefont {Ryan-Anderson}},\ and\ \bibinfo {author} {\bibfnamefont {B.}~\bibnamefont {Neyenhuis}},\ }\bibfield  {title} {\bibinfo {title} {Demonstration of the trapped-ion quantum ccd computer architecture},\ }\href {https://doi.org/10.1038/s41586-021-03318-4} {\bibfield  {journal} {\bibinfo  {journal} {Nature}\
  }\textbf {\bibinfo {volume} {592}},\ \bibinfo {pages} {209} (\bibinfo {year} {2021})}\BibitemShut {NoStop}%
\bibitem [{\citenamefont {Malinowski}\ \emph {et~al.}(2023)\citenamefont {Malinowski}, \citenamefont {Allcock},\ and\ \citenamefont {Ballance}}]{malinowski2023}%
  \BibitemOpen
  \bibfield  {author} {\bibinfo {author} {\bibfnamefont {M.}~\bibnamefont {Malinowski}}, \bibinfo {author} {\bibfnamefont {D.}~\bibnamefont {Allcock}},\ and\ \bibinfo {author} {\bibfnamefont {C.}~\bibnamefont {Ballance}},\ }\bibfield  {title} {\bibinfo {title} {How to wire a $1000$-qubit trapped-ion quantum computer},\ }\href {https://doi.org/10.1103/PRXQuantum.4.040313} {\bibfield  {journal} {\bibinfo  {journal} {PRX Quantum}\ }\textbf {\bibinfo {volume} {4}},\ \bibinfo {pages} {040313} (\bibinfo {year} {2023})}\BibitemShut {NoStop}%
\bibitem [{\citenamefont {Niffenegger}\ \emph {et~al.}(2020)\citenamefont {Niffenegger}, \citenamefont {Stuart}, \citenamefont {Sorace-Agaskar}, \citenamefont {Kharas}, \citenamefont {Bramhavar}, \citenamefont {Bruzewicz}, \citenamefont {Loh}, \citenamefont {Maxson}, \citenamefont {McConnell}, \citenamefont {Reens}, \citenamefont {West}, \citenamefont {Sage},\ and\ \citenamefont {Chiaverini}}]{Niffenegger2020}%
  \BibitemOpen
  \bibfield  {author} {\bibinfo {author} {\bibfnamefont {R.~J.}\ \bibnamefont {Niffenegger}}, \bibinfo {author} {\bibfnamefont {J.}~\bibnamefont {Stuart}}, \bibinfo {author} {\bibfnamefont {C.}~\bibnamefont {Sorace-Agaskar}}, \bibinfo {author} {\bibfnamefont {D.}~\bibnamefont {Kharas}}, \bibinfo {author} {\bibfnamefont {S.}~\bibnamefont {Bramhavar}}, \bibinfo {author} {\bibfnamefont {C.~D.}\ \bibnamefont {Bruzewicz}}, \bibinfo {author} {\bibfnamefont {W.}~\bibnamefont {Loh}}, \bibinfo {author} {\bibfnamefont {R.~T.}\ \bibnamefont {Maxson}}, \bibinfo {author} {\bibfnamefont {R.}~\bibnamefont {McConnell}}, \bibinfo {author} {\bibfnamefont {D.}~\bibnamefont {Reens}}, \bibinfo {author} {\bibfnamefont {G.~N.}\ \bibnamefont {West}}, \bibinfo {author} {\bibfnamefont {J.~M.}\ \bibnamefont {Sage}},\ and\ \bibinfo {author} {\bibfnamefont {J.}~\bibnamefont {Chiaverini}},\ }\bibfield  {title} {\bibinfo {title} {Integrated multi-wavelength control of an ion qubit},\ }\href {https://doi.org/10.1038/s41586-020-2811-x}
  {\bibfield  {journal} {\bibinfo  {journal} {Nature}\ }\textbf {\bibinfo {volume} {586}},\ \bibinfo {pages} {538} (\bibinfo {year} {2020})}\BibitemShut {NoStop}%
\bibitem [{\citenamefont {Mehta}\ \emph {et~al.}(2016)\citenamefont {Mehta}, \citenamefont {Bruzewicz}, \citenamefont {McConnell}, \citenamefont {Ram}, \citenamefont {Sage},\ and\ \citenamefont {Chiaverini}}]{Mehta2016}%
  \BibitemOpen
  \bibfield  {author} {\bibinfo {author} {\bibfnamefont {K.~K.}\ \bibnamefont {Mehta}}, \bibinfo {author} {\bibfnamefont {C.~D.}\ \bibnamefont {Bruzewicz}}, \bibinfo {author} {\bibfnamefont {R.}~\bibnamefont {McConnell}}, \bibinfo {author} {\bibfnamefont {R.~J.}\ \bibnamefont {Ram}}, \bibinfo {author} {\bibfnamefont {J.~M.}\ \bibnamefont {Sage}},\ and\ \bibinfo {author} {\bibfnamefont {J.}~\bibnamefont {Chiaverini}},\ }\bibfield  {title} {\bibinfo {title} {Integrated optical addressing of an ion qubit},\ }\href {https://doi.org/10.1038/nnano.2016.139} {\bibfield  {journal} {\bibinfo  {journal} {Nature Nanotechnology}\ }\textbf {\bibinfo {volume} {11}},\ \bibinfo {pages} {1066} (\bibinfo {year} {2016})}\BibitemShut {NoStop}%
\bibitem [{\citenamefont {Badawi}\ \emph {et~al.}(2025)\citenamefont {Badawi}, \citenamefont {Holz}, \citenamefont {Raffetseder}, \citenamefont {Jungwirth}, \citenamefont {Ulmanis}, \citenamefont {Quenzer}, \citenamefont {Kähler}, \citenamefont {Monz},\ and\ \citenamefont {Schindler}}]{badawi2025}%
  \BibitemOpen
  \bibfield  {author} {\bibinfo {author} {\bibfnamefont {B.}~\bibnamefont {Badawi}}, \bibinfo {author} {\bibfnamefont {P.~C.}\ \bibnamefont {Holz}}, \bibinfo {author} {\bibfnamefont {M.}~\bibnamefont {Raffetseder}}, \bibinfo {author} {\bibfnamefont {N.}~\bibnamefont {Jungwirth}}, \bibinfo {author} {\bibfnamefont {J.}~\bibnamefont {Ulmanis}}, \bibinfo {author} {\bibfnamefont {H.-J.}\ \bibnamefont {Quenzer}}, \bibinfo {author} {\bibfnamefont {D.}~\bibnamefont {Kähler}}, \bibinfo {author} {\bibfnamefont {T.}~\bibnamefont {Monz}},\ and\ \bibinfo {author} {\bibfnamefont {P.}~\bibnamefont {Schindler}},\ }\href {https://arxiv.org/abs/2512.02645} {\bibinfo {title} {Chiplet technology for large-scale trapped-ion quantum processors}} (\bibinfo {year} {2025}),\ \Eprint {https://arxiv.org/abs/2512.02645} {arXiv:2512.02645 [quant-ph]} \BibitemShut {NoStop}%
\bibitem [{\citenamefont {Momenzadeh}\ \emph {et~al.}(2026)\citenamefont {Momenzadeh}, \citenamefont {Sun}, \citenamefont {Wu}, \citenamefont {You}, \citenamefont {Tang}, \citenamefont {H{\"a}ffner},\ and\ \citenamefont {Shcherbakov}}]{Momenzadeh2026}%
  \BibitemOpen
  \bibfield  {author} {\bibinfo {author} {\bibfnamefont {M.}~\bibnamefont {Momenzadeh}}, \bibinfo {author} {\bibfnamefont {K.}~\bibnamefont {Sun}}, \bibinfo {author} {\bibfnamefont {Q.}~\bibnamefont {Wu}}, \bibinfo {author} {\bibfnamefont {B.}~\bibnamefont {You}}, \bibinfo {author} {\bibfnamefont {Y.-L.}\ \bibnamefont {Tang}}, \bibinfo {author} {\bibfnamefont {H.}~\bibnamefont {H{\"a}ffner}},\ and\ \bibinfo {author} {\bibfnamefont {M.~R.}\ \bibnamefont {Shcherbakov}},\ }\bibfield  {title} {\bibinfo {title} {Individual trapped-ion addressing with adjoint-optimized multimode photonic circuits},\ }\href {https://doi.org/10.1038/s44310-025-00102-4} {\bibfield  {journal} {\bibinfo  {journal} {npj Nanophotonics}\ }\textbf {\bibinfo {volume} {3}},\ \bibinfo {pages} {3} (\bibinfo {year} {2026})}\BibitemShut {NoStop}%
\bibitem [{\citenamefont {Mehta}\ \emph {et~al.}(2020)\citenamefont {Mehta}, \citenamefont {Zhang}, \citenamefont {Malinowski}, \citenamefont {Nguyen}, \citenamefont {Stadler},\ and\ \citenamefont {Home}}]{Mehta2020}%
  \BibitemOpen
  \bibfield  {author} {\bibinfo {author} {\bibfnamefont {K.~K.}\ \bibnamefont {Mehta}}, \bibinfo {author} {\bibfnamefont {C.}~\bibnamefont {Zhang}}, \bibinfo {author} {\bibfnamefont {M.}~\bibnamefont {Malinowski}}, \bibinfo {author} {\bibfnamefont {T.-L.}\ \bibnamefont {Nguyen}}, \bibinfo {author} {\bibfnamefont {M.}~\bibnamefont {Stadler}},\ and\ \bibinfo {author} {\bibfnamefont {J.~P.}\ \bibnamefont {Home}},\ }\bibfield  {title} {\bibinfo {title} {Integrated optical multi-ion quantum logic},\ }\href {https://doi.org/10.1038/s41586-020-2823-6} {\bibfield  {journal} {\bibinfo  {journal} {Nature}\ }\textbf {\bibinfo {volume} {586}},\ \bibinfo {pages} {533} (\bibinfo {year} {2020})}\BibitemShut {NoStop}%
\bibitem [{\citenamefont {Eaton}\ \emph {et~al.}(2011)\citenamefont {Eaton}, \citenamefont {Ng}, \citenamefont {Osellame},\ and\ \citenamefont {Herman}}]{Eaton2011}%
  \BibitemOpen
  \bibfield  {author} {\bibinfo {author} {\bibfnamefont {S.~M.}\ \bibnamefont {Eaton}}, \bibinfo {author} {\bibfnamefont {M.~L.}\ \bibnamefont {Ng}}, \bibinfo {author} {\bibfnamefont {R.}~\bibnamefont {Osellame}},\ and\ \bibinfo {author} {\bibfnamefont {P.~R.}\ \bibnamefont {Herman}},\ }\bibfield  {title} {\bibinfo {title} {High refractive index contrast in fused silica waveguides by tightly focused, high-repetition rate femtosecond laser},\ }\href {https://doi.org/https://doi.org/10.1016/j.jnoncrysol.2010.11.082} {\bibfield  {journal} {\bibinfo  {journal} {Journal of Non-Crystalline Solids}\ }\textbf {\bibinfo {volume} {357}},\ \bibinfo {pages} {2387} (\bibinfo {year} {2011})},\ \bibinfo {note} {17th International Symposium on Non-Oxide and New Optical Glasses (XVII ISNOG)}\BibitemShut {NoStop}%
\bibitem [{\citenamefont {Guan}\ \emph {et~al.}(2014)\citenamefont {Guan}, \citenamefont {Scott}, \citenamefont {Qin}, \citenamefont {Fontaine}, \citenamefont {Su}, \citenamefont {Ferrari}, \citenamefont {Cappuzzo}, \citenamefont {Klemens}, \citenamefont {Keller}, \citenamefont {Earnshaw},\ and\ \citenamefont {Yoo}}]{Guan2014}%
  \BibitemOpen
  \bibfield  {author} {\bibinfo {author} {\bibfnamefont {B.}~\bibnamefont {Guan}}, \bibinfo {author} {\bibfnamefont {R.~P.}\ \bibnamefont {Scott}}, \bibinfo {author} {\bibfnamefont {C.}~\bibnamefont {Qin}}, \bibinfo {author} {\bibfnamefont {N.~K.}\ \bibnamefont {Fontaine}}, \bibinfo {author} {\bibfnamefont {T.}~\bibnamefont {Su}}, \bibinfo {author} {\bibfnamefont {C.}~\bibnamefont {Ferrari}}, \bibinfo {author} {\bibfnamefont {M.}~\bibnamefont {Cappuzzo}}, \bibinfo {author} {\bibfnamefont {F.}~\bibnamefont {Klemens}}, \bibinfo {author} {\bibfnamefont {B.}~\bibnamefont {Keller}}, \bibinfo {author} {\bibfnamefont {M.}~\bibnamefont {Earnshaw}},\ and\ \bibinfo {author} {\bibfnamefont {S.~J.~B.}\ \bibnamefont {Yoo}},\ }\bibfield  {title} {\bibinfo {title} {Free-space coherent optical communication with orbital angular, momentum multiplexing/demultiplexing using a hybrid 3d photonic integrated circuit},\ }\href {https://doi.org/10.1364/OE.22.000145} {\bibfield  {journal} {\bibinfo  {journal} {Opt. Express}\ }\textbf
  {\bibinfo {volume} {22}},\ \bibinfo {pages} {145} (\bibinfo {year} {2014})}\BibitemShut {NoStop}%
\bibitem [{\citenamefont {Grüneberg}\ \emph {et~al.}(2023)\citenamefont {Grüneberg}, \citenamefont {Pribošek}, \citenamefont {Llobera}, \citenamefont {Zesar}, \citenamefont {Wahl}, \citenamefont {Preidl}, \citenamefont {Colombe}, \citenamefont {Schüppert}, \citenamefont {Rössler}, \citenamefont {Hurdax}, \citenamefont {Lamprecht},\ and\ \citenamefont {Montagnese}}]{Gruenberg2023}%
  \BibitemOpen
  \bibfield  {author} {\bibinfo {author} {\bibfnamefont {M.}~\bibnamefont {Grüneberg}}, \bibinfo {author} {\bibfnamefont {J.}~\bibnamefont {Pribošek}}, \bibinfo {author} {\bibfnamefont {A.}~\bibnamefont {Llobera}}, \bibinfo {author} {\bibfnamefont {A.}~\bibnamefont {Zesar}}, \bibinfo {author} {\bibfnamefont {J.}~\bibnamefont {Wahl}}, \bibinfo {author} {\bibfnamefont {M.}~\bibnamefont {Preidl}}, \bibinfo {author} {\bibfnamefont {Y.}~\bibnamefont {Colombe}}, \bibinfo {author} {\bibfnamefont {K.}~\bibnamefont {Schüppert}}, \bibinfo {author} {\bibfnamefont {C.}~\bibnamefont {Rössler}}, \bibinfo {author} {\bibfnamefont {P.}~\bibnamefont {Hurdax}}, \bibinfo {author} {\bibfnamefont {B.}~\bibnamefont {Lamprecht}},\ and\ \bibinfo {author} {\bibfnamefont {M.}~\bibnamefont {Montagnese}},\ }\bibfield  {title} {\bibinfo {title} {On-chip laser beam delivery for integrated ion traps},\ }in\ \href@noop {} {\emph {\bibinfo {booktitle} {2023 22nd International Conference on Solid-State Sensors, Actuators and Microsystems
  (Transducers)}}}\ (\bibinfo {year} {2023})\ pp.\ \bibinfo {pages} {1496--1499}\BibitemShut {NoStop}%
\bibitem [{\citenamefont {Auchter}\ \emph {et~al.}(2022)\citenamefont {Auchter}, \citenamefont {Axline}, \citenamefont {Decaroli}, \citenamefont {Valentini}, \citenamefont {Purwin}, \citenamefont {Oswald}, \citenamefont {Matt}, \citenamefont {Aschauer}, \citenamefont {Colombe}, \citenamefont {Holz}, \citenamefont {Monz}, \citenamefont {Blatt}, \citenamefont {Schindler}, \citenamefont {Rössler},\ and\ \citenamefont {Home}}]{Auchter2022}%
  \BibitemOpen
  \bibfield  {author} {\bibinfo {author} {\bibfnamefont {S.}~\bibnamefont {Auchter}}, \bibinfo {author} {\bibfnamefont {C.}~\bibnamefont {Axline}}, \bibinfo {author} {\bibfnamefont {C.}~\bibnamefont {Decaroli}}, \bibinfo {author} {\bibfnamefont {M.}~\bibnamefont {Valentini}}, \bibinfo {author} {\bibfnamefont {L.}~\bibnamefont {Purwin}}, \bibinfo {author} {\bibfnamefont {R.}~\bibnamefont {Oswald}}, \bibinfo {author} {\bibfnamefont {R.}~\bibnamefont {Matt}}, \bibinfo {author} {\bibfnamefont {E.}~\bibnamefont {Aschauer}}, \bibinfo {author} {\bibfnamefont {Y.}~\bibnamefont {Colombe}}, \bibinfo {author} {\bibfnamefont {P.}~\bibnamefont {Holz}}, \bibinfo {author} {\bibfnamefont {T.}~\bibnamefont {Monz}}, \bibinfo {author} {\bibfnamefont {R.}~\bibnamefont {Blatt}}, \bibinfo {author} {\bibfnamefont {P.}~\bibnamefont {Schindler}}, \bibinfo {author} {\bibfnamefont {C.}~\bibnamefont {Rössler}},\ and\ \bibinfo {author} {\bibfnamefont {J.}~\bibnamefont {Home}},\ }\bibfield  {title} {\bibinfo {title} {Industrially
  microfabricated ion trap with 1 ev trap depth},\ }\href {https://doi.org/10.1088/2058-9565/ac7072} {\bibfield  {journal} {\bibinfo  {journal} {Quantum Science and Technology}\ }\textbf {\bibinfo {volume} {7}},\ \bibinfo {pages} {035015} (\bibinfo {year} {2022})}\BibitemShut {NoStop}%
\bibitem [{\citenamefont {Ari}\ \emph {et~al.}(2025)\citenamefont {Ari}, \citenamefont {Heng}, \citenamefont {Cavillon}, \citenamefont {Bernier}, \citenamefont {Dussauze},\ and\ \citenamefont {Lancry}}]{Ari2025}%
  \BibitemOpen
  \bibfield  {author} {\bibinfo {author} {\bibfnamefont {J.}~\bibnamefont {Ari}}, \bibinfo {author} {\bibfnamefont {Y.}~\bibnamefont {Heng}}, \bibinfo {author} {\bibfnamefont {M.}~\bibnamefont {Cavillon}}, \bibinfo {author} {\bibfnamefont {M.}~\bibnamefont {Bernier}}, \bibinfo {author} {\bibfnamefont {M.}~\bibnamefont {Dussauze}},\ and\ \bibinfo {author} {\bibfnamefont {M.}~\bibnamefont {Lancry}},\ }\bibfield  {title} {\bibinfo {title} {Overview of laser imprinted refractive index changes and related thermal stability in mid-infrared optical glasses},\ }\href {https://doi.org/10.1016/j.optmat.2025.116985} {\bibfield  {journal} {\bibinfo  {journal} {Optical Materials}\ }\textbf {\bibinfo {volume} {163}},\ \bibinfo {pages} {116985} (\bibinfo {year} {2025})}\BibitemShut {NoStop}%
\bibitem [{\citenamefont {Snyder}\ and\ \citenamefont {Love}(1983)}]{SnyderLove1983}%
  \BibitemOpen
  \bibfield  {author} {\bibinfo {author} {\bibfnamefont {A.~W.}\ \bibnamefont {Snyder}}\ and\ \bibinfo {author} {\bibfnamefont {J.~D.}\ \bibnamefont {Love}},\ }\href@noop {} {\emph {\bibinfo {title} {Optical Waveguide Theory}}}\ (\bibinfo  {publisher} {Chapman and Hall},\ \bibinfo {address} {London},\ \bibinfo {year} {1983})\BibitemShut {NoStop}%
\bibitem [{\citenamefont {Li}\ \emph {et~al.}(2019)\citenamefont {Li}, \citenamefont {Ertorer},\ and\ \citenamefont {Herman}}]{Li2019}%
  \BibitemOpen
  \bibfield  {author} {\bibinfo {author} {\bibfnamefont {J.}~\bibnamefont {Li}}, \bibinfo {author} {\bibfnamefont {E.}~\bibnamefont {Ertorer}},\ and\ \bibinfo {author} {\bibfnamefont {P.~R.}\ \bibnamefont {Herman}},\ }\bibfield  {title} {\bibinfo {title} {Ultrafast laser burst-train filamentation for non-contact scribing of optical glasses},\ }\href {https://doi.org/10.1364/OE.27.025078} {\bibfield  {journal} {\bibinfo  {journal} {Optics Express}\ }\textbf {\bibinfo {volume} {27}},\ \bibinfo {pages} {25078} (\bibinfo {year} {2019})}\BibitemShut {NoStop}%
\bibitem [{\citenamefont {Siegman}(1998)}]{Siegman98}%
  \BibitemOpen
  \bibfield  {author} {\bibinfo {author} {\bibfnamefont {A.~E.}\ \bibnamefont {Siegman}},\ }\bibfield  {title} {\bibinfo {title} {How to (maybe) measure laser beam quality},\ }in\ \href {https://doi.org/10.1364/DLAI.1998.MQ1} {\emph {\bibinfo {booktitle} {DPSS (Diode Pumped Solid State) Lasers: Applications and Issues}}}\ (\bibinfo  {publisher} {Optica Publishing Group},\ \bibinfo {year} {1998})\ p.\ \bibinfo {pages} {MQ1}\BibitemShut {NoStop}%
\bibitem [{\citenamefont {Tan}\ \emph {et~al.}(2020)\citenamefont {Tan}, \citenamefont {Sun}, \citenamefont {Wang}, \citenamefont {Zhou}, \citenamefont {Liao},\ and\ \citenamefont {Qiu}}]{Tan2020}%
  \BibitemOpen
  \bibfield  {author} {\bibinfo {author} {\bibfnamefont {D.}~\bibnamefont {Tan}}, \bibinfo {author} {\bibfnamefont {X.}~\bibnamefont {Sun}}, \bibinfo {author} {\bibfnamefont {Q.}~\bibnamefont {Wang}}, \bibinfo {author} {\bibfnamefont {P.}~\bibnamefont {Zhou}}, \bibinfo {author} {\bibfnamefont {Y.}~\bibnamefont {Liao}},\ and\ \bibinfo {author} {\bibfnamefont {J.}~\bibnamefont {Qiu}},\ }\bibfield  {title} {\bibinfo {title} {Fabricating low loss waveguides over a large depth in glass by temperature gradient assisted femtosecond laser writing},\ }\href {https://doi.org/10.1364/OL.396861} {\bibfield  {journal} {\bibinfo  {journal} {Optics Letters}\ }\textbf {\bibinfo {volume} {45}},\ \bibinfo {pages} {3941} (\bibinfo {year} {2020})}\BibitemShut {NoStop}%
\bibitem [{\citenamefont {Hnatovsky}\ \emph {et~al.}(2005)\citenamefont {Hnatovsky}, \citenamefont {Taylor}, \citenamefont {Simova}, \citenamefont {Bhardwaj}, \citenamefont {Rayner},\ and\ \citenamefont {Corkum}}]{Hnatovsky2005}%
  \BibitemOpen
  \bibfield  {author} {\bibinfo {author} {\bibfnamefont {C.}~\bibnamefont {Hnatovsky}}, \bibinfo {author} {\bibfnamefont {R.~S.}\ \bibnamefont {Taylor}}, \bibinfo {author} {\bibfnamefont {E.}~\bibnamefont {Simova}}, \bibinfo {author} {\bibfnamefont {V.~R.}\ \bibnamefont {Bhardwaj}}, \bibinfo {author} {\bibfnamefont {D.~M.}\ \bibnamefont {Rayner}},\ and\ \bibinfo {author} {\bibfnamefont {P.~B.}\ \bibnamefont {Corkum}},\ }\bibfield  {title} {\bibinfo {title} {High-resolution study of photoinduced modification in fused silica produced by a tightly focused femtosecond laser beam in the presence of aberrations},\ }\href {https://doi.org/10.1063/1.1944223} {\bibfield  {journal} {\bibinfo  {journal} {Journal of Applied Physics}\ }\textbf {\bibinfo {volume} {98}},\ \bibinfo {pages} {013517} (\bibinfo {year} {2005})}\BibitemShut {NoStop}%
\bibitem [{\citenamefont {Huang}\ \emph {et~al.}(2016)\citenamefont {Huang}, \citenamefont {Salter}, \citenamefont {Payne},\ and\ \citenamefont {Booth}}]{Huang2016}%
  \BibitemOpen
  \bibfield  {author} {\bibinfo {author} {\bibfnamefont {L.}~\bibnamefont {Huang}}, \bibinfo {author} {\bibfnamefont {P.~S.}\ \bibnamefont {Salter}}, \bibinfo {author} {\bibfnamefont {F.}~\bibnamefont {Payne}},\ and\ \bibinfo {author} {\bibfnamefont {M.~J.}\ \bibnamefont {Booth}},\ }\bibfield  {title} {\bibinfo {title} {Aberration correction for direct laser written waveguides in a transverse geometry},\ }\href {https://doi.org/10.1364/OE.24.010565} {\bibfield  {journal} {\bibinfo  {journal} {Optics Express}\ }\textbf {\bibinfo {volume} {24}},\ \bibinfo {pages} {10565} (\bibinfo {year} {2016})}\BibitemShut {NoStop}%
\bibitem [{\citenamefont {Bisch}\ \emph {et~al.}(2019)\citenamefont {Bisch}, \citenamefont {Guan}, \citenamefont {Booth},\ and\ \citenamefont {Salter}}]{Bisch2019}%
  \BibitemOpen
  \bibfield  {author} {\bibinfo {author} {\bibfnamefont {N.}~\bibnamefont {Bisch}}, \bibinfo {author} {\bibfnamefont {J.}~\bibnamefont {Guan}}, \bibinfo {author} {\bibfnamefont {M.~J.}\ \bibnamefont {Booth}},\ and\ \bibinfo {author} {\bibfnamefont {P.~S.}\ \bibnamefont {Salter}},\ }\bibfield  {title} {\bibinfo {title} {Adaptive optics aberration correction for deep direct laser written waveguides in the heating regime},\ }\href {https://doi.org/10.1007/s00339-019-2635-4} {\bibfield  {journal} {\bibinfo  {journal} {Applied Physics A}\ }\textbf {\bibinfo {volume} {125}},\ \bibinfo {pages} {364} (\bibinfo {year} {2019})}\BibitemShut {NoStop}%
\bibitem [{\citenamefont {Ferreira}\ \emph {et~al.}(2021)\citenamefont {Ferreira}, \citenamefont {Almeida},\ and\ \citenamefont {Mendon\c{c}a}}]{Ferreira2021}%
  \BibitemOpen
  \bibfield  {author} {\bibinfo {author} {\bibfnamefont {P.~H.~D.}\ \bibnamefont {Ferreira}}, \bibinfo {author} {\bibfnamefont {G.~F.~B.}\ \bibnamefont {Almeida}},\ and\ \bibinfo {author} {\bibfnamefont {C.~R.}\ \bibnamefont {Mendon\c{c}a}},\ }\bibfield  {title} {\bibinfo {title} {A simple strategy for increasing optical waveguide performance using spherical aberration},\ }\href {https://doi.org/10.1016/j.optlastec.2021.107235} {\bibfield  {journal} {\bibinfo  {journal} {Optics and Laser Technology}\ }\textbf {\bibinfo {volume} {142}},\ \bibinfo {pages} {107235} (\bibinfo {year} {2021})}\BibitemShut {NoStop}%
\bibitem [{\citenamefont {Eaton}\ \emph {et~al.}(2008{\natexlab{a}})\citenamefont {Eaton}, \citenamefont {Ng}, \citenamefont {Bonse}, \citenamefont {Mermillod-Blondin}, \citenamefont {Zhang}, \citenamefont {Rosenfeld},\ and\ \citenamefont {Herman}}]{Eaton2008a}%
  \BibitemOpen
  \bibfield  {author} {\bibinfo {author} {\bibfnamefont {S.~M.}\ \bibnamefont {Eaton}}, \bibinfo {author} {\bibfnamefont {M.~L.}\ \bibnamefont {Ng}}, \bibinfo {author} {\bibfnamefont {J.}~\bibnamefont {Bonse}}, \bibinfo {author} {\bibfnamefont {A.}~\bibnamefont {Mermillod-Blondin}}, \bibinfo {author} {\bibfnamefont {H.}~\bibnamefont {Zhang}}, \bibinfo {author} {\bibfnamefont {A.}~\bibnamefont {Rosenfeld}},\ and\ \bibinfo {author} {\bibfnamefont {P.~R.}\ \bibnamefont {Herman}},\ }\bibfield  {title} {\bibinfo {title} {Low-loss waveguides fabricated in bk7 glass by high repetition rate femtosecond fiber laser},\ }\href {https://doi.org/10.1364/AO.47.002098} {\bibfield  {journal} {\bibinfo  {journal} {Applied Optics}\ }\textbf {\bibinfo {volume} {47}},\ \bibinfo {pages} {2098} (\bibinfo {year} {2008}{\natexlab{a}})}\BibitemShut {NoStop}%
\bibitem [{\citenamefont {Eaton}\ \emph {et~al.}(2008{\natexlab{b}})\citenamefont {Eaton}, \citenamefont {Zhang}, \citenamefont {Ng}, \citenamefont {Li}, \citenamefont {Chen}, \citenamefont {Ho},\ and\ \citenamefont {Herman}}]{Eaton2008b}%
  \BibitemOpen
  \bibfield  {author} {\bibinfo {author} {\bibfnamefont {S.~M.}\ \bibnamefont {Eaton}}, \bibinfo {author} {\bibfnamefont {H.}~\bibnamefont {Zhang}}, \bibinfo {author} {\bibfnamefont {M.~L.}\ \bibnamefont {Ng}}, \bibinfo {author} {\bibfnamefont {J.}~\bibnamefont {Li}}, \bibinfo {author} {\bibfnamefont {W.-J.}\ \bibnamefont {Chen}}, \bibinfo {author} {\bibfnamefont {S.}~\bibnamefont {Ho}},\ and\ \bibinfo {author} {\bibfnamefont {P.~R.}\ \bibnamefont {Herman}},\ }\bibfield  {title} {\bibinfo {title} {Transition from thermal diffusion to heat accumulation in high repetition rate femtosecond laser writing of buried optical waveguides},\ }\href {https://doi.org/10.1364/OE.16.009443} {\bibfield  {journal} {\bibinfo  {journal} {Optics Express}\ }\textbf {\bibinfo {volume} {16}},\ \bibinfo {pages} {9443} (\bibinfo {year} {2008}{\natexlab{b}})}\BibitemShut {NoStop}%
\bibitem [{\citenamefont {Chen}\ \emph {et~al.}(2008)\citenamefont {Chen}, \citenamefont {Eaton}, \citenamefont {Zhang},\ and\ \citenamefont {Herman}}]{Chen2008}%
  \BibitemOpen
  \bibfield  {author} {\bibinfo {author} {\bibfnamefont {W.-J.}\ \bibnamefont {Chen}}, \bibinfo {author} {\bibfnamefont {S.~M.}\ \bibnamefont {Eaton}}, \bibinfo {author} {\bibfnamefont {H.}~\bibnamefont {Zhang}},\ and\ \bibinfo {author} {\bibfnamefont {P.~R.}\ \bibnamefont {Herman}},\ }\bibfield  {title} {\bibinfo {title} {Broadband directional couplers fabricated in bulk glass with high repetition rate femtosecond laser pulses},\ }\href {https://doi.org/10.1364/OE.16.011470} {\bibfield  {journal} {\bibinfo  {journal} {Optics Express}\ }\textbf {\bibinfo {volume} {16}},\ \bibinfo {pages} {11470} (\bibinfo {year} {2008})}\BibitemShut {NoStop}%
\bibitem [{\citenamefont {Marcuse}(1978)}]{Marcuse1978}%
  \BibitemOpen
  \bibfield  {author} {\bibinfo {author} {\bibfnamefont {D.}~\bibnamefont {Marcuse}},\ }\bibfield  {title} {\bibinfo {title} {Gaussian approximation of the fundamental modes of graded-index fibers},\ }\href {https://doi.org/10.1364/JOSA.68.000103} {\bibfield  {journal} {\bibinfo  {journal} {Journal of the Optical Society of America}\ }\textbf {\bibinfo {volume} {68}},\ \bibinfo {pages} {103} (\bibinfo {year} {1978})}\BibitemShut {NoStop}%
\bibitem [{\citenamefont {Marcuse}(1976)}]{Marcuse1976}%
  \BibitemOpen
  \bibfield  {author} {\bibinfo {author} {\bibfnamefont {D.}~\bibnamefont {Marcuse}},\ }\bibfield  {title} {\bibinfo {title} {Curvature loss formula for optical fibers},\ }\href {https://doi.org/10.1364/JOSA.66.000216} {\bibfield  {journal} {\bibinfo  {journal} {Journal of the Optical Society of America}\ }\textbf {\bibinfo {volume} {66}},\ \bibinfo {pages} {216} (\bibinfo {year} {1976})}\BibitemShut {NoStop}%
\bibitem [{\citenamefont {Arriola}\ \emph {et~al.}(2013)\citenamefont {Arriola}, \citenamefont {Gross}, \citenamefont {Jovanovic}, \citenamefont {Charles}, \citenamefont {Tuthill}, \citenamefont {Olaizola}, \citenamefont {Fuerbach},\ and\ \citenamefont {Withford}}]{Arriola2013}%
  \BibitemOpen
  \bibfield  {author} {\bibinfo {author} {\bibfnamefont {A.}~\bibnamefont {Arriola}}, \bibinfo {author} {\bibfnamefont {S.}~\bibnamefont {Gross}}, \bibinfo {author} {\bibfnamefont {N.}~\bibnamefont {Jovanovic}}, \bibinfo {author} {\bibfnamefont {N.}~\bibnamefont {Charles}}, \bibinfo {author} {\bibfnamefont {P.~G.}\ \bibnamefont {Tuthill}}, \bibinfo {author} {\bibfnamefont {S.~M.}\ \bibnamefont {Olaizola}}, \bibinfo {author} {\bibfnamefont {A.}~\bibnamefont {Fuerbach}},\ and\ \bibinfo {author} {\bibfnamefont {M.~J.}\ \bibnamefont {Withford}},\ }\bibfield  {title} {\bibinfo {title} {Low bend loss waveguides enable compact, efficient 3d photonic chips},\ }\href {https://doi.org/10.1364/OE.21.029785} {\bibfield  {journal} {\bibinfo  {journal} {Optics Express}\ }\textbf {\bibinfo {volume} {21}},\ \bibinfo {pages} {29785} (\bibinfo {year} {2013})}\BibitemShut {NoStop}%
\bibitem [{\citenamefont {Dyakonov}\ \emph {et~al.}(2016)\citenamefont {Dyakonov}, \citenamefont {Vetchinnikov},\ and\ \citenamefont {Lipovskii}}]{Dyakonov2016}%
  \BibitemOpen
  \bibfield  {author} {\bibinfo {author} {\bibfnamefont {I.~V.}\ \bibnamefont {Dyakonov}}, \bibinfo {author} {\bibfnamefont {S.~A.}\ \bibnamefont {Vetchinnikov}},\ and\ \bibinfo {author} {\bibfnamefont {A.~A.}\ \bibnamefont {Lipovskii}},\ }\bibfield  {title} {\bibinfo {title} {Low-loss single-mode integrated waveguides in soda-lime glass},\ }\href {https://doi.org/10.1364/OL.41.004498} {\bibfield  {journal} {\bibinfo  {journal} {Optics Letters}\ }\textbf {\bibinfo {volume} {41}},\ \bibinfo {pages} {4498} (\bibinfo {year} {2016})}\BibitemShut {NoStop}%
\bibitem [{\citenamefont {Wang}\ \emph {et~al.}(2024)\citenamefont {Wang}, \citenamefont {Zhong}, \citenamefont {Lau}, \citenamefont {Han}, \citenamefont {Yang}, \citenamefont {Hu}, \citenamefont {Firstov}, \citenamefont {Chen}, \citenamefont {Ma}, \citenamefont {Tong}, \citenamefont {Chiang}, \citenamefont {Tan},\ and\ \citenamefont {Qiu}}]{Wang2024}%
  \BibitemOpen
  \bibfield  {author} {\bibinfo {author} {\bibfnamefont {Y.}~\bibnamefont {Wang}}, \bibinfo {author} {\bibfnamefont {L.}~\bibnamefont {Zhong}}, \bibinfo {author} {\bibfnamefont {K.~Y.}\ \bibnamefont {Lau}}, \bibinfo {author} {\bibfnamefont {X.}~\bibnamefont {Han}}, \bibinfo {author} {\bibfnamefont {Y.}~\bibnamefont {Yang}}, \bibinfo {author} {\bibfnamefont {J.}~\bibnamefont {Hu}}, \bibinfo {author} {\bibfnamefont {S.}~\bibnamefont {Firstov}}, \bibinfo {author} {\bibfnamefont {Z.}~\bibnamefont {Chen}}, \bibinfo {author} {\bibfnamefont {Z.}~\bibnamefont {Ma}}, \bibinfo {author} {\bibfnamefont {L.}~\bibnamefont {Tong}}, \bibinfo {author} {\bibfnamefont {K.~S.}\ \bibnamefont {Chiang}}, \bibinfo {author} {\bibfnamefont {D.}~\bibnamefont {Tan}},\ and\ \bibinfo {author} {\bibfnamefont {J.}~\bibnamefont {Qiu}},\ }\bibfield  {title} {\bibinfo {title} {Precise mode control of laser-written waveguides for broadband, low-dispersion 3d integrated optics},\ }\href {https://doi.org/10.1038/s41377-024-01372-7} {\bibfield
  {journal} {\bibinfo  {journal} {Light: Science \& Applications}\ }\textbf {\bibinfo {volume} {13}},\ \bibinfo {pages} {31} (\bibinfo {year} {2024})}\BibitemShut {NoStop}%
\bibitem [{\citenamefont {Ross-Adams}\ \emph {et~al.}(2024)\citenamefont {Ross-Adams}, \citenamefont {Mills}, \citenamefont {Withford},\ and\ \citenamefont {Gross}}]{RossAdams2024}%
  \BibitemOpen
  \bibfield  {author} {\bibinfo {author} {\bibfnamefont {A.}~\bibnamefont {Ross-Adams}}, \bibinfo {author} {\bibfnamefont {B.}~\bibnamefont {Mills}}, \bibinfo {author} {\bibfnamefont {M.~J.}\ \bibnamefont {Withford}},\ and\ \bibinfo {author} {\bibfnamefont {S.}~\bibnamefont {Gross}},\ }\bibfield  {title} {\bibinfo {title} {Low bend loss, high index, composite morphology ultra-fast laser written waveguides for photonic integrated circuits},\ }\href {https://doi.org/10.37188/lam.2024.009} {\bibfield  {journal} {\bibinfo  {journal} {Light: Advanced Manufacturing}\ }\textbf {\bibinfo {volume} {5}},\ \bibinfo {pages} {9} (\bibinfo {year} {2024})}\BibitemShut {NoStop}%
\bibitem [{\citenamefont {Anmasser}\ \emph {et~al.}(2026)\citenamefont {Anmasser}, \citenamefont {Zahra}, \citenamefont {Schüppert}, \citenamefont {Pototschnig}, \citenamefont {Wahl}, \citenamefont {Dietl}, \citenamefont {Pfeifer}, \citenamefont {Colombe}, \citenamefont {Repp}, \citenamefont {Brandl}, \citenamefont {Schindler},\ and\ \citenamefont {Rössler}}]{Anmasser2026}%
  \BibitemOpen
  \bibfield  {author} {\bibinfo {author} {\bibfnamefont {F.}~\bibnamefont {Anmasser}}, \bibinfo {author} {\bibfnamefont {M.~A.}\ \bibnamefont {Zahra}}, \bibinfo {author} {\bibfnamefont {K.}~\bibnamefont {Schüppert}}, \bibinfo {author} {\bibfnamefont {M.}~\bibnamefont {Pototschnig}}, \bibinfo {author} {\bibfnamefont {J.}~\bibnamefont {Wahl}}, \bibinfo {author} {\bibfnamefont {M.}~\bibnamefont {Dietl}}, \bibinfo {author} {\bibfnamefont {M.}~\bibnamefont {Pfeifer}}, \bibinfo {author} {\bibfnamefont {Y.}~\bibnamefont {Colombe}}, \bibinfo {author} {\bibfnamefont {J.}~\bibnamefont {Repp}}, \bibinfo {author} {\bibfnamefont {M.}~\bibnamefont {Brandl}}, \bibinfo {author} {\bibfnamefont {P.}~\bibnamefont {Schindler}},\ and\ \bibinfo {author} {\bibfnamefont {C.}~\bibnamefont {Rössler}},\ }\href {https://arxiv.org/abs/2605.16010} {\bibinfo {title} {Demonstration of a multiplexing trapped ion quantum processing unit}} (\bibinfo {year} {2026}),\ \Eprint {https://arxiv.org/abs/2605.16010} {arXiv:2605.16010 [quant-ph]}
  \BibitemShut {NoStop}%
\bibitem [{\citenamefont {Teller}\ \emph {et~al.}(2021)\citenamefont {Teller}, \citenamefont {Fioretto}, \citenamefont {Holz}, \citenamefont {Schindler}, \citenamefont {Messerer}, \citenamefont {Sch\"uppert}, \citenamefont {Zou}, \citenamefont {Blatt}, \citenamefont {Chiaverini}, \citenamefont {Sage},\ and\ \citenamefont {Northup}}]{Teller2021}%
  \BibitemOpen
  \bibfield  {author} {\bibinfo {author} {\bibfnamefont {M.}~\bibnamefont {Teller}}, \bibinfo {author} {\bibfnamefont {D.~A.}\ \bibnamefont {Fioretto}}, \bibinfo {author} {\bibfnamefont {P.~C.}\ \bibnamefont {Holz}}, \bibinfo {author} {\bibfnamefont {P.}~\bibnamefont {Schindler}}, \bibinfo {author} {\bibfnamefont {V.}~\bibnamefont {Messerer}}, \bibinfo {author} {\bibfnamefont {K.}~\bibnamefont {Sch\"uppert}}, \bibinfo {author} {\bibfnamefont {Y.}~\bibnamefont {Zou}}, \bibinfo {author} {\bibfnamefont {R.}~\bibnamefont {Blatt}}, \bibinfo {author} {\bibfnamefont {J.}~\bibnamefont {Chiaverini}}, \bibinfo {author} {\bibfnamefont {J.}~\bibnamefont {Sage}},\ and\ \bibinfo {author} {\bibfnamefont {T.~E.}\ \bibnamefont {Northup}},\ }\bibfield  {title} {\bibinfo {title} {Heating of a trapped ion induced by dielectric materials},\ }\href {https://doi.org/10.1103/PhysRevLett.126.230505} {\bibfield  {journal} {\bibinfo  {journal} {Phys. Rev. Lett.}\ }\textbf {\bibinfo {volume} {126}},\ \bibinfo {pages} {230505} (\bibinfo
  {year} {2021})}\BibitemShut {NoStop}%
\bibitem [{\citenamefont {Freund}\ \emph {et~al.}(2024)\citenamefont {Freund}, \citenamefont {Marciniak},\ and\ \citenamefont {Monz}}]{Freund2024}%
  \BibitemOpen
  \bibfield  {author} {\bibinfo {author} {\bibfnamefont {R.}~\bibnamefont {Freund}}, \bibinfo {author} {\bibfnamefont {C.~D.}\ \bibnamefont {Marciniak}},\ and\ \bibinfo {author} {\bibfnamefont {T.}~\bibnamefont {Monz}},\ }\bibfield  {title} {\bibinfo {title} {A self-referenced optical phase noise analyzer for quantum technologies},\ }\href {https://doi.org/10.1063/5.0183502} {\bibfield  {journal} {\bibinfo  {journal} {Review of Scientific Instruments}\ }\textbf {\bibinfo {volume} {95}},\ \bibinfo {pages} {063005} (\bibinfo {year} {2024})}\BibitemShut {NoStop}%
\bibitem [{\citenamefont {Leibfried}\ \emph {et~al.}(2003)\citenamefont {Leibfried}, \citenamefont {Blatt}, \citenamefont {Monroe},\ and\ \citenamefont {Wineland}}]{Leibfried2003}%
  \BibitemOpen
  \bibfield  {author} {\bibinfo {author} {\bibfnamefont {D.}~\bibnamefont {Leibfried}}, \bibinfo {author} {\bibfnamefont {R.}~\bibnamefont {Blatt}}, \bibinfo {author} {\bibfnamefont {C.}~\bibnamefont {Monroe}},\ and\ \bibinfo {author} {\bibfnamefont {D.}~\bibnamefont {Wineland}},\ }\bibfield  {title} {\bibinfo {title} {Quantum dynamics of single trapped ions},\ }\href {https://doi.org/10.1103/RevModPhys.75.281} {\bibfield  {journal} {\bibinfo  {journal} {Rev. Mod. Phys.}\ }\textbf {\bibinfo {volume} {75}},\ \bibinfo {pages} {281} (\bibinfo {year} {2003})}\BibitemShut {NoStop}%
\bibitem [{\citenamefont {Allcock}\ \emph {et~al.}(2010)\citenamefont {Allcock}, \citenamefont {Sherman}, \citenamefont {Stacey}, \citenamefont {Burrell}, \citenamefont {Curtis}, \citenamefont {Imreh}, \citenamefont {Linke}, \citenamefont {Szwer}, \citenamefont {Webster}, \citenamefont {Steane},\ and\ \citenamefont {Lucas}}]{Allcock2010}%
  \BibitemOpen
  \bibfield  {author} {\bibinfo {author} {\bibfnamefont {D.~T.~C.}\ \bibnamefont {Allcock}}, \bibinfo {author} {\bibfnamefont {J.~A.}\ \bibnamefont {Sherman}}, \bibinfo {author} {\bibfnamefont {D.~N.}\ \bibnamefont {Stacey}}, \bibinfo {author} {\bibfnamefont {A.~H.}\ \bibnamefont {Burrell}}, \bibinfo {author} {\bibfnamefont {M.~J.}\ \bibnamefont {Curtis}}, \bibinfo {author} {\bibfnamefont {G.}~\bibnamefont {Imreh}}, \bibinfo {author} {\bibfnamefont {N.~M.}\ \bibnamefont {Linke}}, \bibinfo {author} {\bibfnamefont {D.~J.}\ \bibnamefont {Szwer}}, \bibinfo {author} {\bibfnamefont {S.~C.}\ \bibnamefont {Webster}}, \bibinfo {author} {\bibfnamefont {A.~M.}\ \bibnamefont {Steane}},\ and\ \bibinfo {author} {\bibfnamefont {D.~M.}\ \bibnamefont {Lucas}},\ }\bibfield  {title} {\bibinfo {title} {Implementation of a symmetric surface-electrode ion trap with field compensation using a modulated raman effect},\ }\href {https://doi.org/10.1088/1367-2630/12/5/053026} {\bibfield  {journal} {\bibinfo  {journal} {New Journal of
  Physics}\ }\textbf {\bibinfo {volume} {12}},\ \bibinfo {pages} {053026} (\bibinfo {year} {2010})}\BibitemShut {NoStop}%
\bibitem [{\citenamefont {Oliveira}\ and\ \citenamefont {Miranda}(2001)}]{Oliveira_2001}%
  \BibitemOpen
  \bibfield  {author} {\bibinfo {author} {\bibfnamefont {M.~H.}\ \bibnamefont {Oliveira}}\ and\ \bibinfo {author} {\bibfnamefont {J.~A.}\ \bibnamefont {Miranda}},\ }\bibfield  {title} {\bibinfo {title} {Biot-savart-like law in electrostatics},\ }\href {https://doi.org/10.1088/0143-0807/22/1/304} {\bibfield  {journal} {\bibinfo  {journal} {European Journal of Physics}\ }\textbf {\bibinfo {volume} {22}},\ \bibinfo {pages} {31} (\bibinfo {year} {2001})}\BibitemShut {NoStop}%
\bibitem [{\citenamefont {Dehmelt}(1968)}]{Dehmelt1968}%
  \BibitemOpen
  \bibfield  {author} {\bibinfo {author} {\bibfnamefont {H.}~\bibnamefont {Dehmelt}},\ }\bibfield  {title} {\bibinfo {title} {Radiofrequency spectroscopy of stored ions i: Storage**part ii: Spectroscopy is now scheduled to appear in volume v of this series.}\ }(\bibinfo  {publisher} {Academic Press},\ \bibinfo {year} {1968})\ pp.\ \bibinfo {pages} {53--72}\BibitemShut {NoStop}%
\end{thebibliography}%
\end{document}